\documentclass[12pt,preprint]{aastex} 

\newcommand*{\logg}{$\log~g$}
\newcommand*{\feh}{[Fe/H]}

\newcommand*{\teff}{$T_{\rm eff}$}

\newcommand*{\gr}{$(g-r)_{0}$}

\usepackage{graphics,graphicx,amsmath,amssymb,amsthm,epsfig}

\shorttitle{The SEGUE Stellar Parameter Pipeline. IV.}

\shortauthors{Smolinski et al.}

\begin{document}

\title{The SEGUE Stellar Parameter Pipeline. IV. Validation with an Extended 
Sample of Galactic Globular and Open Clusters}

\author{Jason P. Smolinski, Young Sun Lee, Timothy C. Beers}
\affil{Department of Physics \& Astronomy and JINA: Joint Institute for Nuclear
Astrophysics, Michigan State University, East Lansing, MI 48824,
USA} \email{smolin19@msu.edu, lee@pa.msu.edu, beers@pa.msu.edu}

\author{Deokkeun An}
\affil{Department of Science Education, Ewha Womans
University, Seoul 120-750, Korea}
\email{deokkeun@ewha.ac.kr}

\author{Steven J. Bickerton}
\affil{Department of Astrophysical Sciences, Princeton University, \\
Princeton, NJ 08544, USA}
\email{bick@astro.princeton.edu}

\author{Jennifer A. Johnson}
\affil{Department of Astronomy, Ohio State University, 140 West 18th
Avenue, Columbus, OH 43210, USA}
\email{jaj@astronomy.ohio-state.edu}

\author{Craig P. Loomis}
\affil{Department of Astrophysical Sciences, Princeton University, \\
Princeton, NJ 08544, USA}
\email{cloomis@astro.princeton.edu}


\author{Constance M. Rockosi}
\affil{UCO/Lick Observatory, University of California, \\
Santa Cruz, CA 95064, USA}
\email{crockosi@ucolick.org}

\author{Thirupathi Sivarani}
\affil{IIAP: Indian Institute of Astrophysics, \\
II Block, Koramangala, Bangalore 560 034, India}
\email{sivarani@iiap.res.in}

\author{Brian Yanny}
\affil{Fermi National Accelerator Laboratory, \\
P.O. Box 500, Batavia, IL 60510, USA}
\email{yanny@fnal.gov}

\begin{abstract}

Spectroscopic and photometric data for likely member stars of five
Galactic globular clusters (M3, M53, M71, M92, and NGC~5053) and
three open clusters (M35, NGC~2158, and NGC~6791) are
processed by the current version of the SEGUE Stellar Parameter
Pipeline (SSPP), in order to determine estimates of metallicities and
radial velocities for the clusters. These results are then compared to
values from the literature. We find that the mean metallicity
($\langle\rm{[Fe/H]}\rangle$) and mean radial velocity ($\langle {\rm
RV}\rangle$) estimates for each cluster are almost all within
$2\sigma$ of the adopted literature values; most are within $1\sigma$.
We also demonstrate that the new version of the SSPP achieves small, but
noteworthy, improvements in $\langle\rm{[Fe/H]}\rangle$ estimates at the
extrema of the cluster metallicity range, as compared to a previous
version of the pipeline software. These results provide additional
confidence in the application of the SSPP for studies of the abundances
and kinematics of stellar populations in the Galaxy.

\end{abstract}


\section{Introduction}\label{secintro}

The Sloan Digital Sky Survey (SDSS), and its extensions, have now obtained
\emph{ugriz} photometry for several hundred million stars (through
DR7; see Abazajian et al. 2009). The Sloan Extension for Galactic
Understanding and Exploration (SEGUE; Yanny et al. 2009), one of three
sub-surveys that collectively formed SDSS-II, obtained $ugriz$ imaging
of some 3500 deg$^{2}$ of sky outside of the SDSS-I footprint
(Fukugita et al. 1996; Gunn et al. 1998, 2006; Stoughton et al. 2002;
Abazajian et al. 2003, 2004, 2005, 2009; Pier et al. 2003; Adelman-McCarthy
et al. 2006, 2007, 2008), with special attention being given to scans
of lower Galactic latitudes ($|b|$ $<$ 35$^{\circ}$) in order to
better probe the disk/halo interface of the Milky Way. SEGUE also
obtained $R$ $\simeq$ 2000 spectroscopy over the wavelength range 
3800$-$9200\,{\AA} for some 240,000 stars in 200 selected areas over the
sky available from Apache Point, New Mexico. When combined with stars
observed during SDSS-I, and the recently completed SEGUE-2 project
within SDSS-III, a total of nearly 500,000 stars exploring the
thin-disk, thick-disk, and halo populations of the Galaxy now have
similar data.

The SEGUE Stellar Parameter Pipeline (SSPP; Lee et al. 2008a,b; Allende
Prieto et al. 2008) processes the wavelength- and flux-calibrated
spectra generated by the standard SDSS spectroscopic reduction pipeline
(Stoughton et al. 2002), obtains equivalent widths and/or line indices
for 85 atomic or molecular absorption lines, and estimates $T_{\rm
eff}$, log $g$, and [Fe/H], along with radial velocities, through the
application of a number of approaches (see Lee et al. 2008a, hereafter
Paper I, for a detailed discussion of the techniques employed by the
SSPP; the appendix of the present paper describes recent changes in the
SSPP). 

A previous validation paper by Lee et al. (2008b, hereafter Paper II)
demonstrated, on the basis of comparisons with a sample of three
Galactic globular clusters (GCs) and two open clusters (OCs), that the SSPP provides
sufficiently accurate estimates of stellar parameters for use in the
analysis of Galactic kinematics and chemistry, at least over the ranges
in parameter space covered by these clusters (in particular, for the
metallicity range $-2.4 <$ [Fe/H] $< 0.0$). However, it was noted in
that paper that the largest outliers in SSPP-derived metallicities were
found for clusters near the extrema of this range. The team of
researchers working on the SSPP have, in the time since publication of
the original validation paper, endeavored to improve the performance of
the SSPP near these extremes. As part of this effort, which is leading
to the production of a version of the SSPP suitable for application to
the DR8 release of results from SDSS-III (including the $\sim$120,000
stars observed during SEGUE-2), we have assembled SDSS photometry and
spectroscopy for an additional sample of five GCs
(including two with [Fe/H] $\sim -2.3$: M92 and NGC~5053, and one
intermediate-metallicity cluster with [Fe/H] $\sim -0.7$: M71), and three
OCs, one of which has been shown in the literature to exhibit
a super-solar metallicity, [Fe/H] $= +0.3$ (NGC~6791). 

This paper, Paper IV in the series describing and testing the SSPP, examines the
derived stellar parameters for our newly added clusters as well as for the
previously reported sample of clusters, based on the most recent version of the
SSPP. From this exercise, it is clear that the low-metallicity behavior of the
SSPP has improved, and that the SSPP is also now capable of obtaining acceptable
parameter estimates for stars up to solar metallicity, or slightly above.
Section \ref{secsample} describes the photometric and spectroscopic data for the
eight clusters in our sample. The procedures for selecting likely true member
stars in each cluster from among stars in the field are described in Section
\ref{secselection}. Section
\ref{secoverall_det} discusses the determination of $\langle\rm{[Fe/H]}\rangle$ and
$\langle\rm{RV}\rangle$ estimates from the selected true member stars;
these are compared to the values obtained by previous studies in Section
\ref{seccomp}. We then process the five clusters from Paper II through
the current version of the SSPP, and compare the results and
improvements in Section \ref{secdrcomp}. Section
\ref{secsummary} provides a summary of our results.  An appendix describes
the changes made in the SSPP since the previous version was released (and used
for stellar parameter estimates in DR7). The present version of the SSPP should
be very similar to that employed for the estimation of stellar
parameters for stellar spectra in the next public release, DR8. 

\section{The Sample}\label{secsample}

We selected five Galactic GCs (M3, M53, M71, M92, and
NGC~5053) and three OCs (M35, NGC~2158, and NGC~6791) which had
already been observed by SDSS and processed by the SSPP. A number of other
clusters were considered, but ultimately had to be rejected due to difficulties
obtaining adequately reduced spectra from fields that were either too crowded or
too heavily reddened. Because the default PHOTO pipeline (Lupton et al. 2001)
was not designed to accurately deal with crowded fields such as those in the
central regions of globular clusters, crowded-field photometric measurements
were obtained using the DAOPHOT/ALLFRAME software package (Stetson 1987; Stetson
1994) for M3, M53, M71, M92, NGC~5053, and NGC~6791 (An et al. 2008). For
the remaining clusters (M35 and NGC~2158) we followed the same procedures as in
An et al. (2008) to obtain crowded-field photometry. Combining the SDSS
photometry of the full field with the crowded-field photometry of the inner
cluster regions, corrected for reddening and extinction using values listed
in Table \ref{tabclusterprops}, resulted in a nearly complete catalog of $ugriz$
photometry for the stars in each cluster region. Table
\ref{tabclusterprops} summarizes the properties of each cluster included in this
study. Metallicity values from the compilation of Harris (1996) are tabulated
as well as values from the recalibrated metallicity scale of 
Carretta et al. (2009).  

The spectroscopic data was obtained during SEGUE observations using the ARC 2.5m
telescope, with stars targeted for spectroscopic follow-up selected from a
photometric color-magnitude diagram (CMD) for each cluster. Stars located on the
diagram in the regions of the main sequence turn-off (MSTO) and red giant branch
(RGB) were then selected as possible cluster members. Other stars in the field
of each cluster were also selected by the default SEGUE target selection
algorithm to fill each plug-plate, many of which ended up being cluster members
themselves. Overall, SDSS spectroscopic data was obtained for 640 targets each
in the regions of M3, M53, and NGC~5053, and 1280 targets each in the regions
of M35, M71, M92, NGC~2158, and NGC~6791, including sky spectra and
calibration objects. Some of these targets had low average signal-to-noise
spectra; for consistency with previous papers in this series, only those spectra
with $\langle \rm{S/N}\rangle > 10/1$ were considered for subsequent analysis.
After processing by the SSPP some targets had no estimates for RV or [Fe/H];
these were excluded as well. After these cuts were made, there remained 487,
495, 579, 1094, 775, 495, 579, and 1087 stars considered for M3, M35, M53,
M71, M92, NGC~2158, NGC~5053, and NGC~6791, respectively.

\section{Cluster Membership Selection}\label{secselection}

Paper I has shown that the stellar spectra processed through the
SSPP have typical uncertainties of 141~K, 0.23 dex, and 0.23 dex for $T_{\rm
eff}$, log $g$, and [Fe/H], respectively. Uncertainties in the radial
velocity depends on the spectral type and apparent magnitude
(and fall in the range 5-20 km s$^{-1}$; for most of the cluster stars
the error is usually much less than 10 km s$^{-1}$.  In this section we discuss
how the adopted true members for each cluster are selected, based in part on their
estimated metallicities and radial velocities.

\subsection{Likely Member Star Selection}\label{subsec_likely_sel}

The procedure for determining the likely members of each cluster is the same as
described by Paper II, and will only be discussed briefly here. Two procedures
were designed for selecting likely true member stars, one for globular clusters
and one for open clusters. The difference is primarily due to the lower number
density of stars on the CMD of an open cluster compared to that of a globular
cluster. However, the techniques are sufficiently different that, due to the
highly evolved nature of NGC~2158 and NGC~6791, the procedure for open clusters
could not be applied to these particular clusters because it relies on a
function fit to the main stellar locus which, in these cases, would be
double-valued around the main-sequence turnoff. Hence, we have employed the
procedure for globular clusters to the open clusters NGC~2158 and NGC~6791, and
describe specific reasons for having done so where applicable.

\begin{figure}
\begin{center}
\plotone{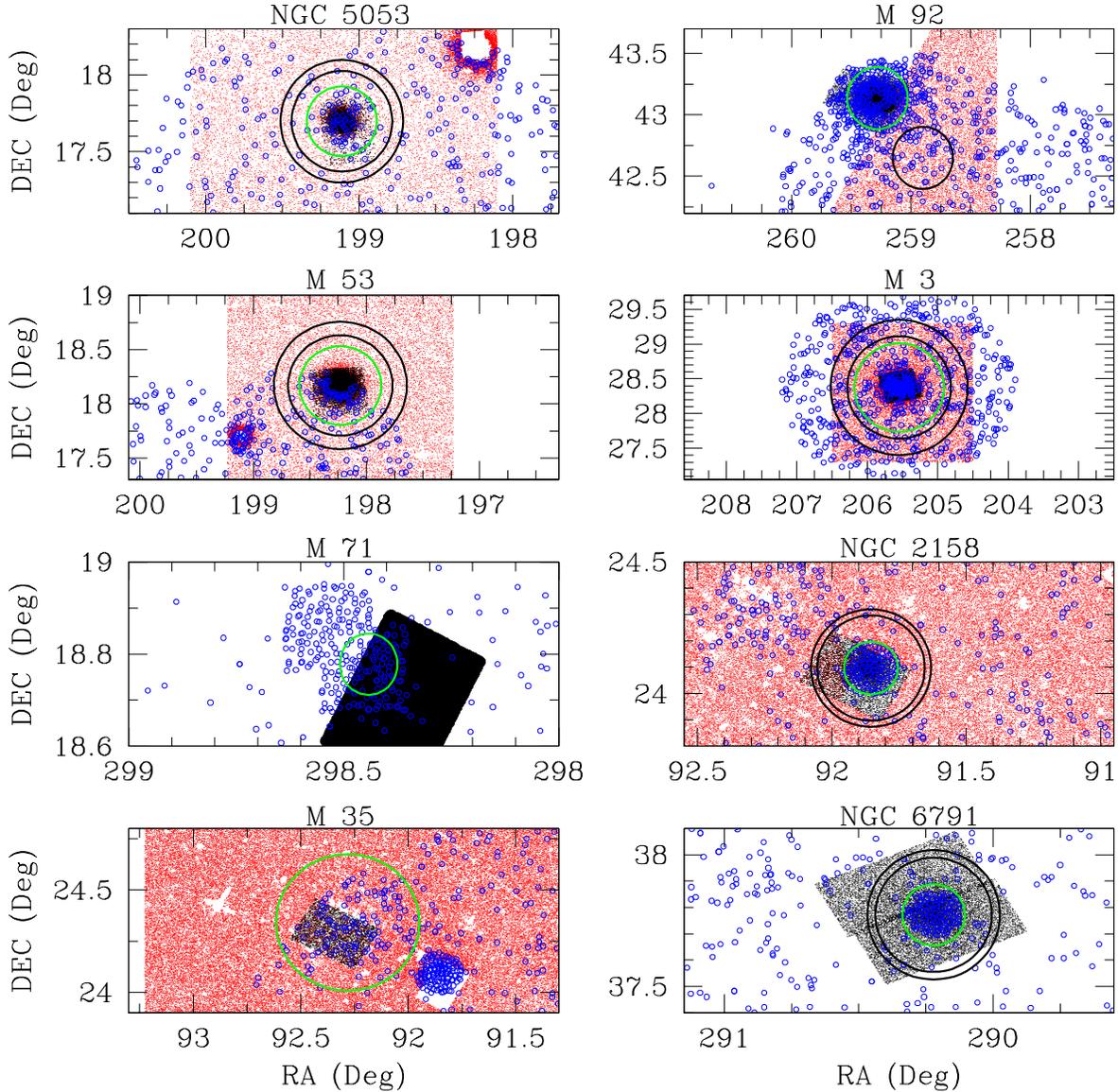}
\caption{ 
Stars with available photometry in the fields of NGC~5053, M92, M53, M3, M71,
NGC~2158, M35, and NGC~6791. The black dots are stars from the crowded-field 
photometric analysis, the red dots are stars with photometry from
the SDSS PHOTO pipeline, and the blue open circles are stars with SDSS
spectroscopy. The green circle is the cluster's tidal radius
(taken here as the cluster region) and the annulus
between the two black circles constitutes the field region.
The green circles are $13.67^{'}, 15.17^{'}, 21.75^{'}, 38.19^{'},
4.0^{'}, 6.0^{'}, 20.0^{'},$ and $7.0^{'}$
in radius, respectively. 
In the
case of M92, the cluster's proximity to the edge of the scan prevented
an adequate annular field region; it was taken adjacent to the cluster
region.  NGC~2158 and NGC~6791 are open
clusters, but due to their evolved nature, they are treated the same as
globular clusters for the identification of likely true members. 
A larger radius was used for these clusters than those listed by Dias et
al. (2002), in order to include as many member stars as possible.}\label{fig8radec}
\end{center}
\end{figure}

Due to the limited number of stars with spectroscopic data, it was necessary to
use the photometry to produce a well-defined CMD, over which the spectroscopic
data were then plotted. The stars inside each cluster's tidal radius (r$_t$)
were selected as the first cut of likely members, indicated by the green circles
in Figure \ref{fig8radec}. Stars inside a concentric annulus (where possible)
were selected as field stars, indicated by the black circles in these figures.
CMDs of both regions were obtained, then divided into sub-grids 0.2 mag wide in
$g_0$ and 0.05 mag wide in $(g-r)_0$ color. Note that the field region of M92
(shown in Figure \ref{fig8radec}) is offset from the cluster center due to its
position at the edge of the photometric scan. This was necessary because an
annular field region around this location would have been inadequately populated
with stars. 

In each sub-grid, the signal-to-noise $(s/n)$ was calculated using:

\begin{equation}
s/n(i,j) = \frac{n_c(i,j)  - gn_f(i,j)}{\sqrt{n_c(i,j)+g^{2}n_f(i,j)}},\label{eqsn}
\end{equation}

\noindent where $n_{c}$ and $n_{f}$ refer to the number of stars counted in each
sub-grid with color index $i$ and magnitude index $j$ within the
cluster region and field region, respectively, and the parameter $g$
is the ratio of the cluster area to the field area. These values were
sorted in descending order in an array with index $l$, then star
counts were obtained in increasingly larger sections of the array. The
area in each section is defined as $a_k = ka_l$, where
$a_l=0.01~\rm{mag}^2$ represents the area of a single sub-grid;
$k$ is the number of sub-grids in the section. Then, the cumulative
signal-to-noise ratio, S/N, as a function of $a_k$, was calculated
using:
\begin{equation}
\rm{S/N}(a_k) = \frac{N_c(a_k) - gN_f(a_k)}{\sqrt{N_c(a_k) + g^2 N_f(a_k)}},\label{eqbigsn}
\end{equation}

\noindent where

\begin{equation}
N_c(a_k) = \sum_{l=1}^k n_c(l), \qquad N_f(a_k) = \sum_{l=1}^k n_f(l).\label{eqNcAndNf}
\end{equation}

\begin{figure}
\begin{center}
\plotone{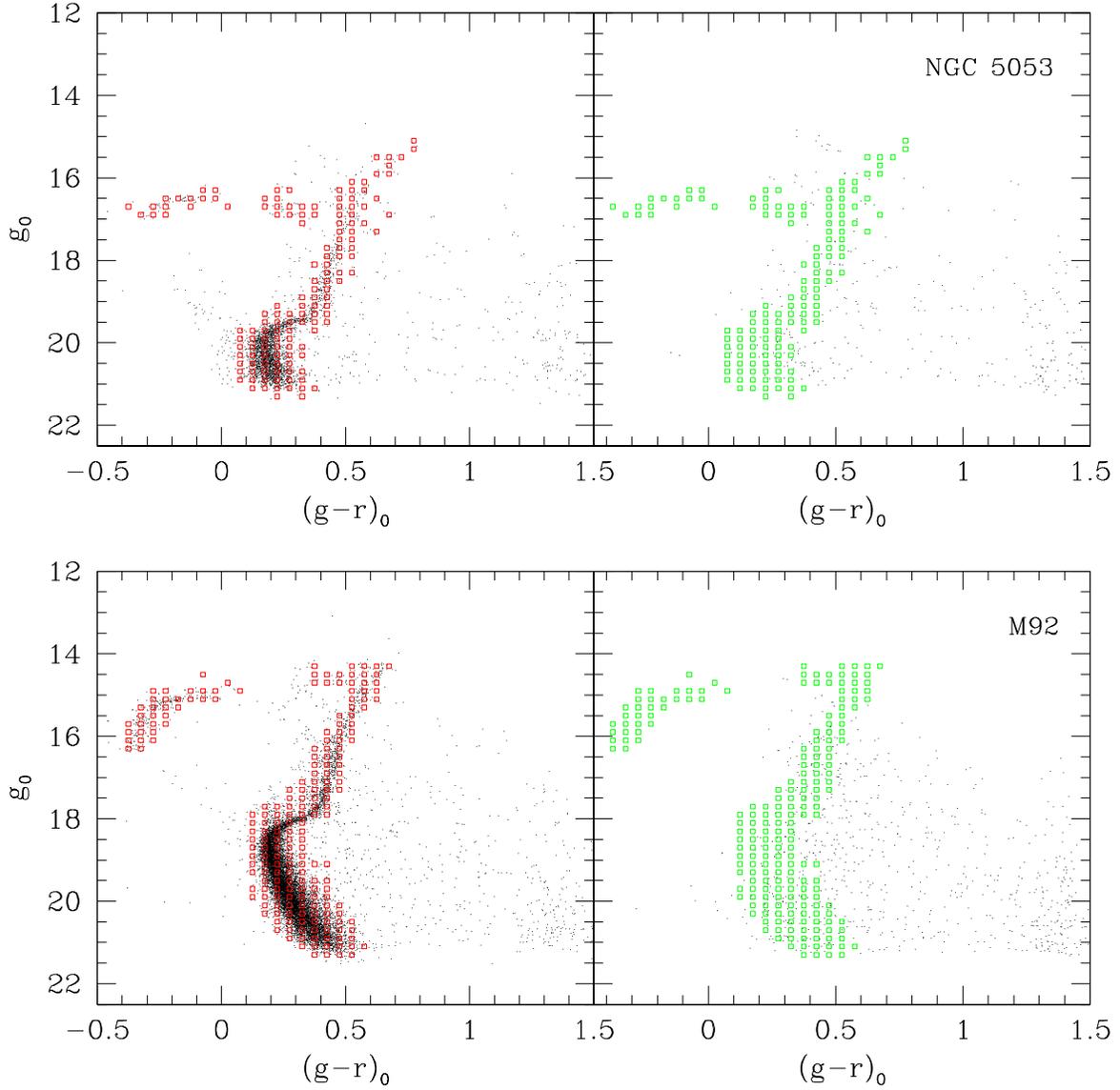}
\caption{Color-Magnitude Diagrams of the stars from NGC~5053 
(upper panels) and M92 (lower panels) 
inside the tidal radius (left-hand panels) and inside the field region
(right-hand panels). The small boxes represent the sub-grids that
were selected in the first cut of the CMD mask algorithm, and contain
the stars used the subsequent analysis.}\label{figm92n5053cmdcut1}
\end{center}
\end{figure}

\begin{figure}
\begin{center}
\plotone{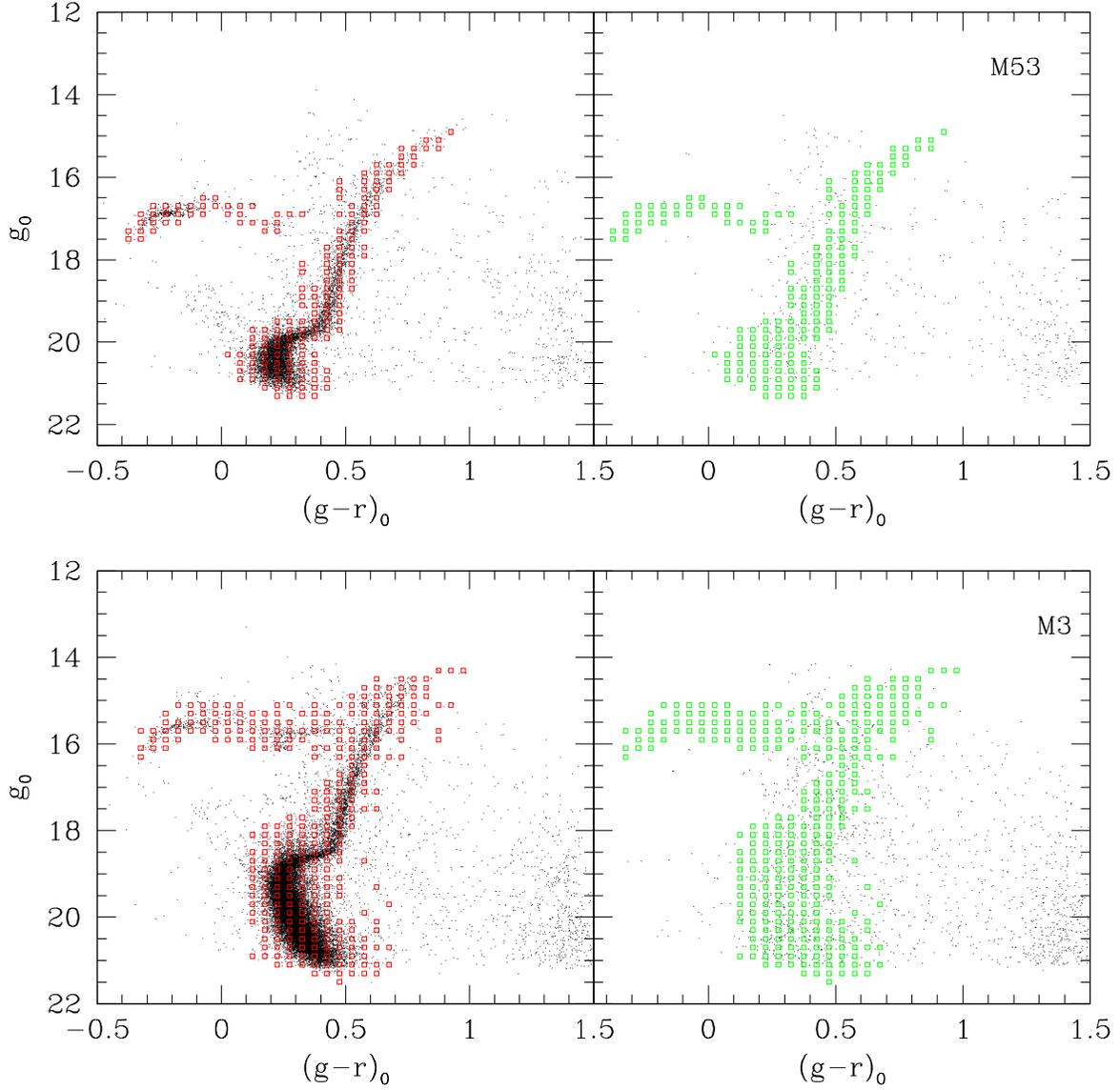}
\caption{Same as Fig. \ref{figm92n5053cmdcut1}, but for M53 (upper
panels) and M3 (lower panels).}\label{figm3m53cmdcut1}
\end{center}
\end{figure}

\begin{figure}
\begin{center}
\plotone{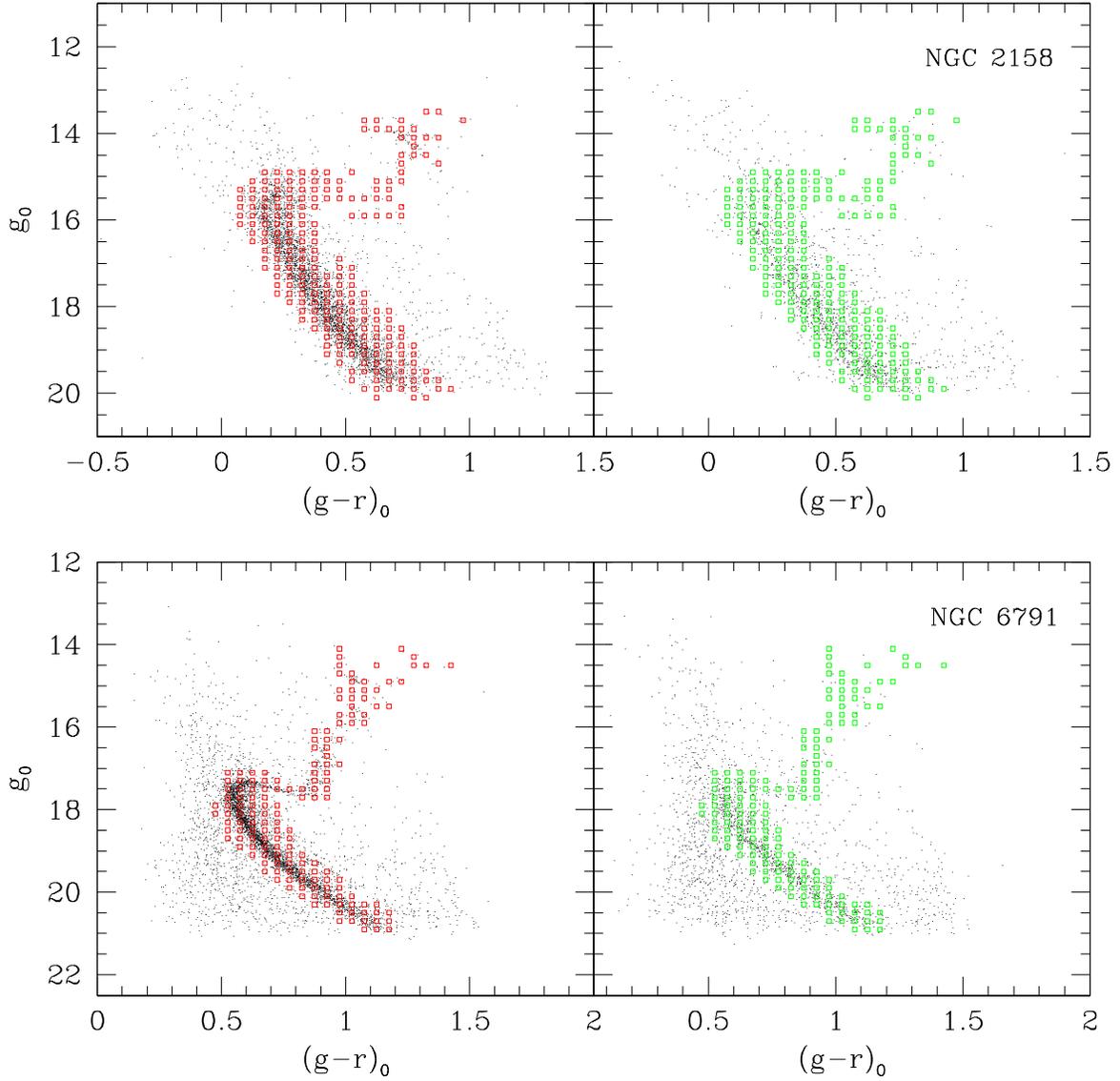}
\caption{Same as Fig. \ref{figm92n5053cmdcut1}, but for NGC~2158 (upper
panels) and NGC~6791 (lower panels).  
Due to the highly-evolved nature of these open clusters,
they were treated in the member selection process as if they were globular
clusters.}\label{fig2ngccmdcut1}
\end{center}
\end{figure}

\noindent Here, $n_c(l)$ represents the number of cluster stars within the
ordered sub-grid array element $l$; $n_f(l)$ represents the same quantity
for the field stars. A threshold value for $s/n$ was adopted, based on the
maximum value of S/N($a_k$), to identify areas of the CMD where the ratio of
cluster stars to field stars was high (rejecting single-star events). These
areas were taken to be sub-grids of likely cluster members, and all sub-grids
with $s/n(i,j)$ greater than this threshold were identified. These sub-grids are
shown as boxes in Figures \ref{figm92n5053cmdcut1} -- \ref{fig2ngccmdcut1}. The
left-hand panels show the stars inside the tidal radius -- the sub-grids with
$s/n$ greater than the threshold value are indicated as red squares.  
The right-hand panels show the stars from the field region with the same
sub-grids, indicated in green.  

\begin{figure}
\begin{center}
\plotone{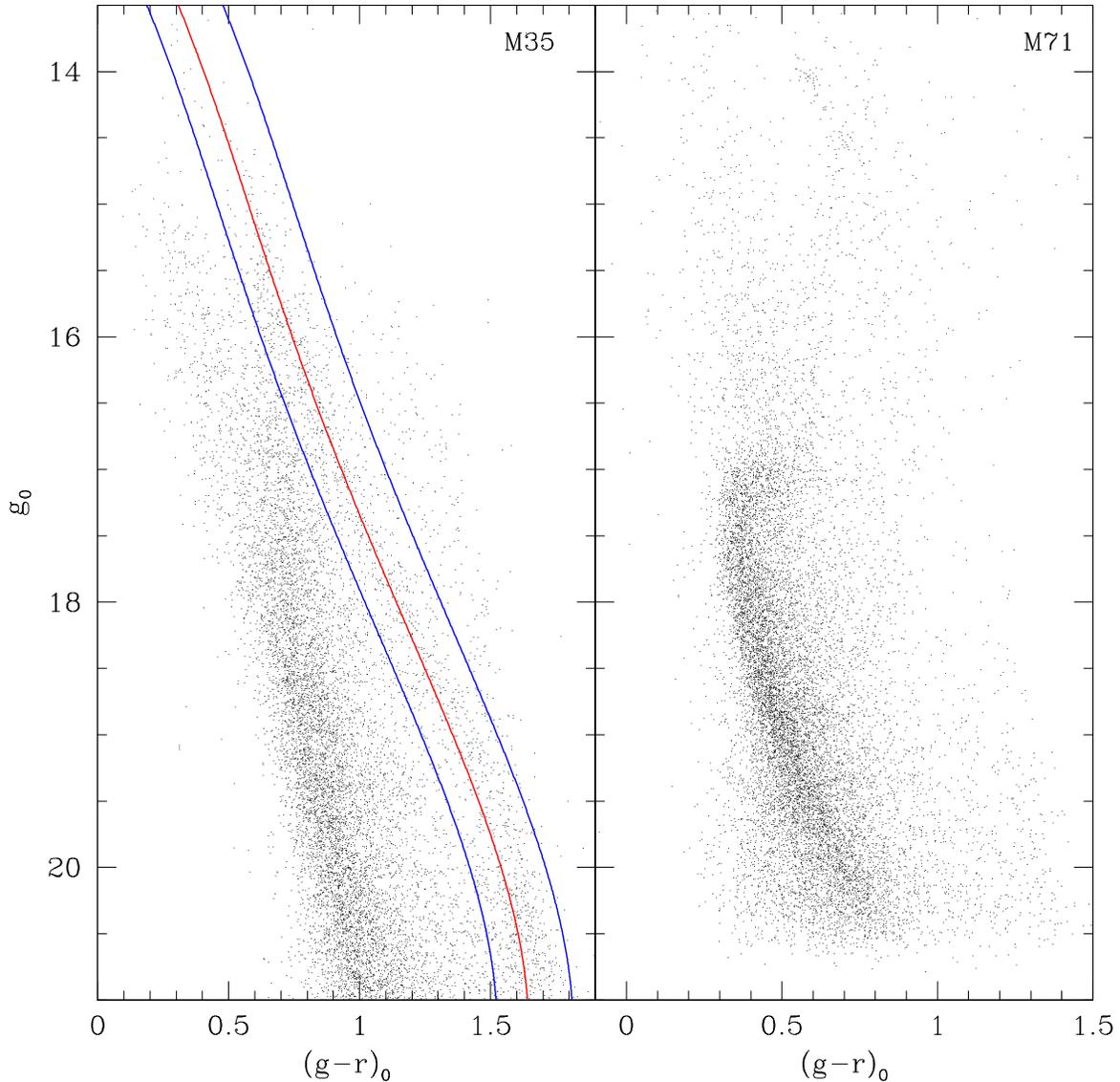}
\caption{Same as Fig. \ref{figm92n5053cmdcut1}, but for M35 (left-hand
panel) and M71 (right-hand panel). The red line in M35 is the fiducial from a
fourth-order polynomial fit, while the blue lines define the offsets of
$^{+0.17}_{-0.12}$ mag inside of which were selected stars regarded as likely members
from the photometric data.  Because of M35's low Galactic latitude, 
the dense stripe of stars on the blue side of the main sequence is 
due to superposed disk stars.  Member stars for M71 were selected
strictly by radial velocity and metallicity cuts rather than by
using the CMD first; no photometry was used for analysis of this 
cluster due to poor calibration.  For this reason,
the CMD for M71 is shown differently from the other globular
clusters.}\label{fig2mcmdcut1}
\end{center}
\end{figure}

The procedures described in Paper II handle OCs
differently from GCs, primarily due to the fact that no field
region is required. Instead of determining sub-grid $s/n$ ratios, a fiducial
line is fit to the open cluster's MS using a polynomial fitting routine,
then a region is picked out by eye corresponding to the MS to represent
the likely member stars. The interested reader is referred to Paper II
for further details on the open cluster member selection procedure. This
procedure works well on young clusters, where no significant evolution
off the MS has occurred. However, NGC~2158 and NGC~6791 are evolved
(older) clusters, and exhibit a distinct MSTO and RGB (see Figure
\ref{fig2ngccmdcut1}). This prevents polynomial fitting of the CMD from
working properly since the function would be double valued, so in this
study NGC~2158 and NGC~6791 are processed (for the purpose of member
assignment) as if they are globular clusters. The usual open
cluster procedure was successfully implemented for M35 (Figure
\ref{fig2mcmdcut1}).

\begin{figure}
\begin{center}
\plotone{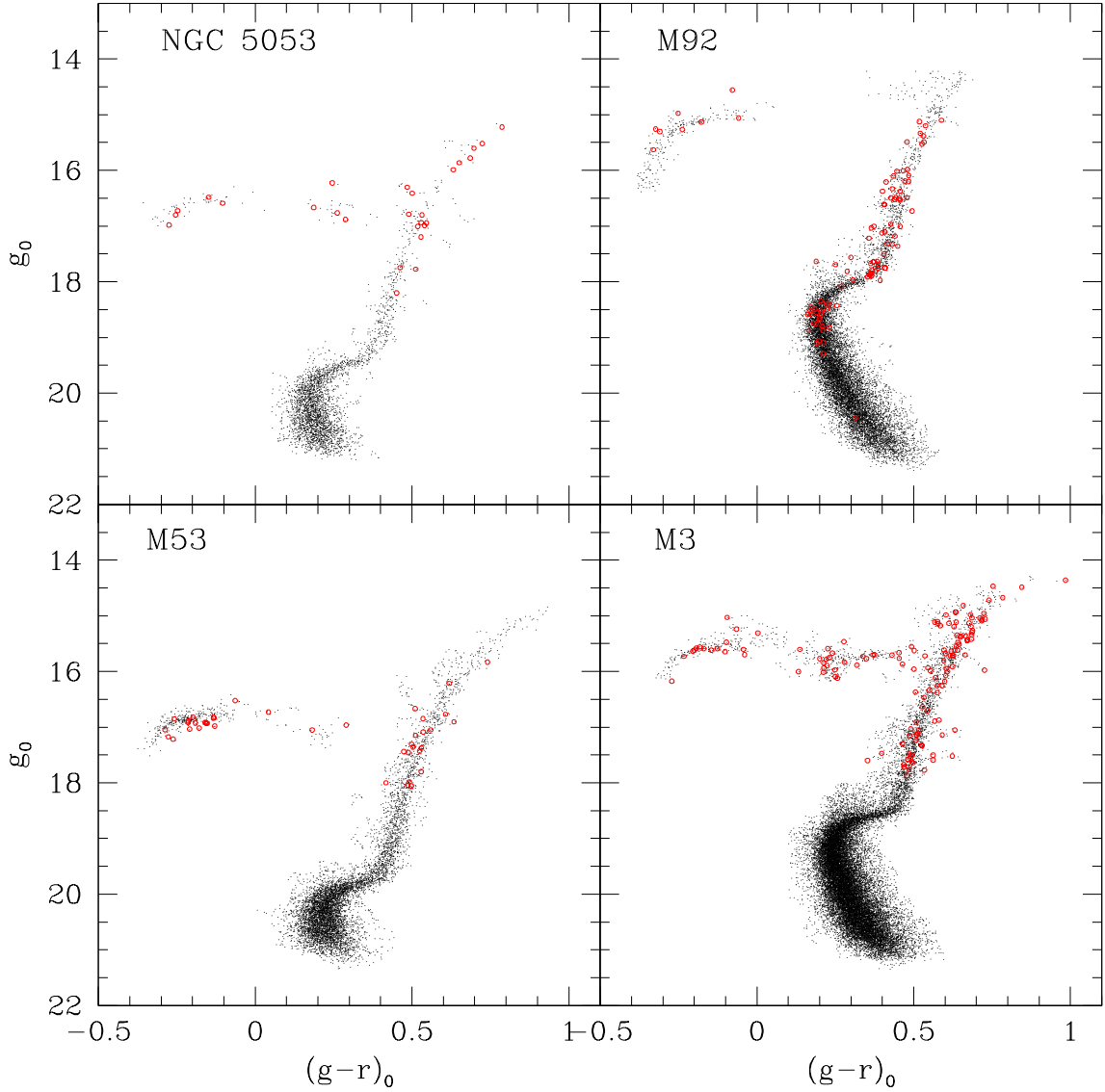}
\caption{The Color-Magnitude Diagram following the second cut of
likely member stars based on the sub-grid selection for NGC~5053 (upper-left
panel), M92 (upper-right panel), M53 (lower-left panel), and M3
(lower-right panel).
Black dots represent stars from the photometric sample, and the red 
open circles represent stars from the spectroscopic sample.}\label{fig1st4cmdcut2}
\end{center}
\end{figure}

\begin{figure}
\begin{center}
\plotone{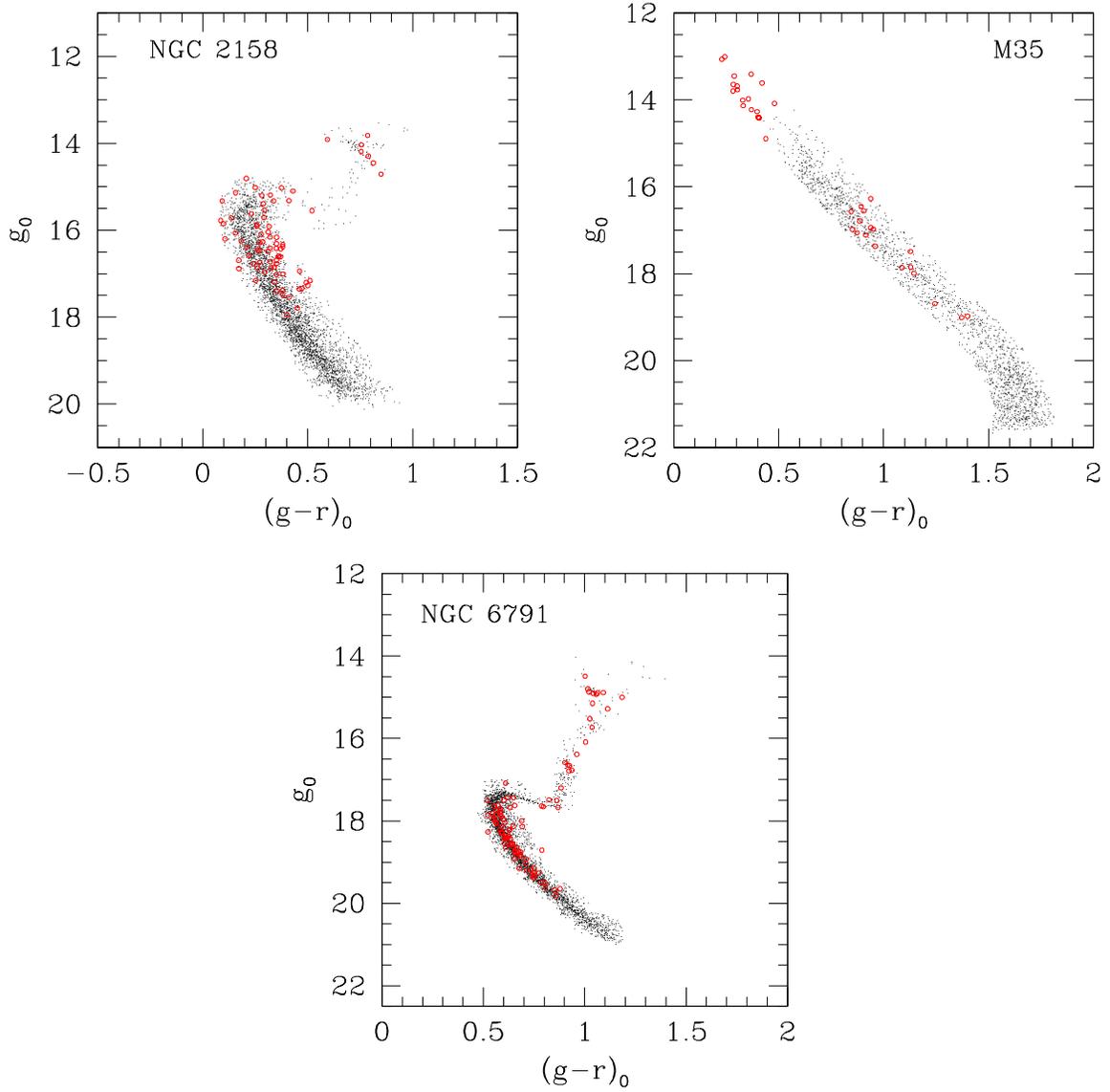}
\caption{The Color-Magnitude Diagram following the second cut of
likely member stars for NGC~2158 (upper-left panel), M35
(upper-right panel),
and NGC~6791 (lower panel).  Black dots represent stars from the photometric sample, and the red 
open circles represent stars from the spectroscopic sample.}\label{fig2ndcmdcut2}
\end{center}
\end{figure}

The cleaned CMDs for our sample are shown in Figures 
\ref{fig1st4cmdcut2} and  \ref{fig2ndcmdcut2}.
The black points are the likely members from the photometry, while the red open
circles are the likely members from the spectroscopic sample. This part of the
procedure could not be carried out for M71 due to difficulties encountered with
the photometry values available for this cluster at the time of our analysis
(see An et al. 2008). Therefore, a first cut was made based on the tidal radius
of the stars, and those stars were passed on to the final step, as outlined in
the following section. Figure \ref{fig2mcmdcut1} shows the first-cut CMD for
M71.

\subsection{Selection of Adopted True Members}\label{subsec_true_sel}

We next determine the true member stars as a subset of the adopted likely member
stars. Figure \ref{fig5053fehrv} shows the distributions of [Fe/H] (left-hand
panel) and radial velocities (RV; right-hand panel) for stars in the field of
NGC~5053 at each culling point in the procedure. The black lines indicate all
579 stars on the original spectroscopic plate (after removing stars with no
parameter estimates from the SSPP or low spectral S/N), while the red lines
indicate only those stars inside r$_t$, and the green lines indicate those stars
that passed the cut using the individual sub-grid $s/n$ and cumulative S/N
calculations. We then performed a Gaussian fit to the highest peak of the
distribution of this final subset (blue line) and obtained estimates of the mean
and standard deviation of [Fe/H] and RV. Finally, outliers were rejected by
applying a 2$\sigma$ cut on both parameters:

\begin{gather}\label{eqgaussrange}
\rm{\langle [Fe/H]\rangle - 2\sigma_{[Fe/H]} \leq [Fe/H]_{\star} \leq \langle [Fe/H]\rangle + 2\sigma_{[Fe/H]}} \\
\rm{\langle RV\rangle - 2\sigma_{RV} \leq RV_{\star} \leq \langle RV\rangle + 2\sigma_{RV}}.
\end{gather}

\noindent $\rm{[Fe/H]_{\star}}$ and $\rm{RV_{\star}}$ correspond to the metallicity and radial velocity of each star
in question. If a star passed both cuts then it was considered a true
member star. The numbers of true member stars determined by this final
cut for each cluster are listed in Table \ref{tabclusterfehrv}.

\section{Determination of Overall Metallicities and Radial Velocities of the Clusters}\label{secoverall_det}

\begin{figure}
\begin{center}
\plotone{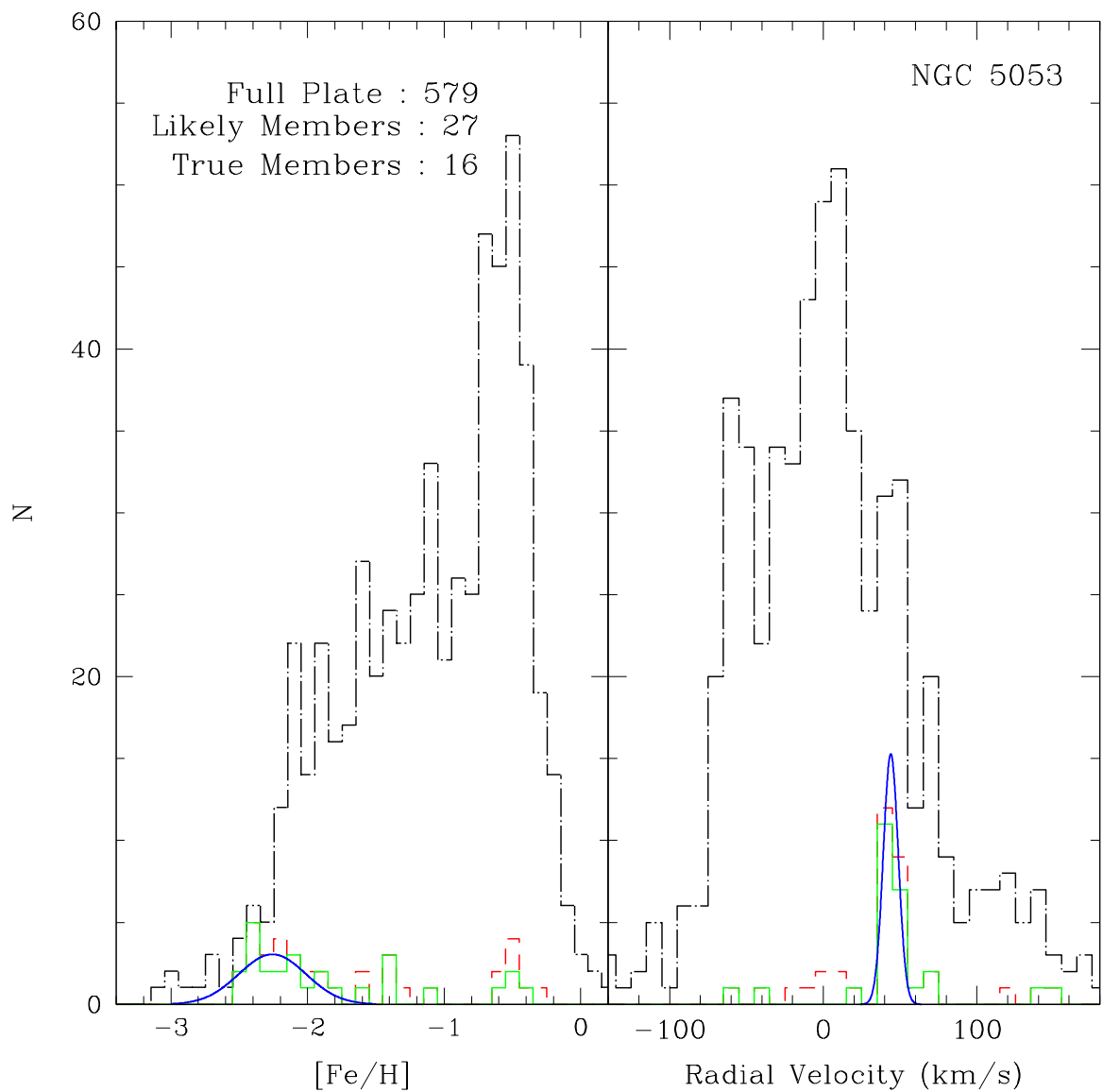}
\caption{Distributions of [Fe/H] and radial velocity for stars in the
field of NGC~5053. The black dot-dashed line corresponds to all the stars on the plate,
the red dashed line corresponds to the stars inside the tidal radius, and
the green solid line corresponds to the stars that were identified as likely
members by the sub-grid $s/n$ procedure described in Section \ref{subsec_likely_sel}.
The blue solid line is a Gaussian fit indicating the region of each distribution in
which the true members are located, as described in 
Section \ref{subsec_true_sel}.}\label{fig5053fehrv}
\end{center}
\end{figure}

\begin{figure}
\begin{center}
\plotone{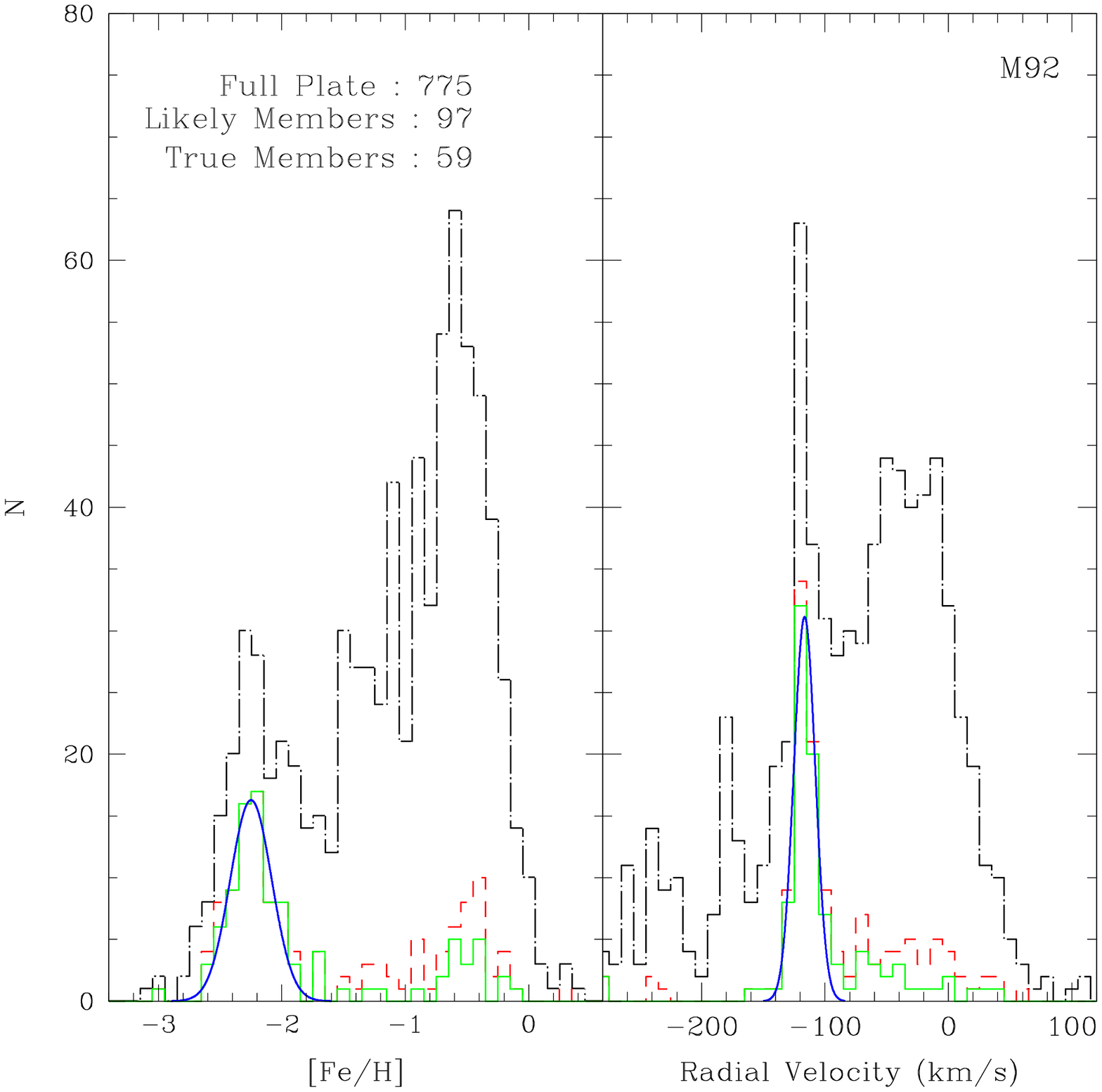}
\caption{Same as Fig. \ref{fig5053fehrv}, but for M92.}\label{figm92fehrv}
\end{center}
\end{figure}

\begin{figure}
\begin{center}
\plotone{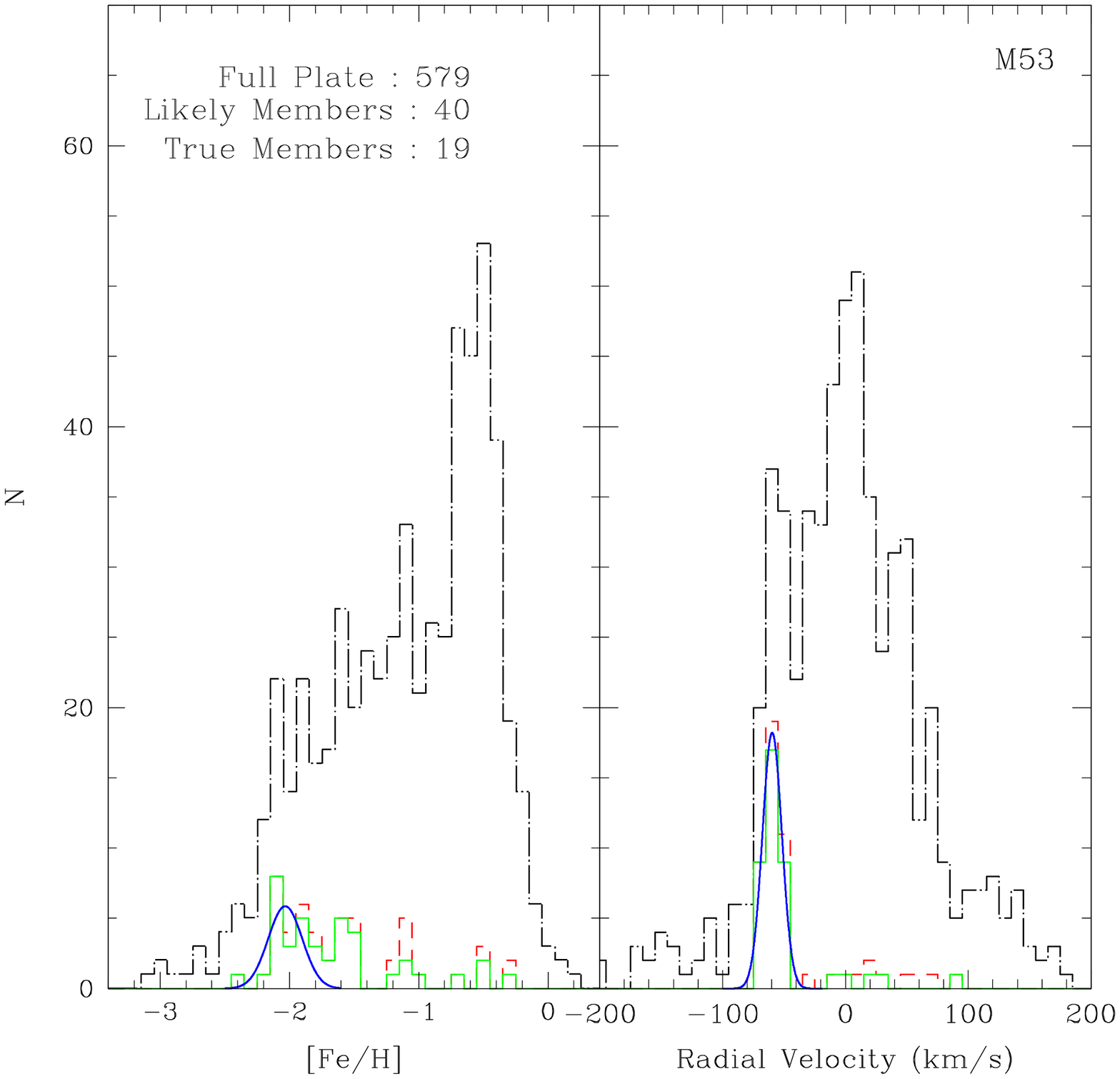}
\caption{Same as Fig. \ref{fig5053fehrv}, but for M53.}\label{figm53fehrv}
\end{center}
\end{figure}

\begin{figure}
\begin{center}
\plotone{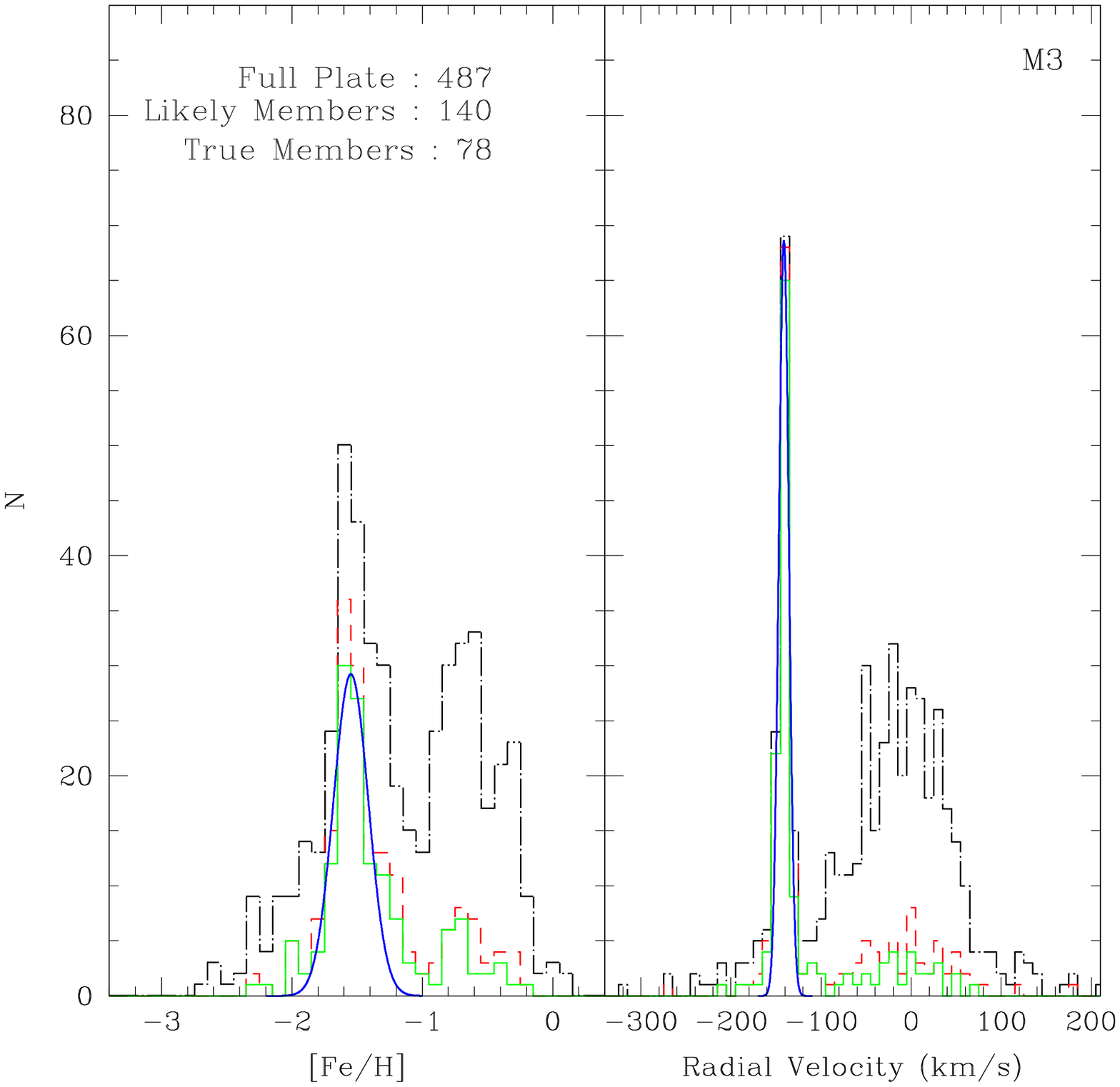}
\caption{Same as Fig. \ref{fig5053fehrv}, but for M3.}\label{figm3fehrv}
\end{center}
\end{figure}

\begin{figure}
\begin{center}
\plotone{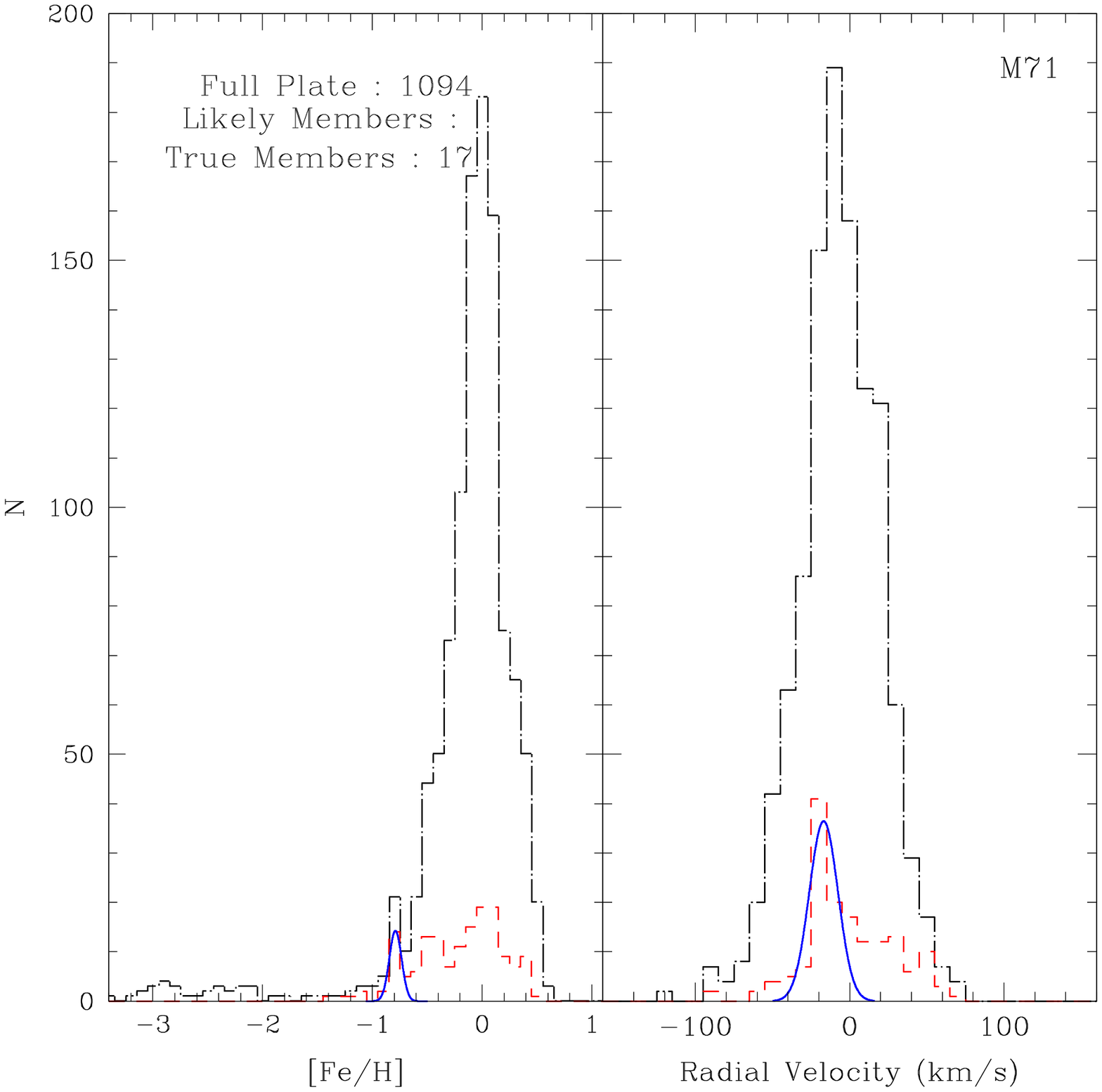}
\caption{Same as Fig. \ref{fig5053fehrv}, but for M71.}\label{figm71fehrv}
\end{center}
\end{figure}

\begin{figure}
\begin{center}
\plotone{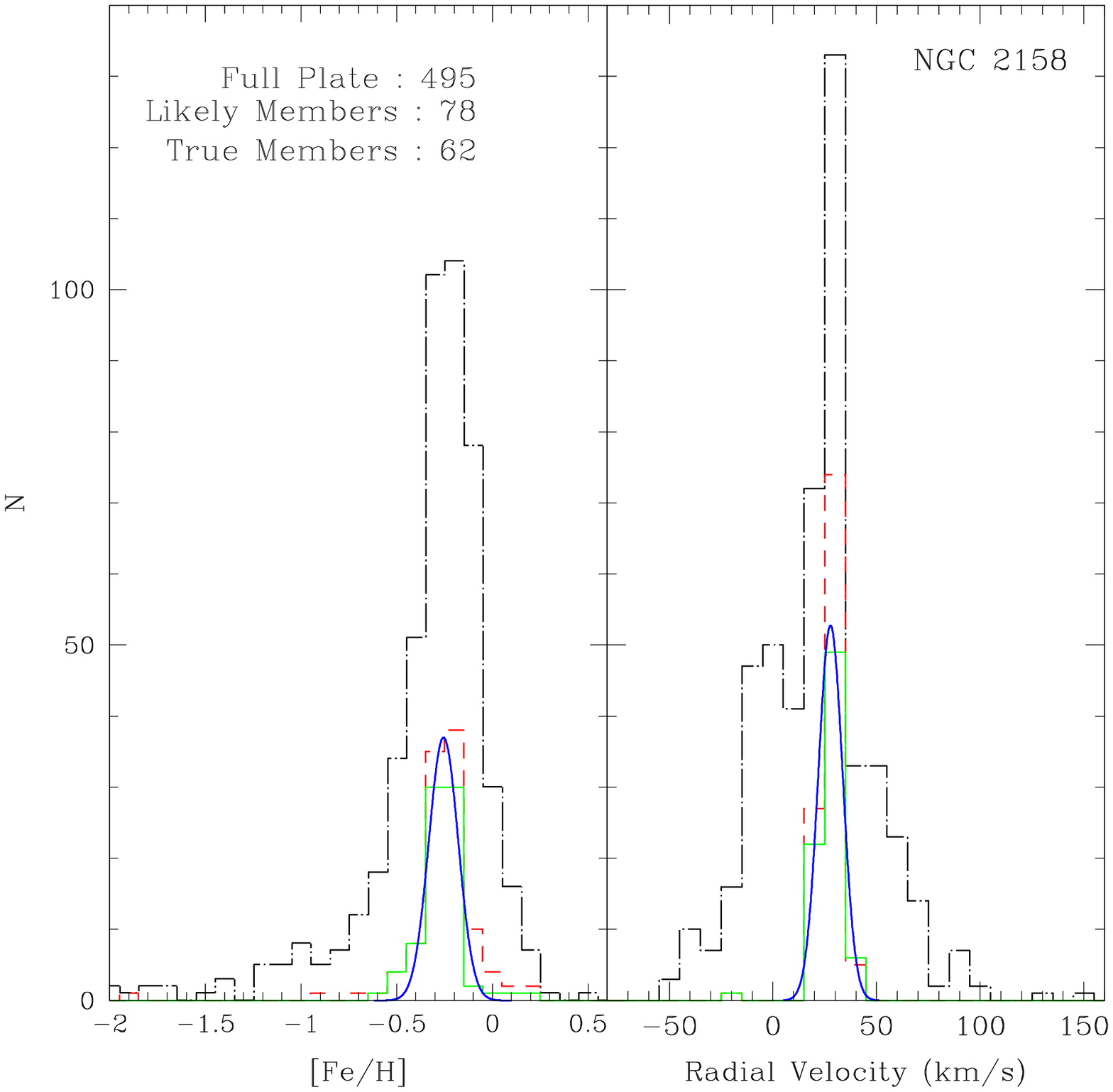}
\caption{Same as Fig. \ref{fig5053fehrv}, but for NGC~2158.}\label{fig2158fehrv}
\end{center}
\end{figure}

\begin{figure}
\begin{center}
\plotone{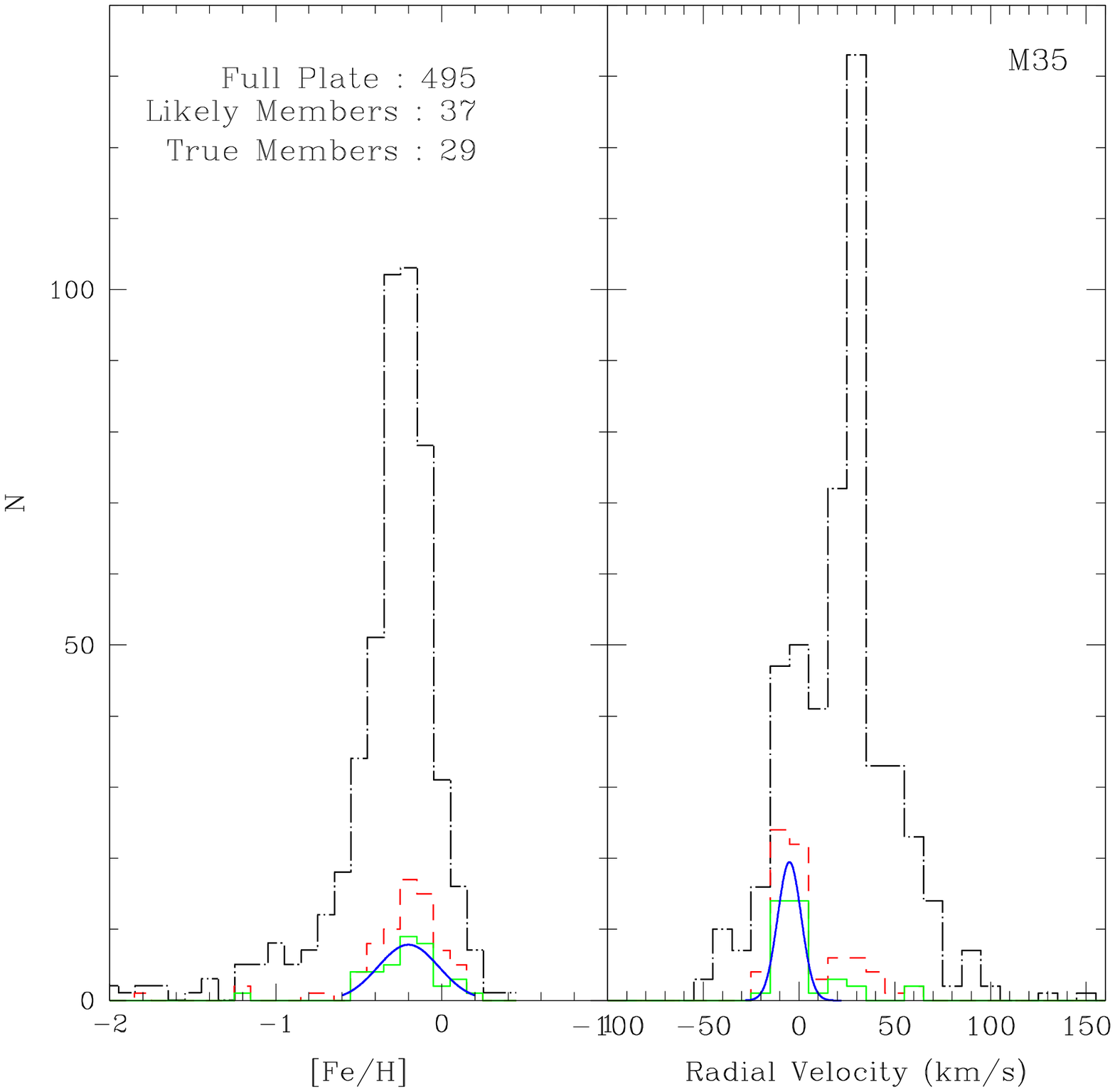}
\caption{Same as Fig. \ref{fig5053fehrv}, but for M35.}\label{figm35fehrv}
\end{center}
\end{figure}

\begin{figure}
\begin{center}
\plotone{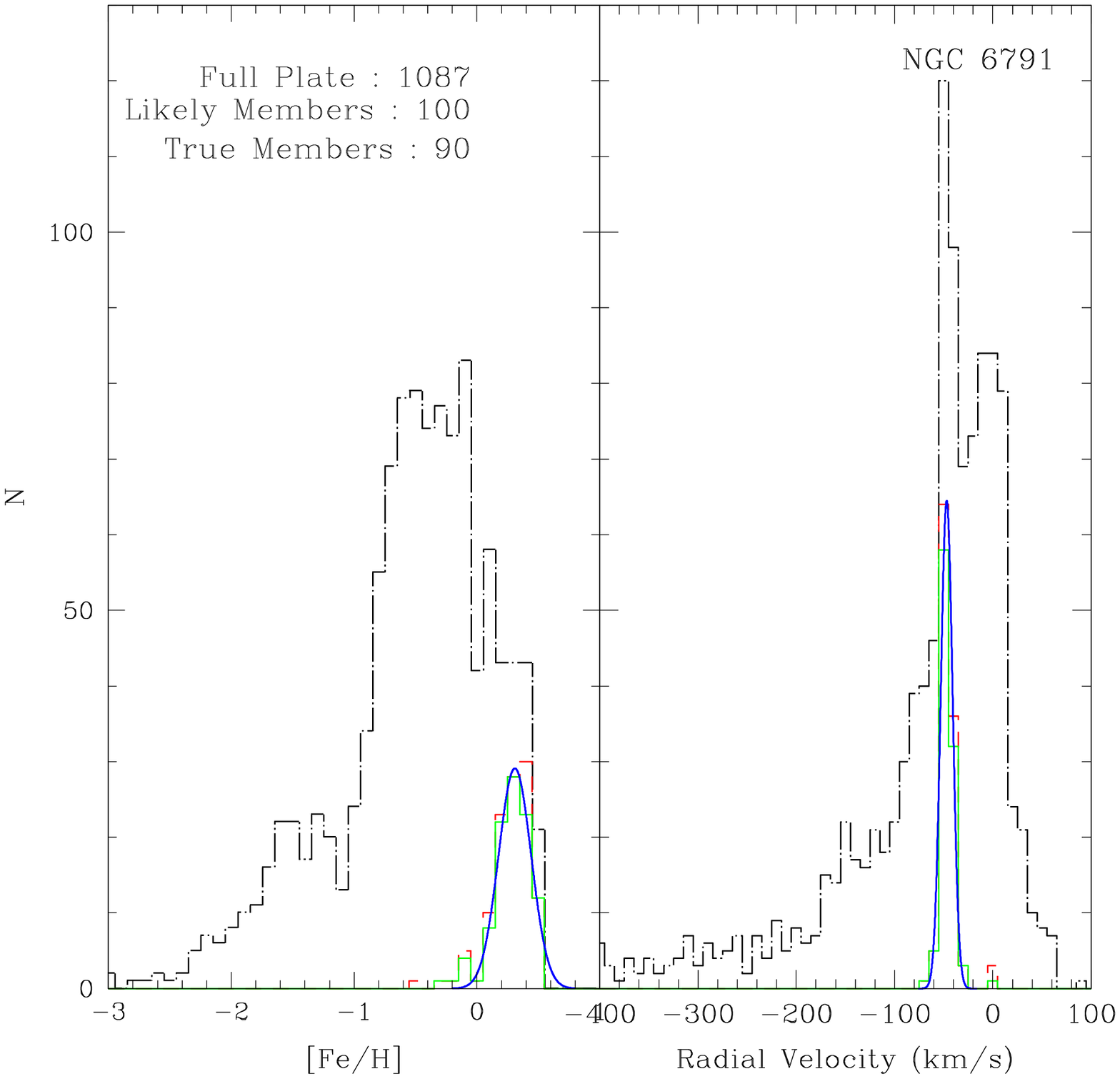}
\caption{Same as Fig. \ref{fig5053fehrv}, but for NGC~6791.}\label{fig6791fehrv}
\end{center}
\end{figure}

Once the true members were selected as described above, final estimates of the
cluster metallicities and radial velocities were obtained. Figures
\ref{fig5053fehrv}--\ref{fig6791fehrv} show binned distributions of
[Fe/H] and RV for each cluster. The black lines in these figures represent the
full distribution of all stars in each cluster's field with available
spectroscopic information, the red lines represent only those stars from the
spectroscopic samples that lie inside each cluster's tidal radius (or a
reasonable radius, for M71 and NGC~6791), and the green lines represent those
stars that passed the sub-grid $s/n$ cut described in Section
\ref{subsec_likely_sel}. Gaussian fits (blue lines) to the highest peak of this
final distribution determined the adopted cluster values, which are listed in
Table \ref{tabclusterfehrv}.  This table also lists the standard error in the
mean ($\sigma_{\mu}$) for the estimates of metallicity and radial velocity for each
cluster; due to the large numbers of true members for each cluster, these are
uniformly small.

\begin{figure}
\begin{center}
\plotone{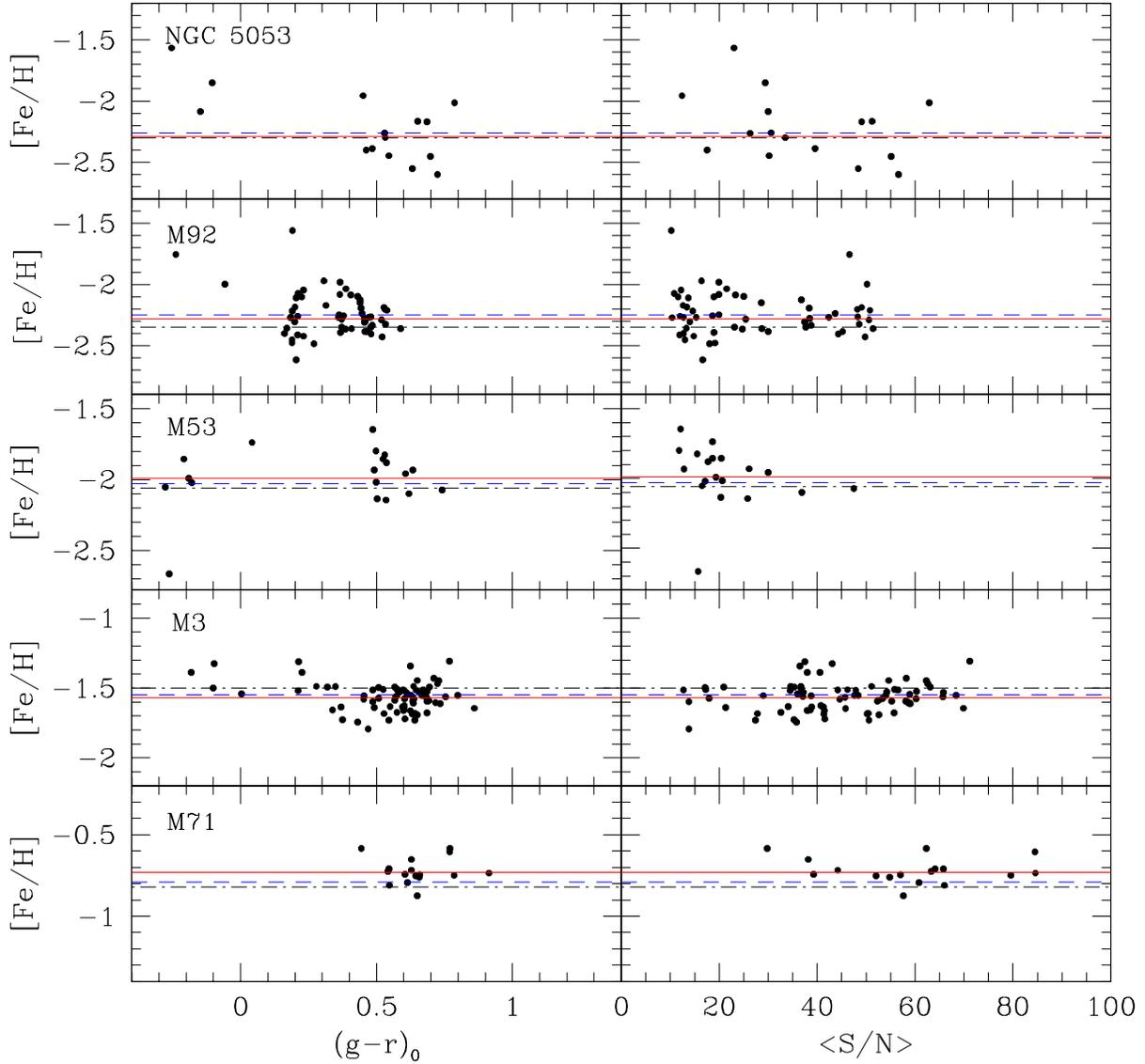}
\caption{Distribution of [Fe/H] as a function of
$(g-r)_0$ (left-hand column) and average signal-to-noise
(right-hand column) for selected true member stars of the globular
clusters NGC~5053, M92,
M53, M3, and M71, ordered from top to bottom on increasing
metallicity.  The red solid line in each panel represents the
adopted value of [Fe/H] for each cluster from the Harris (1996)
catalog, the black dot-dashed line is [Fe/H] from the 
Carretta et al. (2009) recalibration, and
the dashed blue line represents the mean measured value of
each cluster.}\label{figfehgrsng5}
\end{center}
\end{figure}

\begin{figure}
\begin{center}
\plotone{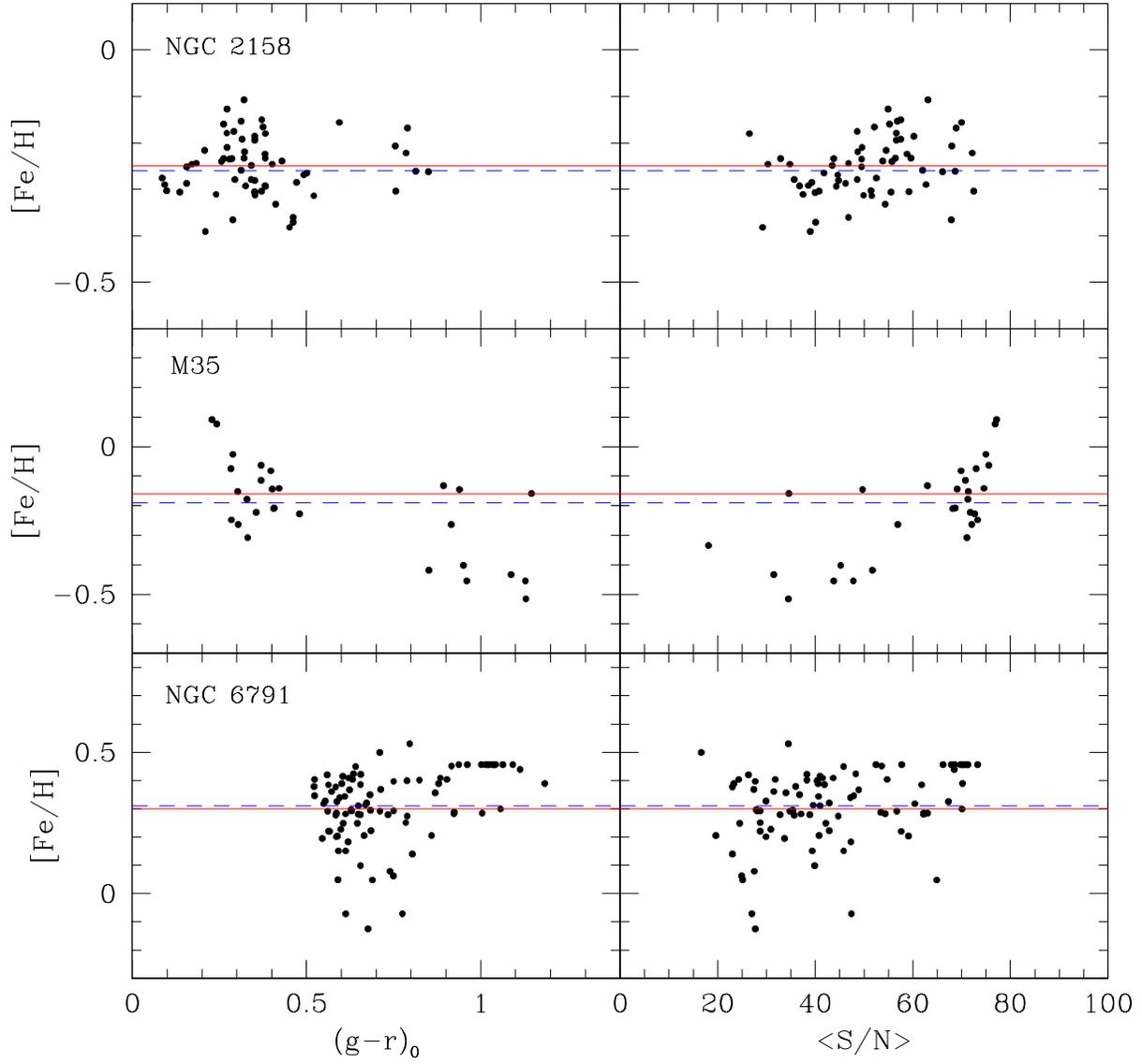}
\caption{Same as Fig. \ref{figfehgrsng5}, but for the open clusters NGC~2158, M35,
and NGC~6791, ordered from top to bottom on increasing
metallicity.}\label{figfehgrsng3}
\end{center}
\end{figure}

No strong trends appear to exist in estimates of
[Fe/H] as a function of color or spectral quality, as shown in Figures
\ref{figfehgrsng5} and \ref{figfehgrsng3}. As a check, we calculated
residuals of [Fe/H] with respect to the values adopted for each cluster from the
literature, using:

\begin{equation}
\rm{Res_{[Fe/H]}} = \rm{[Fe/H]} - \rm{[Fe/H]}_{lit}, \label{eqresid}
\end{equation}

\noindent and performed a linear regression on these values as a function of $(g-r)_0$ color and $\langle\rm{S/N}\rangle$
using models of the form:

\begin{align}
\rm{Res_{[Fe/H]}} &= X \cdot (g-r)_0 + Y\label{eqlinreggr} \\
\rm{Res_{[Fe/H]}} &= X \cdot (S/N) + Y.\label{eqlinregsn}
\end{align}

\begin{figure}
\begin{center}
\plotone{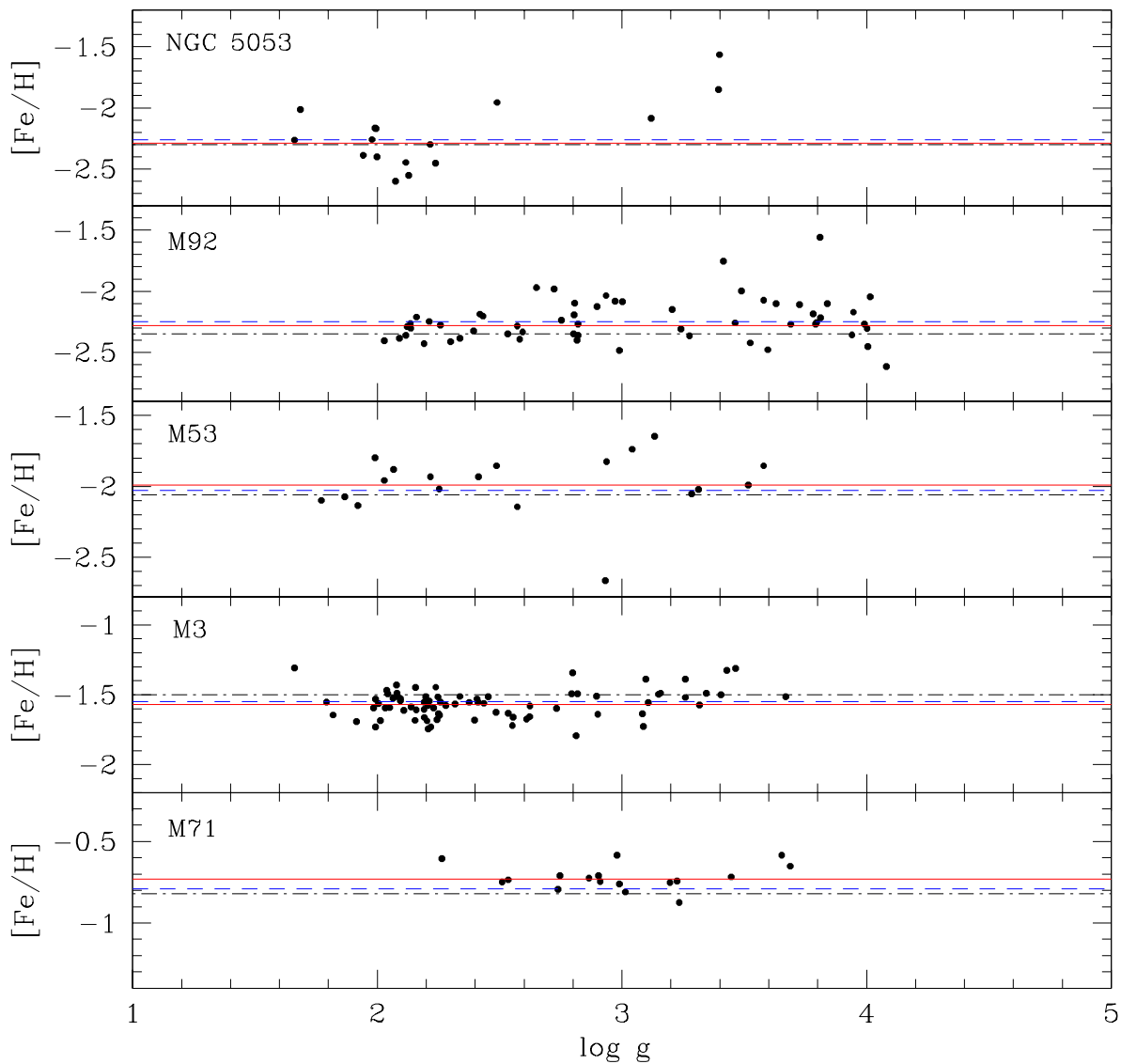}
\caption{Distribution of [Fe/H] as a function of estimated log $g$ for the
selected true member stars of the globular clusters NGC~5053, M92,
M53, M3, and M71, ordered from top to bottom on increasing
metallicity.  As in Fig. \ref{figfehgrsng5}, the red solid line corresponds to the adopted 
value for [Fe/H] for each cluster from Harris (1996), the black dot-dashed is
[Fe/H] from the recalibrated metallicity scale of Carretta et al. (2009),
and the dashed blue is the mean measured value.}\label{figfehloggg5}
\end{center}
\end{figure}

\begin{figure}
\begin{center}
\plotone{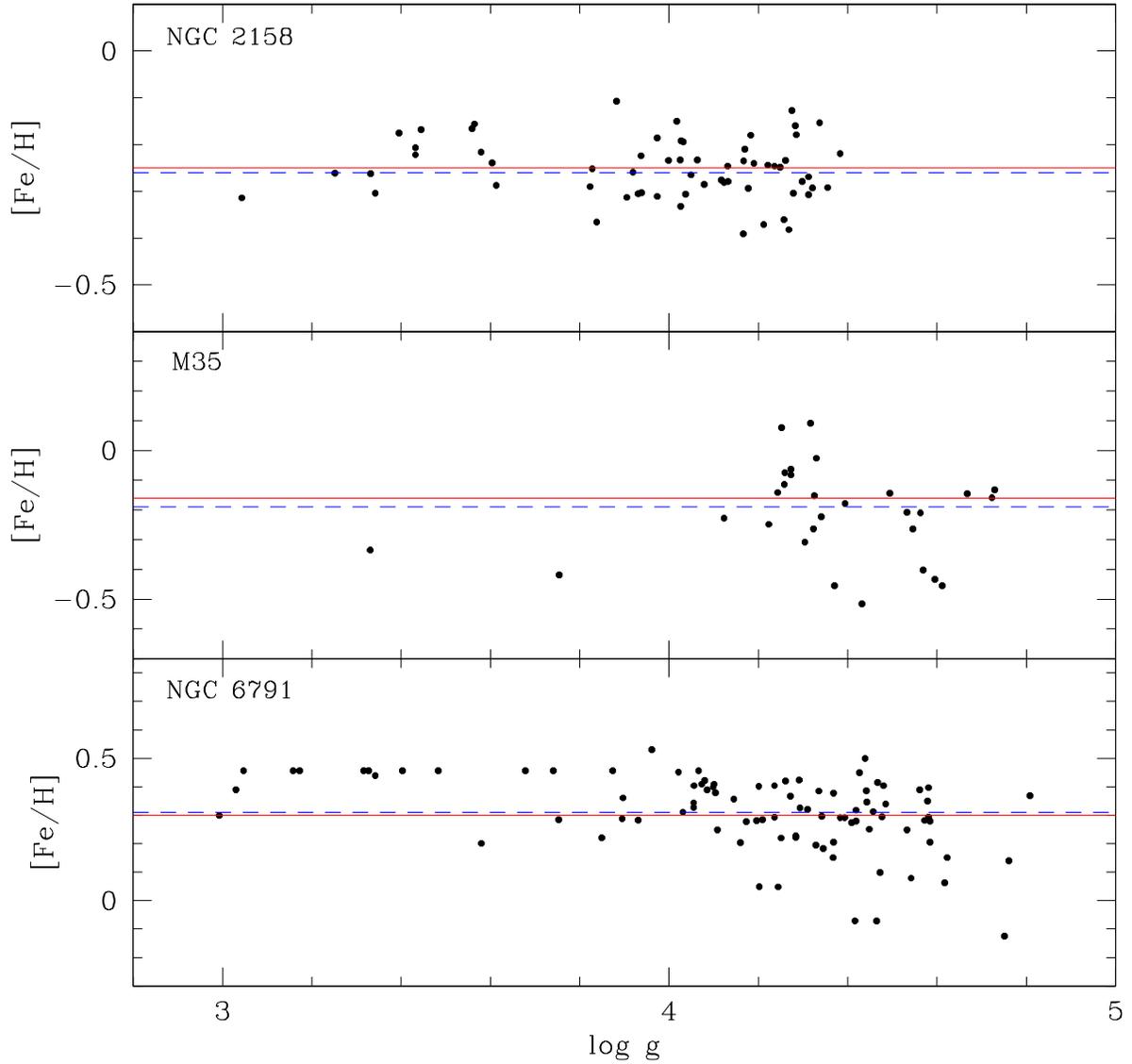}
\caption{Distribution of [Fe/H] as a function of estimated log $g$ for the
selected true member stars of the open clusters NGC~2158, M35, and
NGC~6791, ordered from top to bottom on increasing
metallicity. As in Fig. \ref{figfehgrsng5}, the red solid line
corresponds to the adopted literature value for [Fe/H] for each cluster,
while the dashed blue is the mean measured value. }\label{figfehloggg3}
\end{center}
\end{figure}

The results of the linear regressions are listed in Table
\ref{tabfehreg}. Column (2) lists the number of true member stars used
in the fit, Columns (4) and (6) list the slope and zero-point of the
fit, respectively, while Columns (5) and (7) list the corresponding
uncertainties. Finally, Column (8) lists the $R^2$ value, which
indicates the amount of scatter in the data that can be accommodated by
the regression. Values of $R^2$ close to zero indicate little
dependence on the independent variable (the desired goal), whereas
values of $R^2$ close to one indicate a large dependence on the
independent variable. There are two clusters (NGC~5053 and M35) for which the
$R^2$ values are somewhat high. These appear to have been influenced by stars at
the extrema of the color ranges, but still do not rise to the level of strong
statistical significance. The fits for the rest of the clusters have
sufficiently low values of $R^2$ that the correlations are not statistically
significant; Figures \ref{figfehloggg5} and \ref{figfehloggg3} show the
distribution of metallicity estimates as a function of the estimated surface
gravity. No significant trends are observed, supporting the conclusion of Paper
II that the SSPP is robust and reliable over large ranges in surface gravity
(luminosity) and color, even for spectra with less-than-optimal S/N.

\begin{figure}
\begin{center}
\plotone{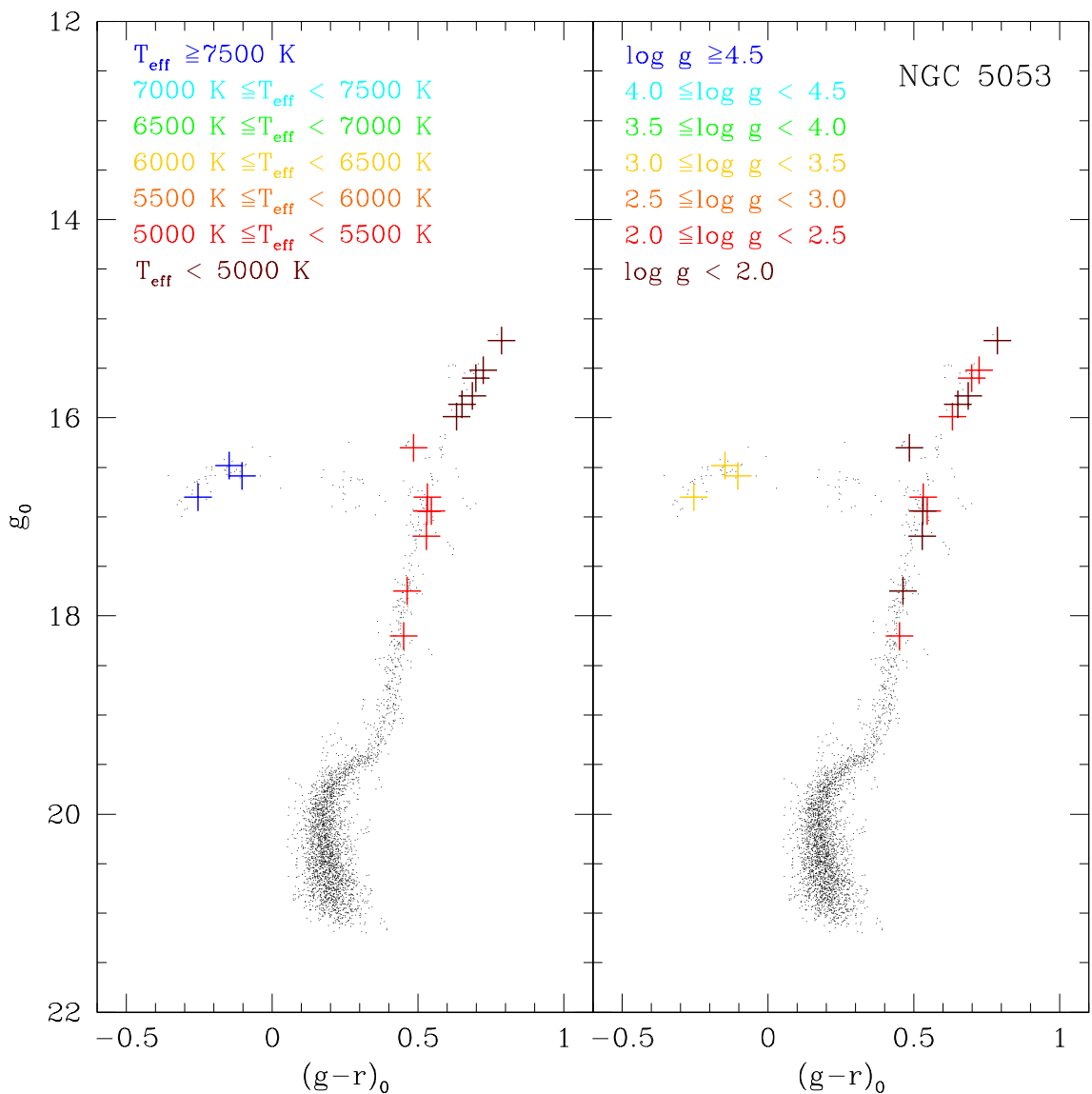}
\caption{Color-Magnitude Diagram of the selected true member stars of NGC~5053.  
The left-hand panel shows the
distribution of effective temperatures, while the right-hand panel shows the
distribution of surface gravity, both based on the spectroscopic sample. The
black dots are the likely member stars from the photometric sample.  Each color
represents a temperature step of width 500~K and a log $g$ step of 0.5
dex, respectively.}\label{fig5053cmdfinal}
\end{center}
\end{figure}

\begin{figure}
\begin{center}
\plotone{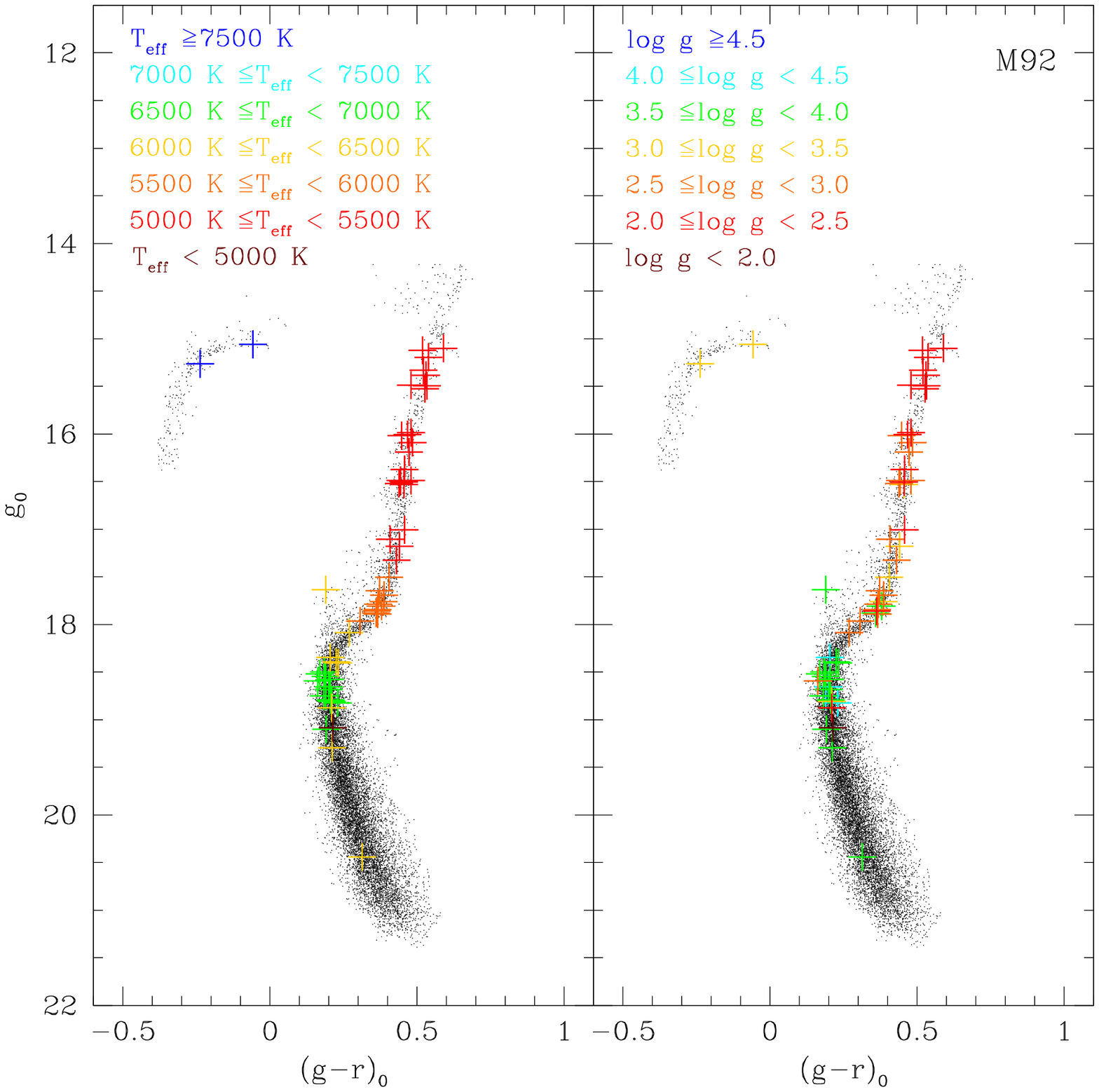}
\caption{Same as Fig. \ref{fig5053cmdfinal}, but for M92.}\label{figm92cmdfinal}
\end{center}
\end{figure}

\begin{figure}
\begin{center}
\plotone{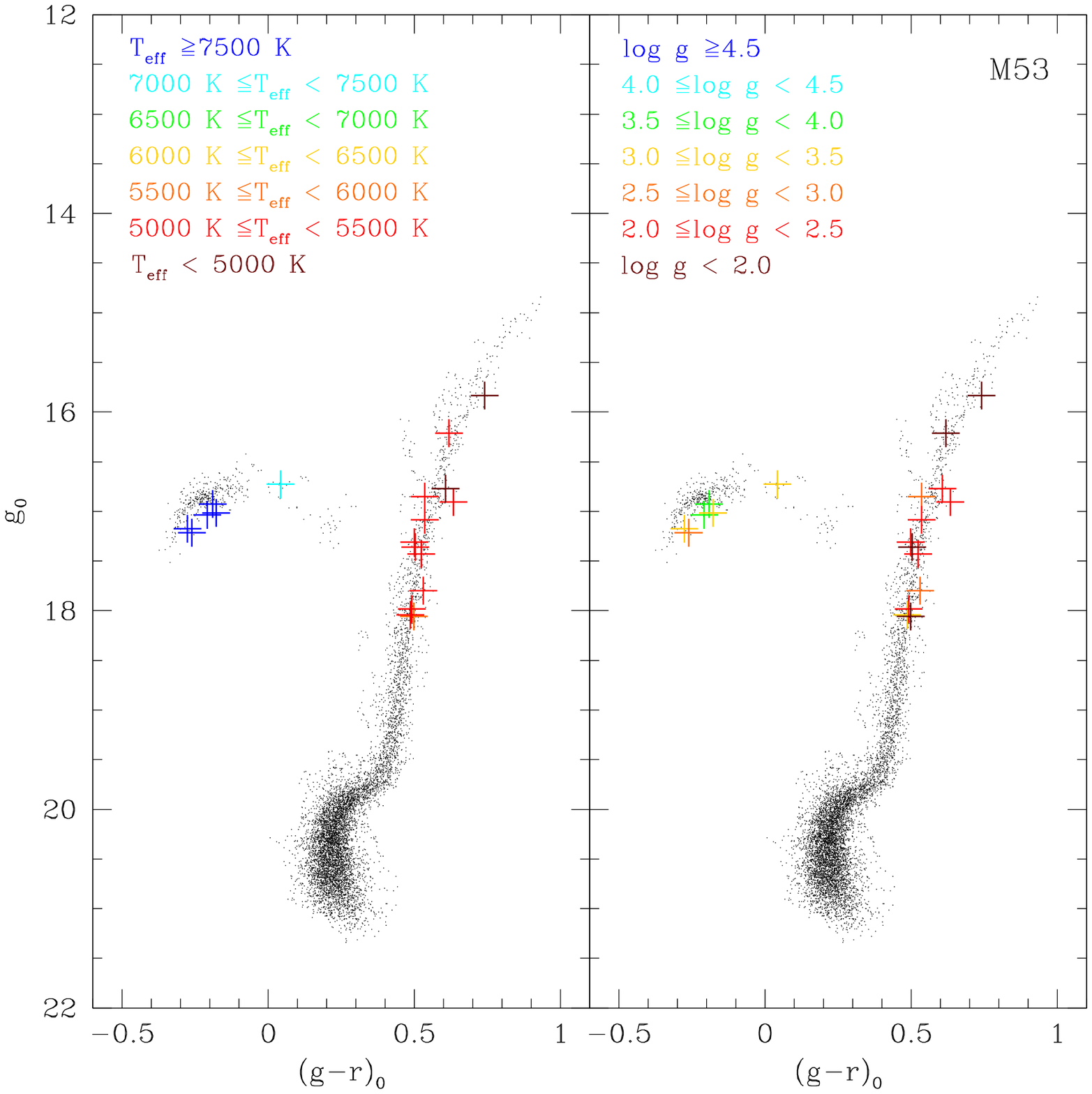}
\caption{Same as Fig. \ref{fig5053cmdfinal}, but for M53.}\label{figm53cmdfinal}
\end{center}
\end{figure}

\begin{figure}
\begin{center}
\plotone{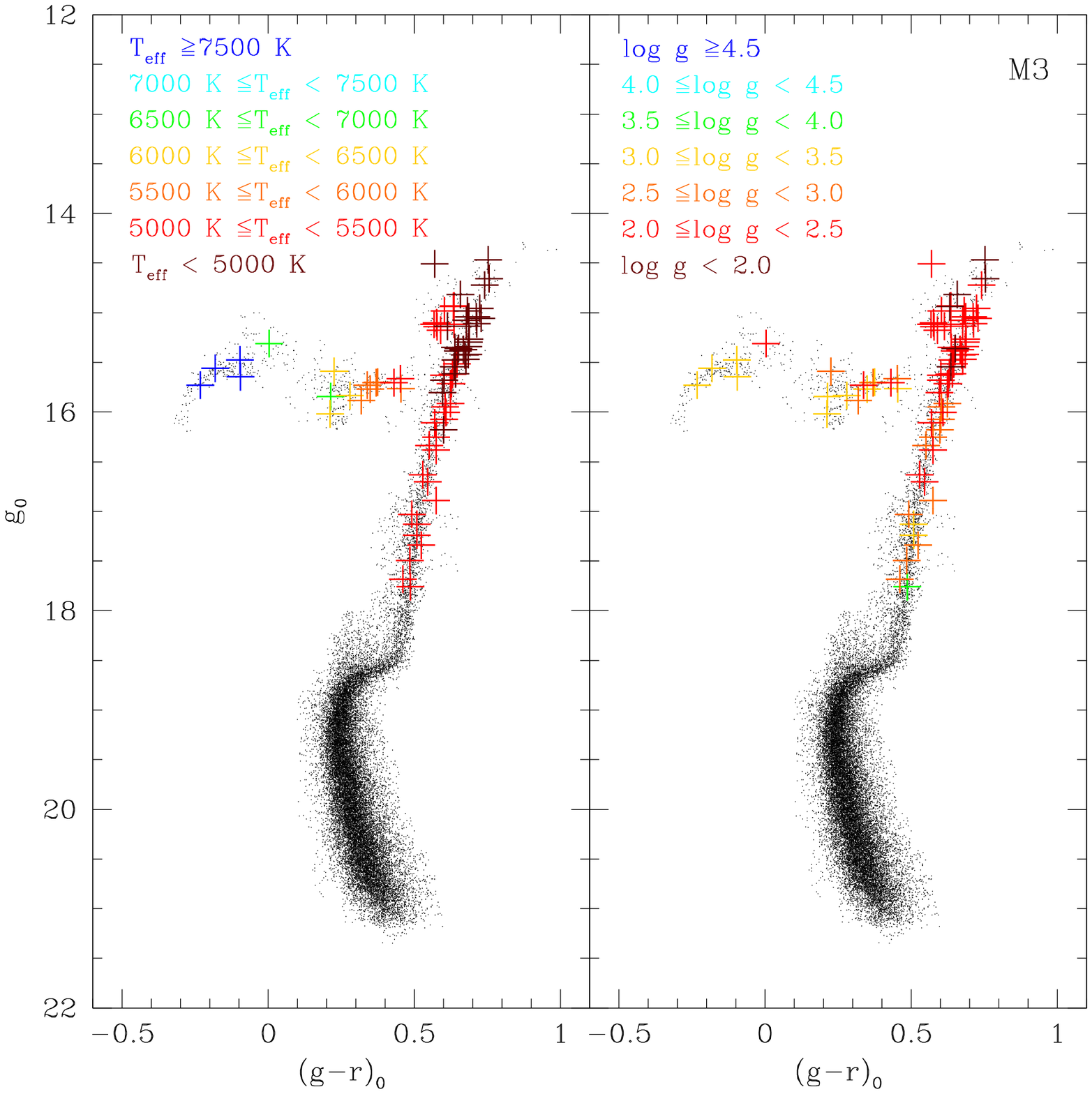}
\caption{Same as Fig. \ref{fig5053cmdfinal}, but for M3.}\label{figm3cmdfinal}
\end{center}
\end{figure}

\begin{figure}
\begin{center}
\plotone{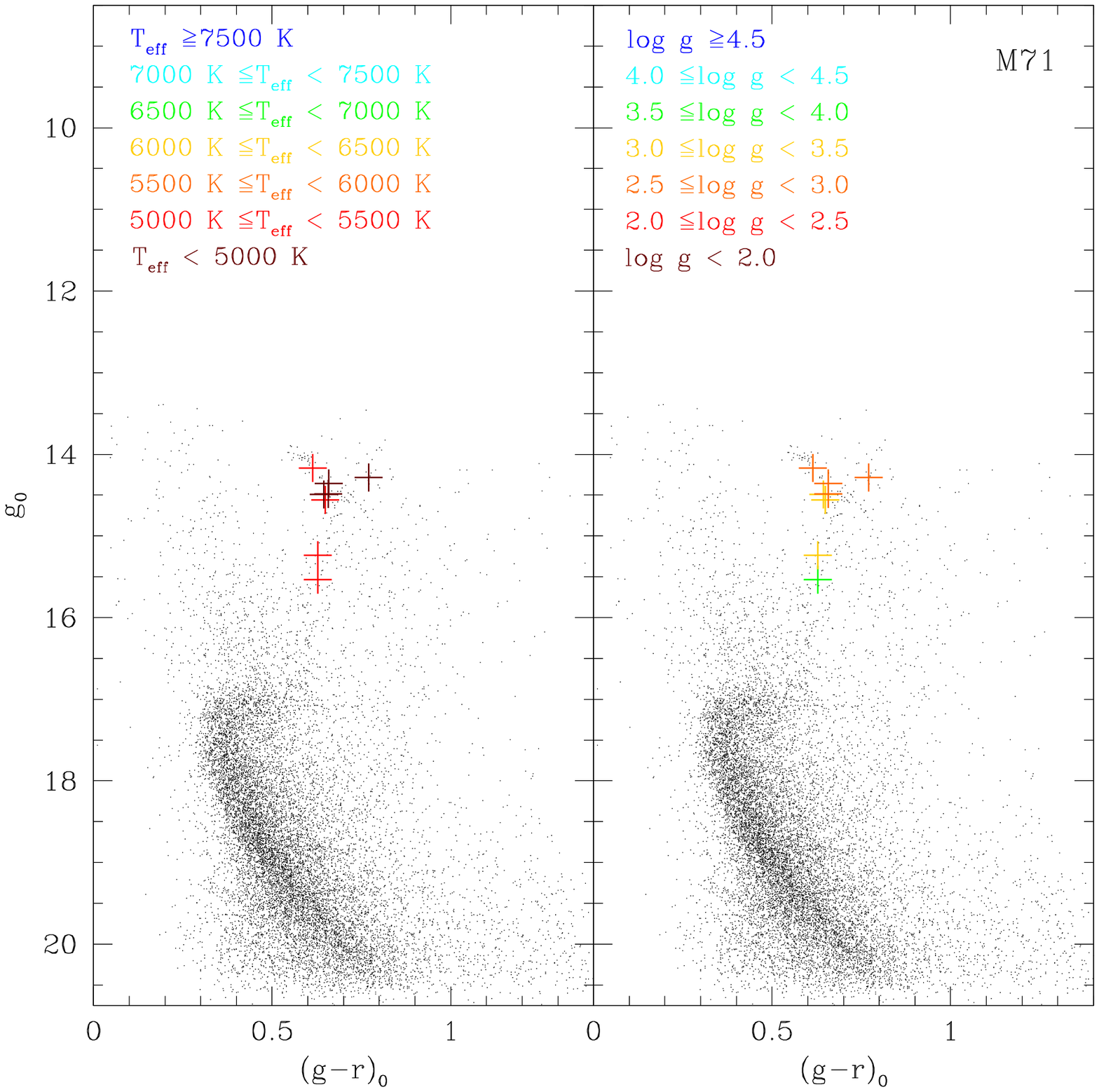}
\caption{Same as Fig. \ref{fig5053cmdfinal}, but for M71.}\label{figm71cmdfinal}
\end{center}
\end{figure}

\begin{figure}
\begin{center}
\plotone{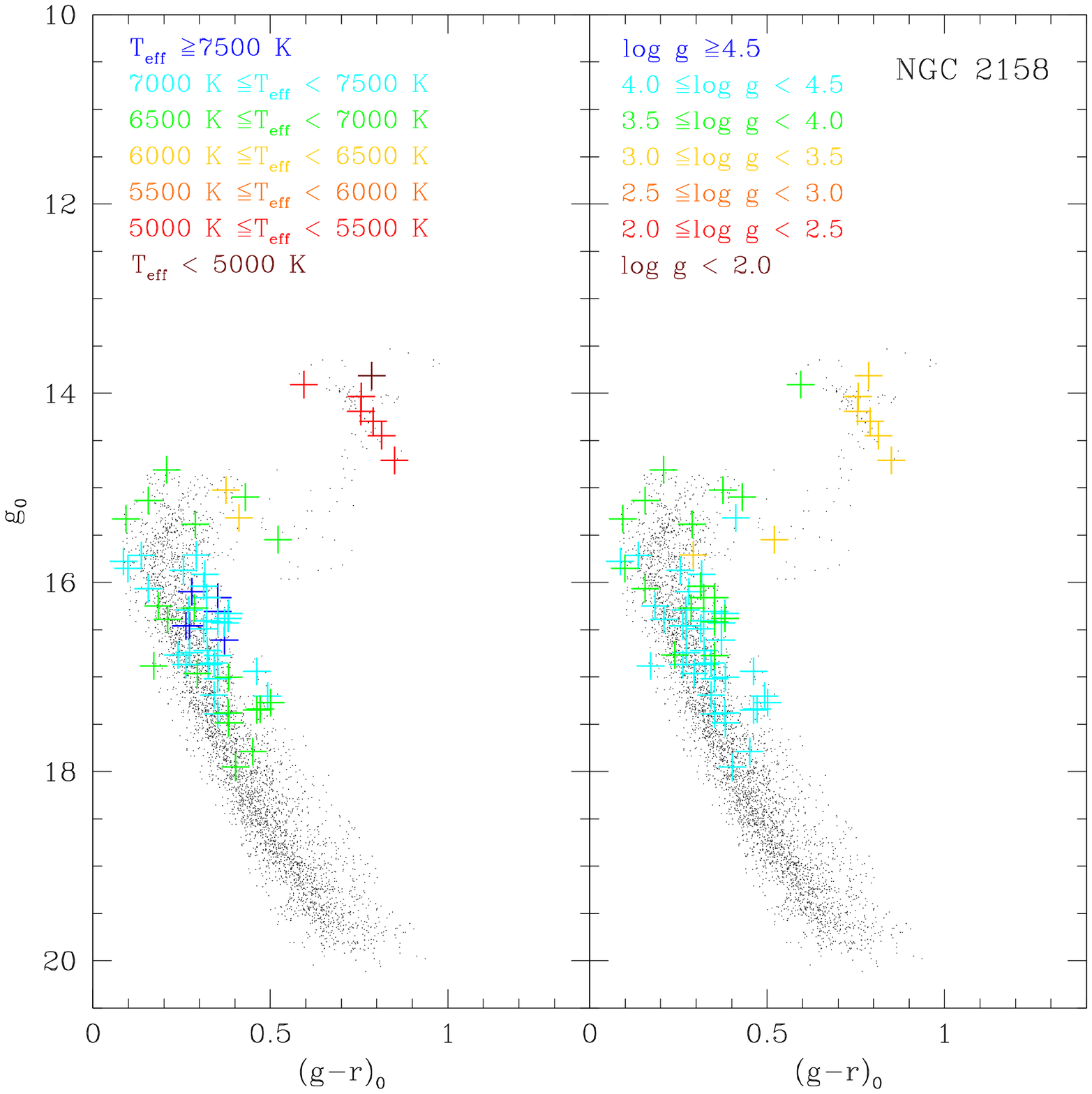}
\caption{Same as Fig. \ref{fig5053cmdfinal}, but for NGC~2158.}\label{fig2158cmdfinal}
\end{center}
\end{figure}

\begin{figure}
\begin{center}
\plotone{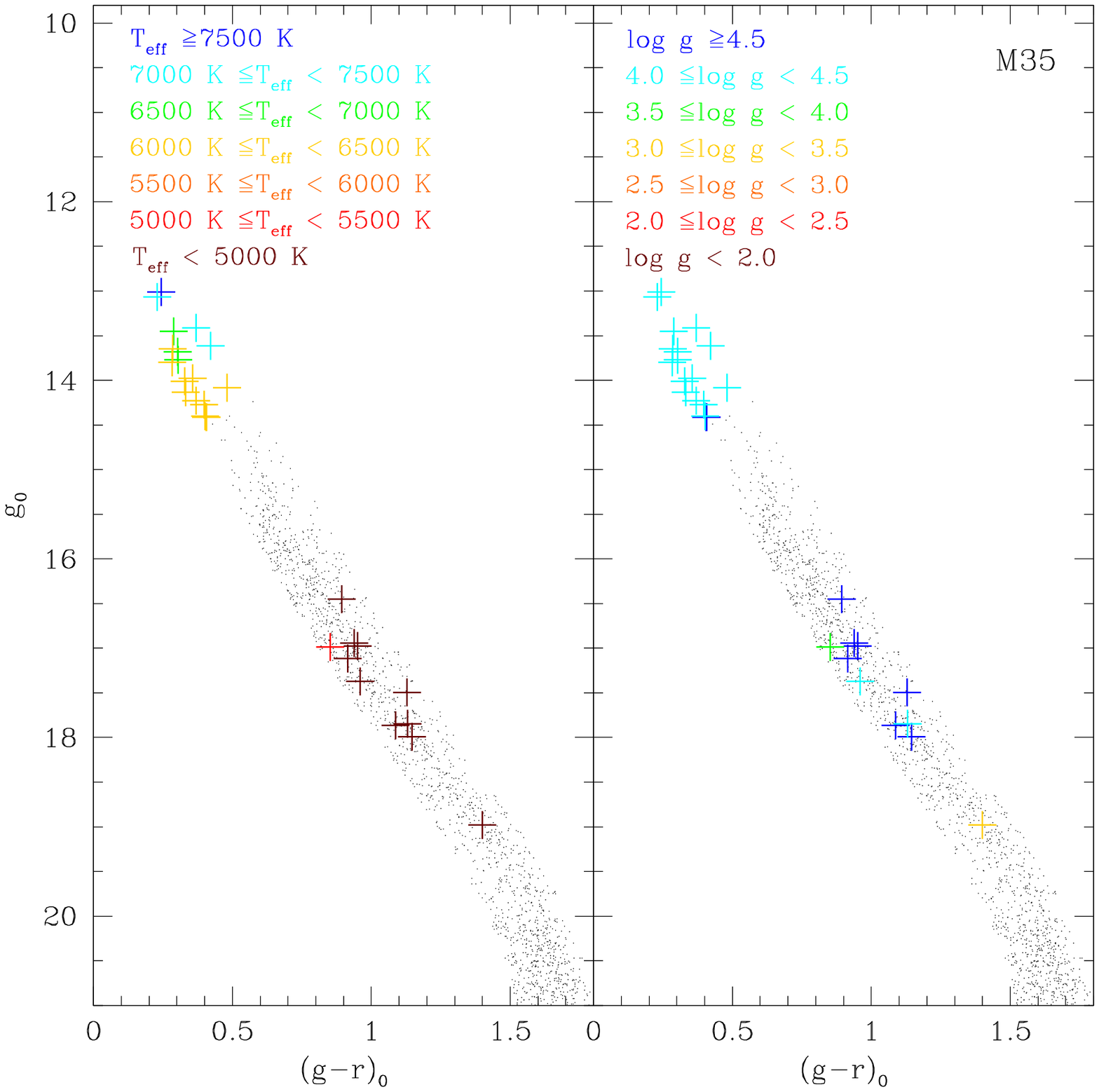}
\caption{Same as Fig. \ref{fig5053cmdfinal}, but for M35.}\label{figm35cmdfinal}
\end{center}
\end{figure}

\begin{figure}
\begin{center}
\plotone{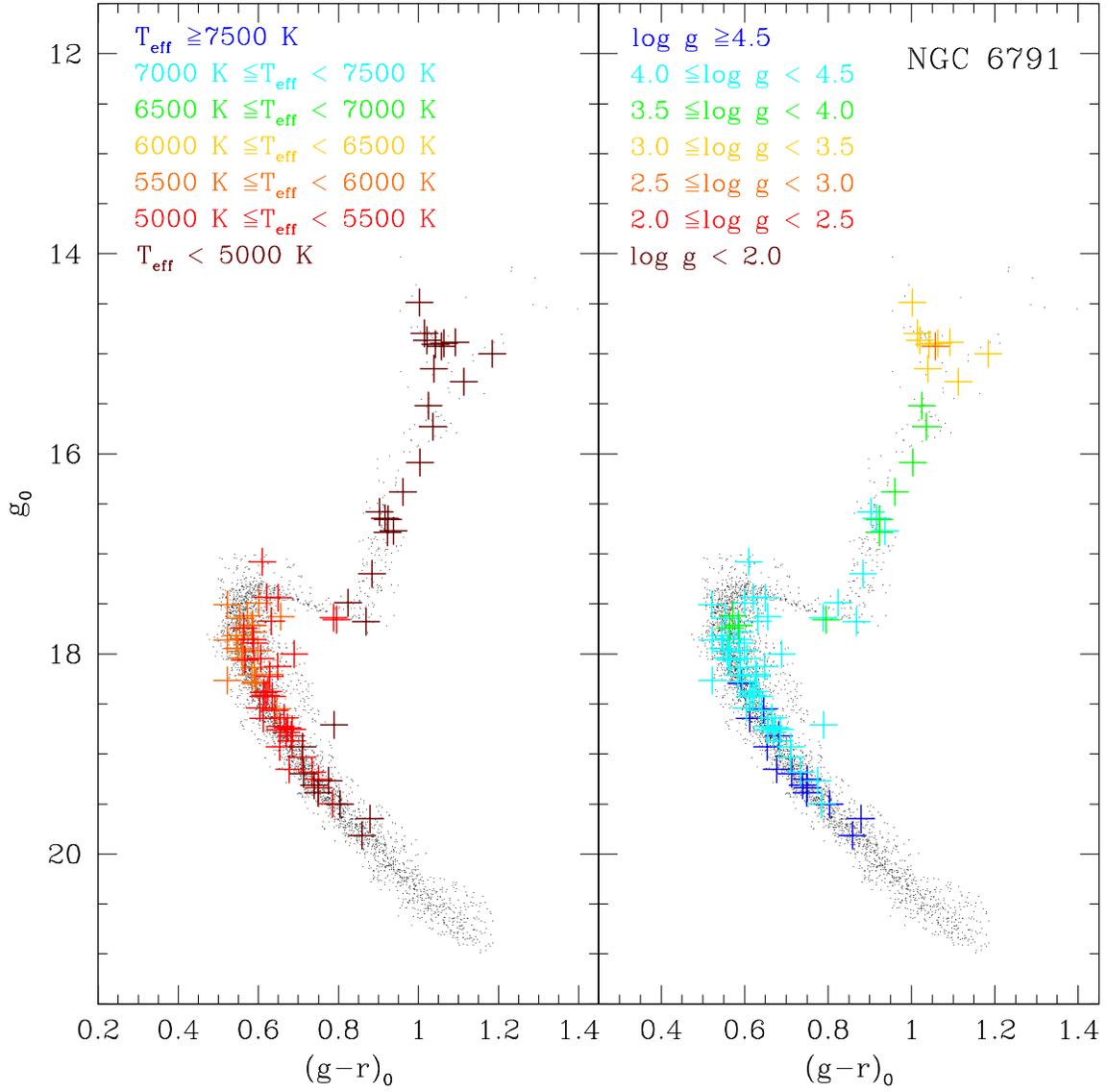}
\caption{Same as Fig. \ref{fig5053cmdfinal}, but for NGC~6791.}\label{fig6791cmdfinal}
\end{center}
\end{figure}

The SSPP-estimated temperatures and surface gravities for true member
stars are plotted in Figures \ref{fig5053cmdfinal}--\ref{fig6791cmdfinal}
over the cleaned CMDs of the likely member stars from the photometric
sample that passed the $s/n$ cut. The spectroscopic data points are
plotted in different colors, in temperature steps of 500 K and log $g$
steps of 0.5 dex. Stars at the top of the MS and on the MSTO have
generally lower S/N than those on the RGB and HB, so the fact that some
non-uniformity is observed in the distribution of $T_{\rm{eff}}$ and log
$g$ in stars near the MSTO is not unexpected.

Table \ref{tabmemprops} lists the SSPP-derived properties for all stars
selected as true cluster members from each cluster, as well as the
extinction-corrected $ugriz$ magnitudes and errors for the photometry
employed.  

\section{Individual Cluster Discussion and Comparison with Previous Studies}\label{seccomp}

Here we examine previous studies of these clusters, and assess how well the
SSPP-derived estimates for cluster metallicity and radial velocity compare with
the values reported in the literature. This section is not intended to be a
comprehensive review, but rather concentrates on high-resolution spectroscopic
results from studies that have been published within the past
decade.\footnote{All references to Harris (1996) refer to the 2003 update on his
web page: {\url http://www.physics.mcmaster.ca/$\sim$harris/mwgc.dat}.} Due to
the relative paucity of radial velocities for some clusters, older studies are
cited where needed. We first consider the globular clusters, followed by the
open clusters, ordered from low metallicity to high metallicity.

\subsection{NGC~5053}\label{subsecngc5053}

NGC~5053 is known to be metal-poor, but has otherwise not been widely studied.
One spectroscopic plug-plate observation produced only 16 true member stars,
with less than optimal coverage inside $r_t$ (see Figure
\ref{fig8radec}). Our estimate of the mean metallicity, $\langle
\rm{[Fe/H]}\rangle = -2.25 \pm 0.25$, is within $1\sigma$ of that
reported by Harris (1996; $-2.29$). The recalibration by Carretta et al. (2009)
reports a value of $-$2.30, with which we are also consistent. 

Our mean radial velocity, $\langle \rm{RV}\rangle = +44.0 \pm 4.9~\rm{km~s^{-1}}$, is
the same as that given by Harris (1996; $+44.0 ~\rm{km~s^{-1}}$).  

\subsection{M92 (NGC~6341)}\label{subsecm92}

Two spectroscopic plug-plate observations of this cluster yielded 
58 true cluster members. Our estimated mean metallicity,
$\langle \rm{[Fe/H]}\rangle = -2.25 \pm 0.17$, is within $1\sigma$ of
the values given by Harris (1996; $-2.28$) and Carretta et al. (2009; 
$-$2.35). While King et al. (1998) obtained a much lower metallicity 
estimate from only \ion{Fe}{1} lines of 6 subgiant stars in their sample
([Fe/H] = $-$2.52), examining the 17 subgiant member stars from this
cluster in our sample reveals a mean metallicity of $-$2.27, in
agreement with our overall mean metallicity as well as with the
metallicities adopted by the Harris and Carretta et al. compilations.
King et al. (1998) acknowledge that their low signal-to-noise spectra
and limited spectral coverage, along with the metal-poor nature of M92
and an uncertain reddening correction, resulted in a degeneracy between
their estimates of \teff ~and microturbulence that may have produced a
lower value for [Fe/H]. In their analysis of literature data, Kraft 
\& Ivans (2003) report abundances from \ion{Fe}{1} and \ion{Fe}{2} lines 
of $-$2.50 and $-$2.38, respectively; both are lower than our result but
consistent with King et al. (1998).

Our SSPP-derived estimate for the radial velocity, $\langle
\rm{RV}\rangle = -116.5 \pm 8.7~\rm{km~s^{-1}}$, is within $1\sigma$ of
that provided by Harris (1996; $-120.3$ km~s$^{-1}$). A recent study by
Drukier et al. (2007) reported a radial velocity of $\rm{RV} = -121.2
~\rm{km~s^{-1}}$, based on a sample of 306 cluster members, which is also in
agreement with our value.  

\subsection{M53 (NGC~5024)}\label{subsecm53}

M53 is located at the edge of the plug-plates for observations of NGC~5053,
resulting in just 50 fibers being placed inside the tidal radius. As a result,
only 19 stars were selected as true members. Our measured mean metallicity,
$\langle \rm{[Fe/H]}\rangle = -2.03 \pm 0.13$, is in agreement with Harris
(1996; $-1.99$) and Carretta et al. (2009; $-$2.06), as well as with most
earlier photometric and spectroscopic abundance studies that indicated a
metallicity lower than $-1.8$ (e.g., Pilachowski et al. 1983). More recently, a
moderate-resolution spectroscopic analysis of member stars from M53 by Lane et
al. (2010) provided a metallicity estimate of $\langle \rm{[Fe/H]}\rangle =
-1.99$, with which our result agrees nicely. Although a recent photometric study
by D\'ek\'any \& Kov\'acs (2009) exhibited a discrepancy in [Fe/H] between
horizontal-branch (variable) stars and stars on the red giant branch, our sample
shows no statistically significant difference between the mean metallicity on
the horizontal branch versus the red-giant branch for this cluster ($\langle
\rm{[Fe/H]}\rangle_{HB} = -2.11 \pm 0.09; \langle \rm{[Fe/H]}\rangle_{RGB} = -1.96 \pm 0.12$). 
Our derived mean metallicity is within $1\sigma$ of their giant-branch mean
metallicity of $-2.12$. 

Radial velocity measurements reported in the literature for this cluster are a
bit more scattered. Harris (1996) reported a value of $-79.1 ~\rm{km~s^{-1}}$,
whereas a more recent medium-resolution spectroscopic study by Lane et al.
(2009), using 180 giant stars, resulted in a mean value of
$-62.8~\rm{km~s^{-1}}$. Our value, $\langle
\rm{RV}\rangle = -59.6 \pm 7.9 ~\rm{km~s^{-1}}$, from 19 RGB and HB
stars, is consistent with the Lane et al. (2009) result.

\subsection{M3 (NGC~5272)}\label{subsecm3}

One spectroscopic plug-plate observation for this cluster produced 77 true
member stars. Our measured value of $\langle\rm{[Fe/H]}\rangle = -1.55 \pm 0.13$
is well within $1\sigma$ of that reported by Harris (1996; $-1.57$) and the
recalibrated scale by Carretta et al. (2009; $-1.50$). A high-resolution
spectroscopic study by Cavallo \& Nagar (2000) of 6 giants at the tip of the RGB
produced an estimate of [Fe/H] = $-$1.54, and an analysis of literature data
performed by Kraft \& Ivans (2003) yielded metallicity estimates from both
\ion{Fe}{1} and \ion{Fe}{2} lines of $-$1.58 and $-$1.50, respectively.
Furthermore, a recent study of 23 RGB stars using high-resolution spectroscopy
from Keck yielded [Fe/H] = $-$1.58 from \ion{Fe}{2} lines (Sneden et al. 2004).
Finally, while our value is only barely within $1\sigma$ of the estimated iron
abundance for M3 from Cohen \& Mel\'endez (2005), who obtained a somewhat
higher value of [Fe/H] = $-$1.39 based on Keck/HIRES spectroscopy, it should be
kept in mind that recent results from Cohen and collaborators adopt a
temperature scale that is several hundred Kelvin warmer than most other
researchers, which could easily accomodate the 0.16 dex offset with respect to
their reported value of metallicity. Thus, our SSPP-derived estimate for [Fe/H]
is in excellent agreement with all of these previous studies, while spanning the
entire length of the RGB, including stars on the horizontal branch as well.

Our estimate of the cluster's mean radial velocity, $\langle
\rm{RV}\rangle = -141.2 ~\rm{km~s^{-1} \pm 5.6}$, is slightly different
those from Harris (1996) and Cohen \& Mel\'endez (2005), who both report the
same value ($-147.6 ~\rm{km~s^{-1}}$), and Sneden et al. (2004), who reported a
mean radial velocity of $-149.4 ~\rm{km~s^{-1}}$. However, it is only just
beyond $1\sigma$ of these values; when accounting for the uncertainty in the
literature values the difference is not significant.

\subsection{M71 (NGC~6838)}\label{subsecm71}

M71 is an important cluster for validation of the SSPP, due to its intermediate
metallicity ([Fe/H] $\sim -0.7)$, a regime that was not represented by
previously considered clusters. Unfortunately, a total of 155 fibers inside the
adopted radius of 4.0 arcmin resulted in just 17 true member stars. Literature
values from Harris (1996; $-0.73$) and a Keck/HIRES study by Boesgaard et al.
(2005; $-0.80$) are both consistent with our value of the mean metallicity,
$\langle \rm{[Fe/H]}\rangle = -0.79 \pm 0.06$, at the $1\sigma$ level, as is
that from Carretta et al. (2009; $-$0.82). In an in-depth analyis using
Keck/HIRES spectroscopy of 25 stars from the turnoff to the RHB, Ram\'irez et
al. (2001) measured iron abundances from \ion{Fe}{1} and \ion{Fe}{2} lines
individually, and compared them against each other for various regions of the
CMD. Their values range from $-0.64$ to $-0.86$, with an error-weighted mean of
$-0.71$, in agreement with our value at the $1.5\sigma$ level. Finally, Kraft \&
Ivans (2003) also report consistent abundances from 
\ion{Fe}{1} and \ion{Fe}{2} lines of $-$0.82 and $-$0.81, respectively.

Our mean radial velocity determination, $\langle \rm{RV}\rangle = -16.9
\pm 9.3 ~\rm{km~s^{-1}}$, is within $1\sigma$ of that reported by Harris (1996; $-22.8
~\rm{km~s^{-1}}$). Keck/HIRES data from Cohen et al. (2001) produced a 
mean radial velocity of $-21.7~\rm{km~s^{-1}}$, which is also consistent
with our observation.

\subsection{NGC~2158}\label{subsecngc2158}

A total of 109 fibers located inside the adopted radius for this open cluster
(6.0 arcmin) resulted in a relatively high yield of 62 true member stars. With
this sample, we measured a mean metallicity of $\langle
\rm{[Fe/H]}\rangle = -0.26 \pm 0.08$. While this is in agreement with
the values from Dias et al. (2002; $-0.25$), a high-resolution spectroscopic
study of one giant star by Jacobson et al. (2009) produced a nearly solar mean
metallicity of $-0.03 \pm 0.14$. However, a more recent follow-up study using
WIYN Hydra spectroscopy at $R \sim 21,000$ for 15 stars in NGC 2158 produced 
a metallicity of [Fe/H] = $-0.28 \pm 0.05$ (H. Jacobson et al. 2010, in preparation), 
a value that is consistent not only with prior studies of this cluster, but with
ours as well.

Using moderate-resolution spectroscopy, Scott (1995) reported a mean radial
velocity for NGC~2158 of $+28.1~\rm{km~s^{-1}}$.  This and the value 
reported by Dias et al. (2002) of $+28.0$ are both consistent with
our measurement of $+27.8 \pm 5.9~\rm{km~s^{-1}}$.  

\subsection{M35 (NGC~2168)}\label{subsecm35}

This open cluster is located at the edge of the plug-plates from the spectroscopic
observations and was not heavily targeted with fibers. As a result, only 72
fibers were located inside the adopted radius, yielding 29 true members. The
adopted radius is less than the tidal radius due to its proximity to NGC~2158.
The field region of NGC~2158 does overlap with the tidal radius of M35, but
this was not problematic for several reasons.  First, stars included in a field
region were never considered for membership so no M35 stars would have been picked 
up and included in NGC~2158 as potential members.  Secondly, the rather different radial
velocities of the two clusters would have ensured that even if some NGC~2158 stars were
considered for membership in M35, they would have been dropped during the RV cut
if not previously.  Finally, due to their differing positions 
on the CMD, any potential M35 stars included in the field region of NGC~2158
would only have served to reduce the $s/n$ in those sub-grid boxes on the CMD of
NGC~2158.  These being sufficiently far from the main locus, this would not cause 
any complications to the member selection for NGC~2158.

Our measured mean metallicity for this cluster, $\langle
\rm{[Fe/H]}\rangle = -0.20 \pm 0.18$, is consistent with that from Dias
et al. (2002; $-0.16$), as well as with the study of Barrado Y Navascu\'es et
al. (2001), who obtained $\langle \rm{[Fe/H]}\rangle = -0.21$ from a
high-resolution spectroscopic analysis of 39 probable cluster members.  

Barrado Y Navascu\'es et al. (2001) measured a mean radial velocity from their
sample of $\langle \rm{RV}\rangle = -8.0 ~\rm{km~s^{-1}}$, a value consistent
with our observation ($-5.0 \pm 6.2$ km s$^{-1}$). While our value of $\langle
\rm{RV}\rangle$ is slightly higher, compared to both their sample and the value
from Dias et al. (2002; $-8.2$), it is still within $1\sigma$, and therefore can
be considered reliable. A more recent study by Geller et al. (2010) produced a
radial velocity of $\langle \rm{RV}\rangle = -8.16 ~\rm{km~s^{-1}}$ based on
high-resolution spectroscopy.

\subsection{NGC~6791}\label{subsecngc6791}

NGC~6791 is another important cluster for our validation exercise, because it
explores the super-solar metallicity region. This is another regime that was not
considered with previously observed clusters; it is the most metal-rich cluster
(to date) for which we were able to obtain successful spectroscopic reductions.
There were two spectroscopic plug-plate observations for the region surrounding
this cluster, which yielded a total of 90 true members. While our mean
metallicity estimate, $\langle \rm{[Fe/H]}\rangle = +0.31 \pm 0.13$, is
statistically consistent with that given by Dias et al. (2002; $+0.11$) at the
$2\sigma$ level, their reported value is significantly lower than that reported
by other studies. It is known that NGC~6791 is a metal-rich open cluster, with
some estimates from high-resolution spectroscopy as high as $+0.47$ (Gratton et
al. 2006). A study of 24 giant stars with medium-resolution spectroscopy yielded
a metallicity estimate of [Fe/H] = +0.32 (Worthey \& Jowett 2003), while Origlia
et al. (2006) used medium-high resolution Keck/NIRSPEC spectroscopy to obtain an
iron abundance of $+0.35$. Most recently, a high-resoluytion spectroscopic study
of two MSTO stars by Boesgaard et al. (2009) yielded a value of [Fe/H] =
$+0.30$. It is clear that our estimate is in better agreement with these recent
high-resolution observations.

Our measured value of the mean radial velocity, $\langle
\rm{RV}\rangle = -47.0 \pm 6.0 ~\rm{km~s^{-1}}$, is consistent with
that reported by Dias et al. (2002; $-57$ km s$^{-1}$) at the 1.5$\sigma$ level,
as well as with that found by Origlia et al. (2006; $-52~\rm{km~s^{-1}}$).

\section{Comparison of SSPP-7 with SSPP-P8}\label{secdrcomp}

The SSPP has been modified slightly from the version used to produce atmospheric
parameter estimates for stars in SDSS DR7; for clarity, we refer to that
version as SSPP-7. We refer to the current version as SSPP-P8 (for pre-DR8),
since it is anticipated that a number of additional improvements will be made
prior to its application to SDSS DR8. The updates and improvements that have
been made since SSPP-7 are discussed in detail in Appendix A.  

The spectroscopic data from Paper II for the three Galactic globular
clusters M2, M13, and M15, along with the two open clusters M67 and
NGC~2420, have been analyzed with the new version of the SSPP; results are
listed in Table \ref{tabyoungsclusters} alongside those obtained from application of
SSPP-7. The upper section of this table lists results for a final cut on true
members performed using both [Fe/H] and RV, while the lower section shows
results for a final cut using RV alone. Paper II concluded that an RV
cut is sufficient for stars inside a cluster $r_t$ to obtain 
reliable results; this same conclusion is supported by the SSPP-P8 results.
Inspection of this table also reveals clear improvements at the low-metallicity
end of the scale, as compared to literature values from Harris (1996) and
high-resolution spectroscopy reported by Carretta et al. (2009), Anthony-Twarog 
et al. (2006), and Randich et al. (2006), in particular
for M15. The results for the two higher metallicity clusters are mixed, with
M67 at the high-metallicity end showing moderate improvement.  H. Jacobson et al.
(2010, in preparation) report a higher metallicity for NGC~2420 of 
[Fe/H] = $-0.22 \pm 0.07$, which shows closer agreement with our improved SSPP-P8 value.

\section{Summary}\label{secsummary}

We have used spectroscopic and photometric data from SDSS-I and
SDSS-II/SEGUE to determine mean metallicities and radial velocities for
five Galactic GCs, M3, M53, M71, M92, and NGC~5053, as
well as for three OCs, M35, NGC~2158, and NGC~6791.
The data was run through the current version of the SSPP
(which is similar to that which will be used for the next public
release, DR8), and true member stars were selected from each cluster.
The derived $\langle$[Fe/H]$\rangle$ and $\langle$RV$\rangle$ for the true members were
then compared to the cluster properties reported in the literature.

The mean values of [Fe/H] and RV for each cluster from the SSPP are in
good agreement with those values reported in previous studies. Nearly
all of the SSPP estimates are within $1\sigma$ of the adopted literature
values, with the exceptions almost all falling within $2\sigma$. The mean internal
uncertainties of the SSPP-determined metallicities and radial velocities
for true members in our sample are $\sigma_{\rm{[Fe/H]}} = 0.05$ dex and
$\sigma_{\rm{RV}} = 3.0 ~\rm{km~s^{-1}}$, respectively, while the
scatter about the mean residuals compared to the adopted literature
values are $\sigma_{\rm{[Fe/H]}} = 0.11$ dex and $\sigma_{\rm{RV}} = 5.2
~\rm{km~s^{-1}}$, demonstrating good internal and external consistency,
and indicating that estimates of the atmospheric parameters and radial
velocities for SDSS/SEGUE stellar data are sufficiently accurate for use
in studies of the chemical compositions and kinematics of stellar
populations in the Galaxy. 

\acknowledgements

J.P.S., T.C.B., Y.S.L., and T.S. acknowledge partial funding of this work
from grants PHY 02-16783 and PHY 08-22648: Physics Frontiers
Center/Joint Institute for Nuclear Astrophysics (JINA), awarded by the
U.S. National Science Foundation. 

Funding for the SDSS and SDSS-II has been provided by the Alfred P.
Sloan Foundation, the Participating Institutions, the National
Science Foundation, the U.S. Department of Energy, the National
Aeronautics and Space Administration, the Japanese Monbukagakusho,
the Max Planck Society, and the Higher Education Funding Council for
England. The SDSS Web Site is {\tt http://www.sdss.org/}.

The SDSS is managed by the Astrophysical Research Consortium for the
Participating Institutions. The Participating Institutions are the American
Museum of Natural History, Astrophysical Institute Potsdam, University of Basel,
University of Cambridge, Case Western Reserve University, University of Chicago,
Drexel University, Fermilab, the Institute for Advanced Study, the Japan
Participation Group, Johns Hopkins University, the Joint Institute for Nuclear
Astrophysics, the Kavli Institute for Particle Astrophysics and Cosmology, the
Korean Scientist Group, the Chinese Academy of Sciences (LAMOST), Los Alamos
National Laboratory, the Max-Planck-Institute for Astronomy (MPIA), the
Max-Planck-Institute for Astrophysics (MPA), New Mexico State University, Ohio
State University, University of Pittsburgh, University of Portsmouth, Princeton
University, the United States Naval Observatory, and the University of
Washington.

\clearpage
\begin{deluxetable}{lccccc|ccc}
\tabletypesize{\scriptsize}
\rotate
\tablecolumns{9}
\tablewidth{0pt}
\tablecaption{Literature Properties of Globular and Open Clusters}
\tablehead{ \colhead{Parameter} & \colhead{NGC~5053} & \colhead{M92} & \colhead{M53} & \colhead{M3} & \colhead{M71} & \colhead{NGC 2158} & \colhead{M35} & \colhead{NGC 6791}\\
\colhead{} & \colhead{} & \colhead{(NGC 6341)} & \colhead{(NGC 5024)} & \colhead{(NGC 5272)} & \colhead{(NGC 6838)} & \colhead{} & \colhead{(NGC 2168)} & \colhead{}}
\startdata
\rm RA (J2000)              & 13:16:27.0               & 17:17:07.3                & 13:12:55.2                & 13:42:11.2                & 19:53:46.1                & 06:07:25                  & 06:08:54                 & 19:20:53         \\
\rm DEC (J2000)             & $+$17:41:53              & $+$43:08:11               & $+$18:10:08.4             & $+$28:22:32               & $+$18:46:42               & $+$24:05:48               & $+$24:20:00              & $+$37:46:18      \\ 
$(l, b)$                    & (335.7, $+$78.9)         & (68.3, $+$34.9)           & (333.0, $+$79.8)          & (42.2, $+$78.7)           & (56.7, $-$4.6)            & (186.6, $+$1.8)           & (186.6, $+$2.2)          & (70.0, $+$10.9)  \\
\rm [Fe/H]                  & $-$2.29\tablenotemark{a} & $-$2.28\tablenotemark{a}  & $-$1.99\tablenotemark{a}  & $-$1.57\tablenotemark{a}  & $-$0.73\tablenotemark{a}  & $-$0.25\tablenotemark{b}  & $-$0.16\tablenotemark{b} & $+$0.30\tablenotemark{c} \\
$\rm{[Fe/H]_C}$	 	    & $-$2.30		       & $-$2.35                   & $-$2.06                   & $-$1.50                   & $-$0.82                   & \nodata                   & \nodata                  & \nodata   \\
\rm $(m - M)_{0}$           & 16.12\tablenotemark{d}   & 14.64\tablenotemark{e}    & 16.25\tablenotemark{a}    & 14.95\tablenotemark{f}    & 12.86\tablenotemark{f}    & 12.80\tablenotemark{g}    & 9.80\tablenotemark{h}    & 12.95\tablenotemark{f}   \\
\rm V$_{r}~(\rm km~s^{-1}$) & $+$44.0\tablenotemark{a} & $-$120.3\tablenotemark{a} & $-$79.1\tablenotemark{a}  & $-$147.6\tablenotemark{a} & $-$22.8\tablenotemark{a}  & $+$28.0\tablenotemark{b}  & $-$8.2\tablenotemark{b}  & $-$57.0\tablenotemark{b}  \\
$E(B-V)$                    & 0.017\tablenotemark{e}   & 0.023\tablenotemark{e}    & 0.021\tablenotemark{e}    & 0.013\tablenotemark{e}    & 0.275\tablenotemark{i}    & 0.44\tablenotemark{j}     & 0.20\tablenotemark{h}    & 0.117\tablenotemark{b}    \\
\enddata
\tablecomments{Properties of the clusters in our sample as drawn from the literature, divided into globular clusters (left) and open clusters (right).
The parameter r$_{t}$ is the tidal radius in arc minutes for globular clusters or the apparent radius for open clusters.
Exceptions to this are noted in the Fig. \ref{fig8radec}.  The listed
distance modulus $(m-M)_0$ is extinction corrected.  The parameter
$\rm{[Fe/H]_C}$ is from the recalibrated globular cluster metallicity scale of Carretta et al. (2009).  
References: $^{\rm{a}}$ Harris (1996); $^{\rm{b}}$ Dias et al. (2002); $^{\rm{c}}$ Boesgaard et al. (2009); $^{\rm{d}}$ Arellano Ferro et al. (2010); 
$^{\rm{e}}$ Schlegel et al. (1998); $^{\rm{f}}$ An et al. (2009); $^{\rm{g}}$ Carraro et al. (2002); $^{\rm{h}}$ Kalirai et al. (2003); 
$^{\rm{i}}$ Grundahl et al. (2002); $^{\rm{j}}$ Twarog et al. (1997)}\label{tabclusterprops}
\end{deluxetable}

\clearpage
\begin{deluxetable}{lccccccc}
\tabletypesize{\footnotesize}
\renewcommand{\tabcolsep}{5pt}
\tablecolumns{8}
\tablewidth{0pt}
\tablecaption{Measured Metallicities and Radial Velocities of Globular and Open Clusters}
\tablehead{ \colhead{Cluster} & \colhead{$\langle$[Fe/H]$\rangle$} & \colhead{$\sigma$([Fe/H])} & \colhead{$\sigma_{\mu}$([Fe/H])} & \colhead{$\langle$RV$\rangle$} & \colhead{$\sigma$(RV)} & \colhead{$\sigma_{\mu}$(RV)} & \colhead{$N$} \\
\colhead{} & \colhead{} & \colhead{(dex)} & \colhead{(dex)} & \colhead{(km $\rm{s^{-1}}$)} & \colhead{(km $\rm{s^{-1}}$)} & \colhead{(km $\rm{s^{-1}}$)} & \colhead{}}
\startdata
\rm NGC~5053  &  $-$2.26  &  0.25  &  0.06  &  $+$44.0   &  4.9  &  1.2  &  16  \\
\rm M92      &  $-$2.25  &  0.17  &  0.02  &  $-$116.5  &  8.7  &  1.1  &  58  \\
\rm M53      &  $-$2.03  &  0.13  &  0.03  &  $-$59.6   &  7.9  &  1.8  &  19  \\
\rm M3       &  $-$1.55  &  0.14  &  0.02  &  $-$141.2  &  5.6  &  0.6  &  77  \\
\rm M71      &  $-$0.79  &  0.06  &  0.01  &  $-$16.9   &  9.3  &  2.3  &  17  \\
\hline\\[-5pt]
\rm NGC~2158  &  $-$0.26  &  0.08  &  0.01  &  $+$27.8   &  5.9  &  0.7  &  62  \\
\rm M35      &  $-$0.20  &  0.18  &  0.03  &  $-$5.0    &  6.2  &  1.2  &  29  \\
\rm NGC~6791  &  $+$0.31  &  0.13  &  0.01  &  $-$47.0   &  6.0  &  0.6  &  90  \\
\enddata
\tablecomments{Columns 2 and 5 list the measured mean values of [Fe/H] and RV for each cluster, while columns
3 and 6 list the $1\sigma$ spread of each value.  Columns 4 and 7 are the
standard errors in the mean ($\sigma_{\mu}$) of the estimates.
$N$ lists the number of true member stars for
each cluster determined by the final application of the 2$\sigma$ range to the mean of the Gaussian fits on
[Fe/H] and RV.}\label{tabclusterfehrv}
\end{deluxetable}

\clearpage
\begin{deluxetable}{lccccccc}
\tabletypesize{\footnotesize}
\renewcommand{\tabcolsep}{12pt}
\tablecolumns{8}
\tablewidth{0pt}
\tablecaption{Linear Regression on [Fe/H] Residuals}
\tablehead{ \colhead{Cluster} & \colhead{$N$} & \colhead{Parameter} & \colhead{$X$} & \colhead{$\sigma_X$} & \colhead{$Y$} & \colhead{$\sigma_Y$} & \colhead{$R^2$} \\
\colhead{(1)} & \colhead{(2)} & \colhead{(3)} & \colhead{(4)} & \colhead{(5)} & \colhead{(6)} & \colhead{(7)} & \colhead{(8)}}
\startdata
\rm NGC 5053  &  16  &  $(g-r)_0$ &  $-$0.585  &  0.164  &  +0.347    &  0.090  &  0.475  \\
              &      &  S/N       &  $-$0.007  &  0.005  &  +0.333    &  0.182  &  0.133  \\
\rm M92      &  58  &  $(g-r)_0$ &  $-$0.347  &  0.145  &  $+$0.229  &  0.053  &  0.093  \\
              &      &  S/N       &  $-$0.001  &  0.002  &  $+$0.135  &  0.051  &  0.004  \\
\rm M53      &  19  &  $(g-r)_0$ &  $+$0.166  &  0.138  &  $+$0.027  &  0.066  &  0.078  \\
              &      &  S/N       &  $-$0.005  &  0.006  &  +0.192    &  0.132  &  0.048  \\
\rm M3       &  77  &  $(g-r)_0$ &  $-$0.219  &  0.056  &  +0.059    &  0.032  &  0.071  \\
              &      &  S/N       &  $+$0.001  &  0.001  &  $-$0.085  &  0.044  &  0.022  \\
\enddata
\tablecomments{The variables $X$ and $Y$ are the slope and zero-points, respectively,  of a
linear regression on the residuals in our measured [Fe/H] values and those adopted from the
literature, along with the corresponding uncertainties from the regression. 
The parameter $R^2$ indicates the fraction of the variance accounted for by
the correlations in the variables $(g-r)_0$ and S/N for each cluster.}
\label{tabfehreg}
\end{deluxetable}

\clearpage
\begin{deluxetable}{clccccr|ccccr|ccc}
\tabletypesize{\scriptsize}
\rotate
\renewcommand{\tabcolsep}{3pt}
\tablecolumns{15}
\tablewidth{0pt}
\tablecaption{Comparison of Estimated Cluster Parameters by SSPP-7 and SSPP-P8 }
\tablehead{ 
\colhead{Cut} & \colhead{Cluster} & \colhead{$\langle$[Fe/H]$\rangle_7$} & \colhead{$\sigma$([Fe/H])$_7$} & \colhead{$\langle$RV$\rangle_7$} & \colhead{$\sigma$(RV)$_7$} & \colhead{$N_7$}
   & \colhead{$\langle$[Fe/H]$\rangle_{P8}$} &
\colhead{$\sigma$([Fe/H])$_{P8}$} &
\colhead{$\langle$RV$\rangle_{P8}$} & \colhead{$\sigma$(RV)$_{P8}$} &
\colhead{$N_{P8}$} & \colhead{[Fe/H]$_{\rm{H}}$} & \colhead{RV$_{\rm{H}}$} & \colhead{[Fe/H]$_{\rm{HR}}$} \\
\colhead{} & \colhead{} & \colhead{} & \colhead{(dex)} & \colhead{(km $\rm{s^{-1}}$)} & \colhead{(km $\rm{s^{-1}}$)} & \colhead{} & \colhead{} & \colhead{(dex)} & \colhead{(km $\rm{s^{-1}}$)} & \colhead{(km $\rm{s^{-1}}$)} & \colhead{} & \colhead{} & \colhead{(km $\rm{s^{-1}}$)} & \colhead{} \\
\colhead{} & \colhead{(1)} & \colhead{(2)} & \colhead{(3)} & \colhead{(4)} & \colhead{(5)} & \colhead{(6)} & \colhead{(7)} & \colhead{(8)} & \colhead{(9)} & \colhead{(10)} & \colhead{(11)} & \colhead{(12)} & \colhead{(13)} & \colhead{(14)}
}
\startdata
[Fe/H] $\&$ RV &              &         &      &          &      &     &         &      &          &      &     &         &          &         \\
               & \rm M15      & $-$2.19 & 0.17 & $-$108.2 & 11.7 & 98  & $-$2.31 & 0.21 & $-$109.0 & 11.5 & 98  & $-$2.26 & $-$107.0 & $-$2.33 \\
               & \rm M2       & $-$1.52 & 0.18 & $-$2.1   & 9.8  & 76  & $-$1.61 & 0.13 & $-$2.2   & 10.3 & 71  & $-$1.62 & $-$5.3   & $-$1.66 \\
               & \rm M13      & $-$1.59 & 0.13 & $-$244.8 & 8.8  & 293 & $-$1.63 & 0.13 & $-$244.8 & 8.7  & 293 & $-$1.54 & $-$245.6 & $-$1.58 \\
               & \rm NGC~2420  & $-$0.38 & 0.10 & +75.1    & 5.9  & 163 & $-$0.31 & 0.11 & +75.0    & 5.9  & 164 & \nodata & +75.5    & $-$0.37 \\  
               & \rm M67      & $-$0.08 & 0.07 & +34.9    & 4.1  & 52  & $-$0.01 & 0.08 & +35.0    & 3.4  & 75  & \nodata & +32.9    & +0.05   \\[+5pt]
\hline\\[-5pt]
RV             &               &         &      &          &      &     &         &      &          &      &     &         &          &         \\
\enddata
\tablecomments{Comparison of SSPP-estimated parameters from Paper II,
which used the DR7 version of the SSPP (SSPP-7), with those produced by the
pre-DR8 version (SSPP-P8). Columns 2-6 list parameters yielded by
SSPP-7, and Columns 7-11 list
parameters yielded by SSPP-P8. Columns 12-13 contain literature
values from Harris (1996, Columns 12 and 13), while values from  high-resolution
spectroscopy reported by Carretta et al. (2009; M15, M13, and M2) 
and Randich et al. (2006; M67) are listed in Column 14.  The value
given in Column 14 for NGC~2420 (Anthony-Twarog et al. 2006) is derived
from Stromgren photometry, not high-resolution spectroscopy, whereas 
H. Jacobson et al. (2010, in preparation) reports a metallicity result 
from high-resolution spectroscopy of $-0.22$.  Moderate
improvement in the [Fe/H] estimates is seen at both lower and 
higher metallicities.
The upper section of the table contains estimates based on a final
true member cut using both [Fe/H] estimates as well as radial
velocities, whereas the lower section contains estimates based on
a final cut using radial velocities alone.}\label{tabyoungsclusters}
\end{deluxetable}

\clearpage
\begin{deluxetable}{lccrccccccccccccccccccc}
\tabletypesize{\scriptsize}
\rotate
\renewcommand{\tabcolsep}{1pt}
\tablecolumns{23}
\tablewidth{0in}
\tablecaption{Properties of Adopted True Member Stars}

\tablehead{
\colhead{spSpec name} & \colhead{$\alpha$} & \colhead{$\delta$} & \colhead{RV} &
\colhead{$\sigma_{\rm{RV}}$} & \colhead{$\rm{T_{eff}}$} & \colhead{$\sigma_{\rm{T_{eff}}}$} &
\colhead{log $g$} & \colhead{$\sigma_{\rm{log} \emph{g}}$} & \colhead{[Fe/H]} & \colhead{$\sigma_{\rm{[Fe/H]}}$} &
\colhead{$u$} & \colhead{$\sigma_u$} & \colhead{$g$} & \colhead{$\sigma_g$} & \colhead{$r$} &
\colhead{$\sigma_r$} & \colhead{$i$} & \colhead{$\sigma_i$} & \colhead{$z$} & \colhead{$\sigma_z$} &
\colhead{$\langle \rm{S/N} \rangle$} & \colhead{Tag}\\

\colhead{} & \colhead{(deg)} & \colhead{(deg)} & \colhead{(km $\rm{s^{-1}}$)} &
\colhead{(km $\rm{s^{-1}}$)} & \colhead{(K)} & \colhead{(K)} &
\colhead{} & \colhead{(dex)} & \colhead{} & \colhead{(dex)} &
\colhead{} & \colhead{} & \colhead{} & \colhead{} & \colhead{} &
\colhead{} & \colhead{} & \colhead{} & \colhead{} & \colhead{} &
\colhead{} & \colhead{}}

\startdata
 & & & & & & & & & NGC~5053 & & & & & & & & & & & & & \\
2476-53826-486  &   199.04518  &    17.60554  &    46.8  &  6.4  &  5287  &   101  &  1.99  &  0.47  & $-$2.41  &  0.04  &     18.887  &      0.022  &     17.746  &      0.009  &     17.284  &      0.009  &     17.078  &      0.012  &     16.978  &      0.020  &  17.5  &	D \\
2476-53826-488  &   199.09269  &    17.69851  &    42.5  &  2.2  &  4951  &    87  &  2.00  &  0.21  & $-$2.14  &  0.06  &     17.349  &      0.013  &     15.780  &      0.017  &     15.094  &      0.007  &     14.797  &      0.011  &     14.656  &      0.011  &  49.1  &	D \\
2476-53826-490  &   199.07441  &    17.62914  &    37.0  &  4.3  &  8452  &   171  &  3.08  &  0.28  & $-$2.10  &  0.03  &     17.806  &      0.013  &     16.483  &      0.015  &     16.631  &      0.009  &     16.782  &      0.008  &     16.835  &      0.016  &  30.0  &	D \\
2476-53826-497  &   199.08809  &    17.59394  &    36.2  &  9.8  &  5397  &    87  &  2.46  &  0.10  & $-$1.90  &  0.08  &     19.352  &      0.031  &     18.201  &      0.011  &     17.750  &      0.015  &     17.562  &      0.011  &     17.483  &      0.019  &  12.4  &	D \\
2476-53826-501  &   199.16802  &    17.67369  &    43.4  &  2.6  &  4973  &    52  &  2.11  &  0.25  & $-$2.56  &  0.07  &     17.417  &      0.010  &     15.988  &      0.008  &     15.356  &      0.005  &     15.073  &      0.012  &     14.942  &      0.011  &  48.4  &	D \\
2476-53826-505  &   199.19265  &    17.70156  &    46.8  &  3.3  &  5353  &    63  &  1.93  &  0.28  & $-$2.37  &  0.04  &     17.522  &      0.015  &     16.302  &      0.009  &     15.817  &      0.008  &     15.600  &      0.015  &     15.520  &      0.015  &  39.6  &	D \\
2476-53826-506  &   199.15790  &    17.64537  &    46.8  &  5.0  &  8072  &   116  &  3.51  &  0.23  & $-$1.76  &  0.08  &     17.793  &      0.019  &     16.589  &      0.012  &     16.693  &      0.008  &     16.745  &      0.013  &     16.830  &      0.022  &  29.4  &	D \\
\enddata

\tablecomments{SSPP-derived properties of the true member stars selected from all clusters in our sample.  
Column 1 lists the spSpec name, which identifies the star on the spectral plate in the form of spectroscopic
plug-plate number (four digits), Modified Julian Date (five digits), and fiber used (three digits).  For 
details on how the uncertainties in these parameters are estimated, see Paper I.  Values with an ellipsis
were problematic and have been omitted.  The final column indicates whether photometric values were drawn 
from ``Best'' photometry (B), the ``Uber calibration'' (U), the CASJOBS database (C), or the DAOPHOT 
crowded-field reduction (D).}\label{tabmemprops}
\end{deluxetable}

\clearpage

\clearpage

\clearpage

\clearpage

\clearpage

\clearpage

\clearpage

\clearpage

\clearpage

\clearpage

\clearpage

\clearpage

\clearpage

\clearpage

\clearpage

\clearpage

\clearpage

\clearpage

\clearpage

\clearpage

\clearpage

\clearpage

\clearpage

\clearpage

\clearpage

\clearpage

\clearpage

\clearpage
\setcounter{table}{0}
\setcounter{figure}{0}
\setcounter{equation}{0}
\appendix
\renewcommand*\thetable{\Alph{section}\arabic{table}}
\renewcommand*\thefigure{\Alph{section}\arabic{figure}}
\section{Appendix: Changes in SSPP-7 in Preparation for SSPP-8}\label{secappintro}

In the period since SDSS Data Release 7 (DR7; Abazajian et al. 2009),
the SEGUE Stellar Parameter Pipeline (SSPP; Paper I) has evolved
somewhat, in order to improve our estimates of the stellar parameters
\teff, \logg, and \feh. In the version of the SSPP used for DR7, there were six
primary temperature estimates and an auxiliary set of five empirically
and theoretically determined estimates. For surface gravity estimation,
ten methods were employed. Twelve different methods were employed to
determine [Fe/H]. Depending on a star's \gr~and the S/N of the
spectrum, an indicator variable (taking on values of 0 or 1) was
assigned for each technique used for a given parameter estimate. Following application of a
parameter decision tree, all available estimates from individual methods
for each parameter were combined to yield final adopted values. Details
on each method and the decision tree for each parameter can be found in
Paper I.

At the time the DR7 version of the SSPP was constructed there existed a
dearth of metal-rich (\feh~$> 0.0$) and metal-poor (\feh~$< -3.0$) stars
available as calibrators, hence the metallicity determinations at the
extrema were not well-constrained. Since then, we have obtained data for more
metal-poor and metal-rich clusters, including the important clusters
M92 and NGC~6791, and also secured more SDSS/SEGUE stars with available
high-resolution spectroscopy. These enabled substantial improvement in
parameter estimates for SDSS/SEGUE stellar spectra.

Here we highlight major and minor changes that have been made on the
SSPP since the DR7 version; the new version of the SSPP is referred to
as SSPP-P8, as a version similar to this will be used for application to
Data Release 8 (DR8), scheduled for January 2011. The version of the
SSPP used for DR7 is referred to as SSPP-7. Here, {\it major change}
indicates that the modification described directly affects the parameter
estimation for each method, and hence the final adopted value, whereas
{\it minor change} indicates that the modification doesn't influence
parameter estimation, but helps to more easily identify peculiar
behavior in the observed spectra, or possibly the presence of
ill-measured parameters.

\subsection {Major Changes in the SSPP}\label{subsecssppchgs}

Since there are no substantial changes in the methodology for estimating
$T_{\rm eff}$ and log $g$, or in the averaging scheme employed to obtain
final adopted values, we focus on modifications made to obtain improved
metallicity estimates. However, note that the final adopted value of
$T_{\rm eff}$ and log $g$ estimates are slightly different in
SSPP-P8, due to the re-calibration of the {\tt NGS1} and {\tt NGS2}
approaches, and to some additional changes in the validity ranges of S/N
and \gr. The basic ideas for deciding which estimator goes in the final
averaging stage for those two parameters and the nomenclature for each
method can be found in Paper I.

\subsubsection{Changes in S/N and $(g-r)_{0}$ Ranges for Individual Methods}\label{subsubsecsngrchgs}

The valid ranges of S/N and $(g-r)_{0}$ for each method mostly remain the same
as before, but the color range for application of the {\tt WBG} method is
substantially narrowed, since it is based on a grid of synthetic spectra that
only extends to [Fe/H] = 0.0, thus it is not applicable for the full range
of expected metallicities for metal-rich G- and K-type stars. In SSPP-7, 
its use lowered the overall metallicity estimates for stars with super-solar
metallicity ([Fe/H] $>$ 0.0). Table \ref{tabapptable1} summarizes the current status of the S/N
and $(g-r)_{0}$ ranges for individual methods.

\subsubsection{Re-Calibration of the {\tt NGS1} and {\tt NGS2} Methods}\label{subsubsecrecal}

The {\tt NGS2} method implements a dense and extended grid of synthetic spectra,
spanning from 4000 K $\leq T_{\rm eff} \leq$ 8000 K in steps of 250 K, 0.0 $\leq
\log g \leq $ 5.0 in steps of 0.2 dex, and $-4.0 \leq \rm [Fe/H]
\leq$ +0.4 in steps of 0.2 dex. The [$\alpha$/Fe] ratio covers
from $-0.1 \leq [\alpha/{\rm Fe}] \leq +0.6$ at each node of $T_{\rm eff}$, log
$g$, and [Fe/H]. Details on the models used to generate the synthetic spectra
are described in Paper I.

A linear flux interpolation routine has been added to the {\tt NGS1} approach in order
to generate synthetic spectra in finer steps of 125 K, 0.125 dex, and 0.1 dex
for $T_{\rm eff}$, log $g$, and [Fe/H], respectively, using the existing {\tt
NGS1} grid, before $\chi^{2}$ minimization calculations are carried out. This
provides a tighter parameter search space for the $\chi^{2}$ minimization scheme for
the {\tt NGS1} technique than previously.

Following these adjustments, metallicity estimation of the {\tt NGS1} and {\tt
NGS2} methods is re-calibrated using likely member stars of globular (M92, 
M15, M13, and M2) and open clusters (NGC~2420, M67, and NGC~6791), by fitting a
simple linear function of [Fe/H] to the residuals between recent literature
values and the metallicity estimates from the {\tt NGS1} and {\tt NGS2}
methods, after
adding a metal-poor globular cluster (M92) and a super solar metal-rich open
cluster (NGC~6791), which were not available at the time of the SSPP-7
calibration. The calibration procedure adopts the following
metallicities: M92 ([Fe/H] $=-2.35$), M15 ([Fe/H] $=-2.33$), M13 ([Fe/H] $=-1.58$), 
and M2 ([Fe/H] $=-1.66$) from Table A1 in Carretta et al. (2009), NGC~2420 ([Fe/H]
$=-0.37$) from Anthony-Twarog, et al. (2006), M67 ([Fe/H] $= +0.05$) from Randich
et al. (2006), and NGC~6791 ([Fe/H] $= +0.30$) from Boesgaard et al. (2009). After
re-calibration, we have obtained the following correction functions of the
metallicity scale compared to the un-calibrated values:

\begin{equation}
\rm [Fe/H]_{\tt NGS1} = \rm [Fe/H] + 0.178 \cdot \rm [Fe/H] + 0.406,
\end{equation}
\begin{equation}
\rm [Fe/H]_{\tt NGS2} = \rm [Fe/H] + 0.212 \cdot \rm [Fe/H] + 0.417.
\end{equation}

\noindent Along with the extended grid for the {\tt NGS2}, this re-calibration
has improved the final adopted metallicity in SSPP-P8 at both
the low-metallicity ($< -3.0$) and high-metallicity ($>$ 0.0)
extrema.

\subsubsection{A New Decision Tree for [Fe/H] Estimates}\label{subsubsecnewtree}

Although the basic idea of averaging the various metallicity estimates follows
the decision tree implemented in SSPP-7, we have added to the averaging scheme a few
more criteria to reject likely outliers.

There are 12 estimates of [Fe/H] in the SSPP-P8, as was also the case for the
SSPP-7. We adopt the validity ranges of S/N and \gr~listed in Table \ref{tabapptable1} 
to assign 1 or 0 as an indicator variable for each method. We then proceed
as follows. First, we generate a
synthetic spectrum for each estimate of [Fe/H] that has an indicator variable of
1 (using the adopted $T_{\rm eff}$ and log~$g$) by interpolating within the
pre-existing grid of synthetic spectra from the {\tt NGS1} approach. Next, we
calculate a correlation coefficient ({\tt CC}) and the mean of the absolute
residuals ({\tt MAR}) between the observed and the generated synthetic spectrum
in two different wavelength regions: 3850$-$4250 \AA~ and 4500$-$5500 \AA,
where the \ion{Ca}{2} K and H lines, as well as numerous metallic lines, are
present, yielding two values of {\tt CC} and {\tt MAR} for each metallicity
estimator. We then select between the two values by choosing the one with {\tt
CC} closest to unity, and with {\tt MAR} closest to zero.  This applies for all
estimates of [Fe/H] from the individual methods. At the end of this process, we
have $N$ values of the {\tt CC} and {\tt MAR} (maximum of $N=12$) for the $N$
estimates of [Fe/H] with indicator variables of 1. There are thus two arrays
with $N$ elements: one from the {\tt CC} and the other one from the {\tt MAR}
values.

We then sort the {\tt CC} array in descending order, and select the
metallicity estimate corresponding to the first and second element of the sorted
array. The same procedure is carried out for the {\tt MAR} array, after sorting
in ascending order. The reason for implementing calculations involving the {\tt
MAR}s is that, although we may have a correlation coefficient close to unity
between the observed and the synthetic spectrum, from time to time there are
large residuals between the two spectra, indicating a poor match. Thus, the
computations involving the {\tt MAR} provide additional security that the
methods are producing reasonable abundance estimates at this stage.

At this point we have two metallicity estimates with the highest {\tt CC}s, and
two metallicity estimates with the lowest {\tt MAR}s. We then take an average of
the four metallicities, and use this average to select from among the full set
of metallicity estimates with an indicator variable of 1 and within $\pm$0.5
dex of the average. We carry along the {\tt CC}s and {\tt MAR}s for the selected
metallicity estimates for further processing.

In the next step we obtain an average $\mu_{\tt CC}$ ($\mu_{\tt MAR}$) and standard
deviation $\sigma_{\tt CC}$ ($\sigma_{\tt MAR}$) of the {\tt CC}s ({\tt MAR}s)
for the surviving metallicity estimates from the previous step. As a final step
to reject likely outliers, we select from the surviving metallicity estimates
the ones with the {\tt CC} greater than ($\mu_{\tt CC}$ - $\sigma_{\tt CC}$) and
the {\tt MAR} less than ($\mu_{\tt MAR}$ + $\sigma_{\tt MAR}$). The 
metallicity estimators that remain after this step are assigned indicator variables of 2.
This procedure effectively ignores metallicity estimates that produce poor
matches with the synthetic spectra. The final adopted value of [Fe/H] is
computed by taking a biweight average of the remaining values of [Fe/H] (those
with indicator variables of 2).

Figure \ref{figappfig1} shows comparisons of metallicity estimates from individual methods with
those from the high-resolution spectroscopic analysis, and confirms how well the
new outlier rejection algorithm works.  When inspecting such plots, it is well
to keep in mind that one can assume that the high-resolution predictions of
metallicity carry {\it at least} an internal error on the order of 0.1 dex, and
(since they were not obtained from a uniform analysis), a user-to-user error
that may be of similar magnitude when various samples are combined as we have
done.

\subsection{Minor Changes in the SSPP}\label{subsecssppminorchgs}

Although the minor additions (or subtractions) to the SSPP do not alter
the parameter estimations, they greatly assist the user interested in
being made aware of peculiarities in the spectra or poorly determined
parameters.

\subsubsection{A New Color-based Temperature Estimate}\label{subsubsecnewTest}

It has proven useful to provide a new estimated $T_{\rm eff}$ based on
$(g-z)_{0}$, as it has a longer baseline than other colors (e.g., \gr).
This is especially useful for redder stars. After a careful calibration
procedure using likely cluster members and the high-resolution
calibration stars, we have derived the following two color-temperature
relations, which are applicable over two different metallicity ranges.

\noindent For [Fe/H] $< -1.5$,
\begin{equation}
    T_{\rm eff} = 6993 - 2573\cdot(g-z)_{0} + 530.9\cdot(g-z)_{0}^{2}
\end{equation}

\noindent and for [Fe/H] $ \ge -1.5$,

\begin{equation}
    T_{\rm eff} = 6947 - 2480\cdot(g-z)_{0} + 509.3\cdot(g-z)_{0}^{2}.
\end{equation}

\noindent The typical error in $T_{\rm eff}$ is less than 200 K for a dwarf with an uncertainty of
0.1 mag in $(g-z)_0$.


\subsubsection{$\chi^{2}$ Minimization with Fixed $T_{\rm eff}$ from $(g-z)_{0}$}\label{subsubsecchisq}

Another set of [Fe/H] and log $g$ estimates are obtained from {\tt
NGS1}, {\tt NGS2}, and {\tt CaIIK1}, by minimizing $\chi^{2}$ over [Fe/H]
and log $g$ after fixing the temperature determined from the $(g-z)_{0}$
approach described above. In this procedure, the H$\beta$ line is masked
out for the {\tt NGS1} and {\tt NGS2} methods. These parameters are
derived as a check on the parameters of the metal-poor cool giants, for
which the SSPP derives slightly higher temperatures (about 200~K) and
higher metallicities (about 0.3 dex), as compared to the high-resolution
analysis of the ESI spectra (Lai et al. 2009). For now, these parameters
are not considered in the final averaging step.

\subsubsection{Flux Interpolation Scheme}\label{subsubsecfluxinterp}

Since spline interpolation exhibits finer absorption features than the linear
interpolation approach, the former is employed to obtain fluxes in the linear wavelength
scale used by the SSPP than is derived from the SDSS logarithmic wavelength
scale. The synthetic spectra for the {\tt NGS1} and {\tt NGS2} grids are also
treated in the same fashion.

\subsubsection{Spectroscopy-based Parameters}\label{subsubsecspecpars}

We have chosen to output another set of adopted $T_{\rm eff}$, log $g$, and
[Fe/H] estimates, following the same decision tree as before, but only including
individual estimates for which reported colors (e.g. $(g-r)_{0}$) are not involved in
the process of their determination. These parameters are useful to compare with
the final adopted parameters for cases where the reported colors are
suspicious, just wrong, or highly reddened.

\subsubsection{New Flags for Visual Inspection}\label{subsubsecnewflags}

A flag based on a six letter combination is added to speed up the visual inspection of the
stellar spectra. Those spectra where one or more of these flags are raised are
visually inspected, while those with no flags raised (`nnnnnn') can be safely assumed to be
OK.

Definitions for each flag are as follows:

\begin{itemize}
  \item  `n' : This flag indicates nominal behavior
  \item  `F' : This flag is raised if there are no parameters or no radial velocity determined
  \item  `T' : This flag is raised if the difference in $T_{\rm eff}$ between the adopted and
               $(g-z)_{0}$ color-based $T_{\rm eff} $ is $ > $ 500~K
  \item  `t' : This flag is raised if the difference in  $T_{\rm eff}$ between the adopted and
               the spectroscopic-based  $T_{\rm eff}$ is $ >$ 500~K
  \item  `M' : This flag is raised if the difference in [Fe/H] between the adopted and
               spectroscopic-based [Fe/H] is $>$ 0.3 dex
  \item  `m' : This flag is raised if the error of the adopted metallicity is $>$ 0.3 dex
  \item  `C' : This flag is raised if the correlation coefficient is $<$ 0.4
\end{itemize}

\subsubsection{Changes on Raising Flags}\label{subsubsecflagchgs}

There have been some flags added and some dropped among the conventional SSPP flags. Table 
\ref{tabapptable2} lists
the flag definitions used in SSPP-P8. Refer to the sixth column of the table to see if a
flag is added or has dropped out. Note that the `P' flag now has a different
meaning than in SSPP-7, and the `N' flag is replaced by `X'.

\subsection{Comparison with High-Resolution Spectroscopic Observations}\label{subsechirescomp}

In addition to the high-resolution sample used to validate SSPP-7 (Allende
Prieto et al. 2008), we have continued to add to the sample of SDSS/SEGUE stars
that have been observed with high-dispersion spectrographs on various large
telescopes, such as HET, KECK, SUBARU, and the VLT. Table \ref{tabapptable3}
summarizes the current sample of the high-resolution spectroscopy for SDSS and
SEGUE stars. The ESI, SUBARU, and VLT data obtained since DR7 were analyzed by
David Lai, Wako Aoki, and Piercarlo Bonifacio, respectively, who kindly provided
their derived parameters in advance of publication.

Among about 340 stars, after removing problematic spectra, for example, those
with low $S/N (< 20/1)$, we have 244 stars to compare with the parameters
derived from SSPP-P8. Figure \ref{figappfig2} shows a summary of these results
(based on the adopted parameters only), including a comparison with SSPP-7. The
grey dots and lines denote the comparisons with SSPP-P8, while the black dots and
lines indicate comparisons with SSPP-7. The different total number of the stars
to compare arises from the different number of high-resolution spectra available
at the time of running each version of the SSPP. The reason for the much lower
number of stars in the gravity comparison is that most of the SUBARU spectra
were analyzed under the assumption of \logg~$=4.0$ (as they are mostly turn-off
stars). Therefore, they were removed in order to obtain a fair comparison.
Inspection of the plots shows that there is not much change in the
\teff~and \logg~estimates between the SSPP-7 and SSPP-P8 versions, even with the much larger
sample size now available, although the overall gravity determination is
shifted by about 0.1 dex toward higher values.

Even though there are some outliers below \feh~$< -2.0$, we can see that the
scatter above \feh~$> -1.0$ and the offset below \feh$< -2.5$ in SSPP-P8 are
smaller than those of SSPP-7. Only considering the stars with \feh~$> -1.0$ in
the comparison with the high-resolution results, we obtain a scatter of 0.14 dex
for SSPP-7 and 0.12 dex for SSPP-P8, whereas for the stars with \feh~$< -2.5$,
the offsets are 0.27 dex for SSPP-7 and 0.05 dex for SSPP-P8, with a similar scatter
of about 0.24 dex. The much smaller scatter and offset found for SSPP-P8 arises
mainly from the extended grid for the {\tt NGS2}, the re-calibration of the
metallicity scale for the {\tt NGS1} and the {\tt NGS2} methods, and the new
outlier rejection algorithm for computing the final adopted metallicity. It is
worth noting that the ``waves'' in the residuals for [Fe/H] could in principle
be empirically fit and calibrated out, but we have hesitated to do this until a
more uniform and homogeneous set of high-resolution analyses has been carried
out.

\begin{figure}
\centering
\includegraphics[scale=0.65]{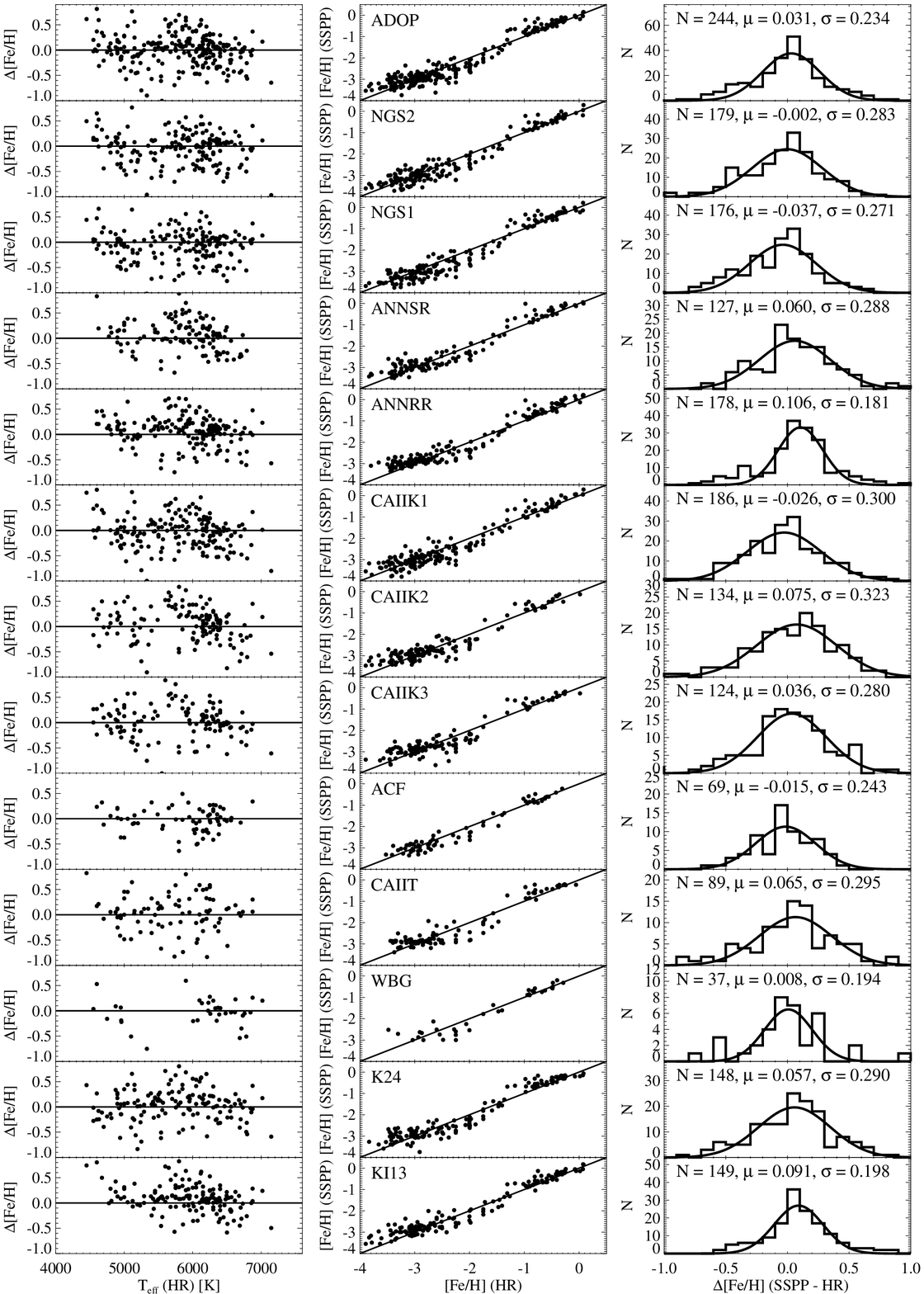}
\caption{Metallicity comparison of the individual methods in SSPP-P8 with
the metallicities obtained from high-resolution spectra. Because only the
estimators with indicator variables set to 2 are considered (except in the case
of the adopted value {\tt ADOP}), the total number of the stars differs from
method to method. These plots show how well the outlier rejection
routine works -- there are few large outliers in the individual comparisons.}\label{figappfig1}
\end{figure}

As mentioned in Section \ref{subsubsecnewtree}, Figure \ref{figappfig1} shows a comparison 
of the metallicity estimates
for each method used in the present SSPP, as a function of the
high-resolution estimates of temperature and metallicity. In this figure one can
clearly also see the evidence of very similar ``waves'' in the metallicity
residuals shown in the middle column of panels, which makes us
suspicious
that the problem lies in the high-resolution determinations, not in the
individual methods themselves, which go back to very different
individual calibration approaches.
 
\subsection{Comparison with Likely Cluster Member Stars}

Two OCs (NGC~2420, M67) and three GCs (M15, M13, and
M2) were used to calibrate and validate the parameters derived by
SSPP-7 (Paper II). Since there was only one metal-rich cluster near
solar metallicity (M67) and one metal-poor cluster (M15) included, at
the high-metallicity and low-metallicity ends SSPP-7 was not well-calibrated, as
can be seen in Figure \ref{figappfig3}. However, thanks to adding two more clusters to the list from
Paper II (NGC~6791 and M92), one super metal-rich open cluster ([Fe/H] $=
+0.3$), and another metal-poor globular cluster ([Fe/H] $=
-2.35$), respectively, we are able to re-calibrate the individual
pipelines in the SSPP, with the help of the high-resolution spectra for
many stars with [Fe/H] $< -3.0$. Figure \ref{figappfig4} shows the results of the
calibration and the comparison. One can see that at both the metal-poor
and metal-rich ends, SSPP-P8 reproduces the literature values very well,
throughout the entire metallicity range shown in the figure.


\subsection{Summary}\label{subsecsum}

We have described major and minor changes made to the SSPP since the DR7 version.
There are three major changes: 1) an extended grid for {\tt NGS2} has
been added, 2) a re-calibration for {\tt NGS1} and {\tt NGS2} has been
performed, including four GCs (M92, M15, M13, and M2) and three OCs 
(NGC~2420, M67, and NGC~6791), along with the aid of SDSS/SEGUE stars for which high-resolution
spectra were obtained, and 3) a new outlier rejection scheme has been introduced.

\begin{figure}
\centering
\includegraphics[scale=0.8]{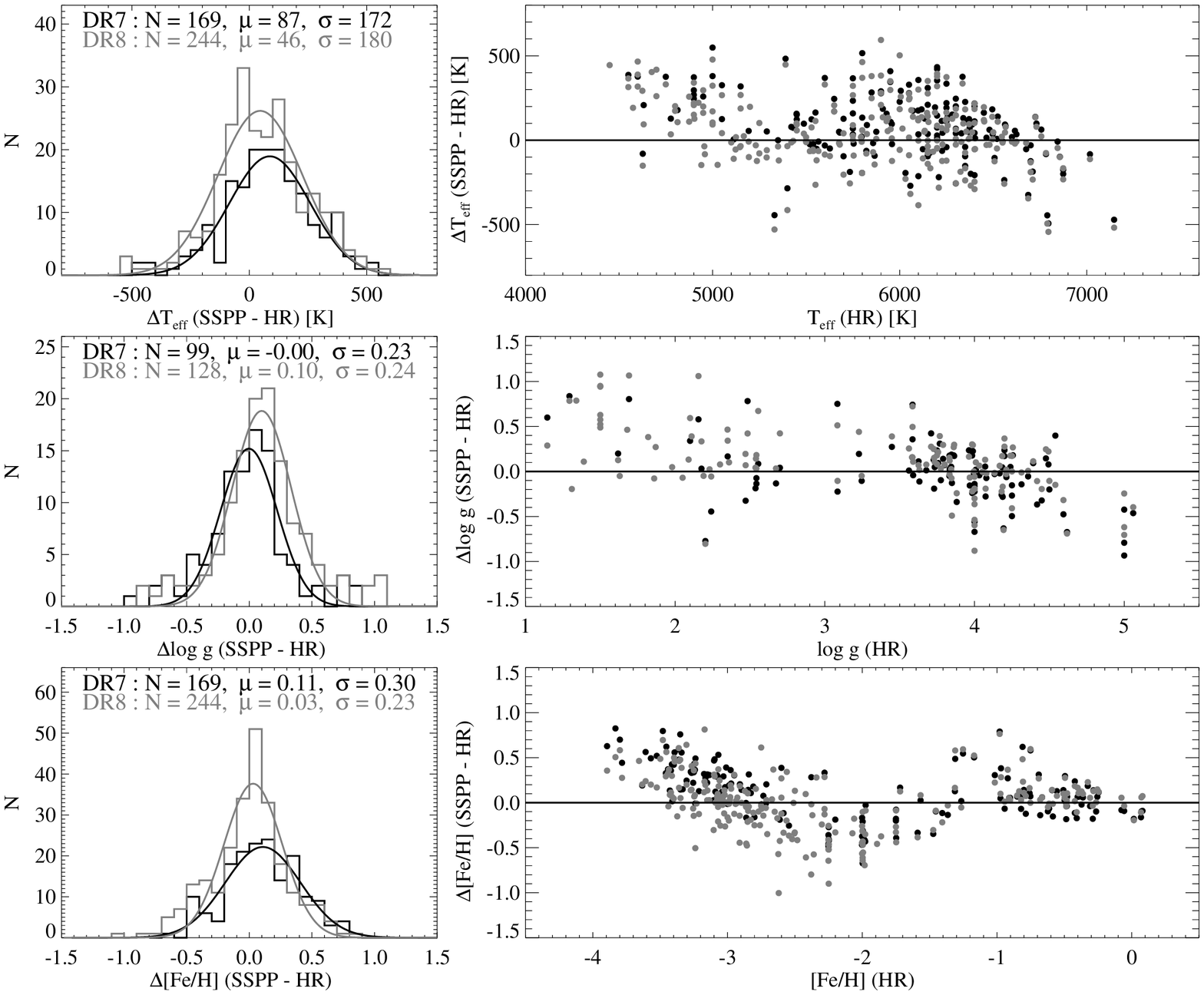}
\caption{Comparison of SSPP-7 with SSPP-P8 for the high-resolution
calibration stars. The grey dots and lines are associated with SSPP-P8, while
the black dots and lines correspond to SSPP-7. Although there are still
outliers, overall there is substantial improvement in estimation of [Fe/H]
in SSPP-P8, as can be seen at the top of the lower-right panel.  The offset is reduced
by 0.08 dex and the scatter by 0.07 dex from SSPP-7 to SSPP-P8.  Note that the 
high-resolution results are (unfortunately) not all derived
in a homogeneous manner, a defect that hopefully will be remedied soon, based on
work in progress.  In particular, we believe that the  ``waves'' in the
metallicity estimates arise, not due to inconsistencies in the SSPP, but
rather, due to the inhomogenous high-resolution analyses. }\label{figappfig2}
\end{figure}

\begin{figure}
\centering
\includegraphics[scale=0.9]{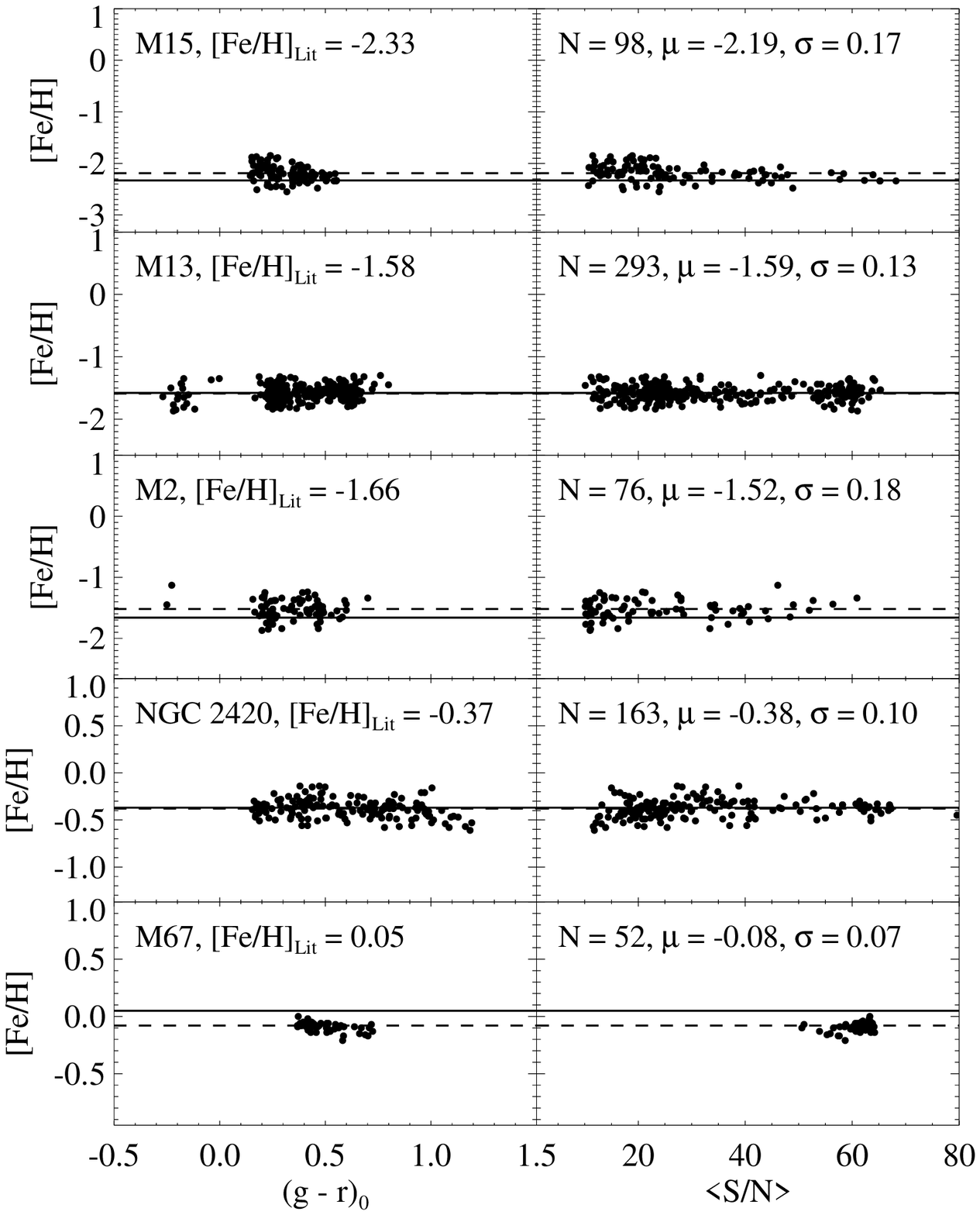}
\caption{Comparison with true cluster member stars based on SSPP-7. The
solid line indicates the literature value, while the dashed line is the
average value reported by SSPP-7 for a given cluster. Note that there exist slight
offsets between the overall mean of SSPP-7 estimates and the literature values for
M15, M2, and M67.}\label{figappfig3}
\end{figure}

\begin{figure}
\centering
\includegraphics[scale=0.9]{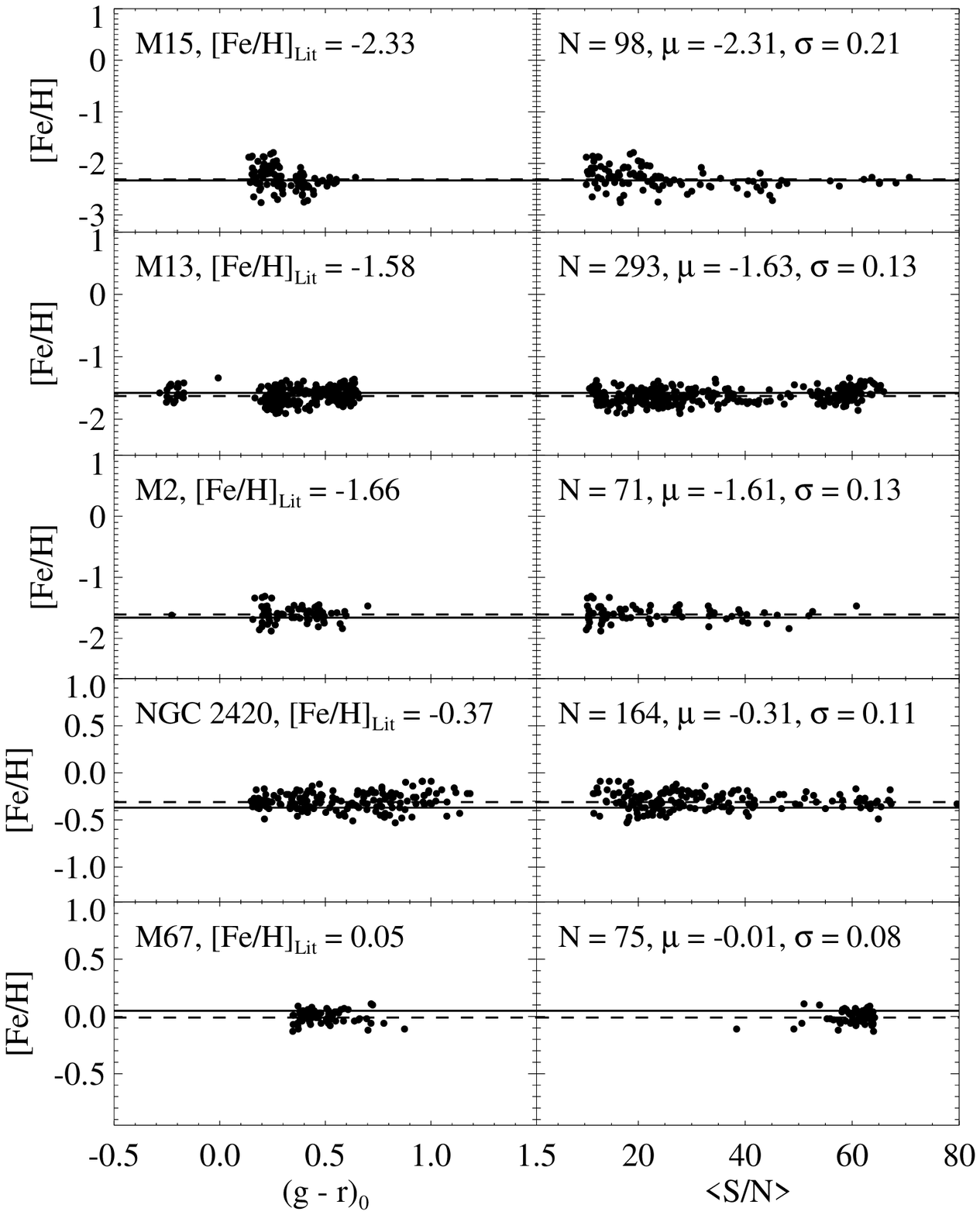}
\caption{Comparison with true cluster member stars based on SSPP-P8.
The solid line indicates the literature value, while the dashed line is the
average value reported by SSPP-P8 for a given cluster. Comparing with the
SSPP-7 plot shown in Figure \ref{figappfig3}, note that the slight
offsets between the
overall means of SSPP-P8 and the literature values for M15, M2, and
M67 are much smaller. 
}\label{figappfig4}
\end{figure}

With the implementation of these major changes, an overall improvement
for estimation of [Fe/H] has been obtained for SSPP-P8. In particular,
estimates at high and low metallicities have been much improved,
compared to SSPP-7. Adopting the intrinsic error in [Fe/H] for the HET
data described in Paper II as a typical internal error for the
high-resolution analysis (0.049 dex), and 0.23 dex in the lower panel
of Figure \ref{figappfig2} as the SSPP-P8 metallicity error, this results with an error
of 0.225 dex for the metallicity after subtracting the errors in
quadrature. Similarly, for gravity estimates, the error of the HET
high-resolution spectra is 0.129 dex; accepting 0.24 dex as the SSPP-P8
error, and taking a quadratic subtraction of the two errors, including the 0.1 dex
offset in SSPP-P8 as shown in the second panel of Figure \ref{figappfig2}, we obtain
an expected error of 0.225 dex. Considering that the SDSS/SEGUE spectra are
rather low resolution, these error estimates are remarkably good. They
would be even lower if we had a more uniform analysis of the
high-resolution spectra available, a process that is now underway. 


There are also various minor changes made on the SSPP. These changes
help identify peculiar spectra and those with ill-measured parameters.

The calibration effort to improve parameter estimation of the SSPP
will continue, focusing in particular on super metal-rich dwarfs, very
low-gravity stars, low-metallicity stars, and cooler stars.

\clearpage
\begin{deluxetable}{llrllrllrrrl}
\tablecolumns{10}  \tablewidth{0in}
\tabletypesize{\scriptsize}
\renewcommand{\tabcolsep}{3pt} \tablecaption{Valid Ranges of $g-r$ and $S/N$ for
Individual Methods in the SSPP-P8}
\tablehead{\multicolumn{3}{c}{$T_{\rm eff}$} &
\multicolumn{3}{c}{$\log g$} & \multicolumn{3}{c}{[Fe/H]} & \multicolumn{2}{c}{$S/N$} & \colhead{Reference} \\
\colhead{Name} & \colhead{Method} & \colhead{$g-r$} &
\colhead{Name} & \colhead{Method} & \colhead{$g-r$} &
\colhead{Name} & \colhead{Method} & \colhead{$g-r$} & \colhead{} &  \colhead{} & \colhead{}}
\startdata
T1    & \tt ki13    &    0.0 $-$ 0.8 &  G1  & \tt ki13    &    0.0 $-$ 0.8   & M1   & \tt ki13   &    0.0 $-$ 0.8 & & $\geq$ 15 & \S 4.1\\
T2    & \tt k24     &    0.0 $-$ 0.8 &  G2  & \tt k24     &    0.0 $-$ 0.8   & M2   & \tt k24    &    0.0 $-$ 0.8 & & $\geq$ 15 & Allende Prieto et al. (2006)\\
T3    & \tt WBG     & $-$0.3 $-$ 0.3$^{*}$ &  G3  & \tt WBG     & $-$0.3 $-$ 0.3$^{*}$   & M3   & \tt WBG    & $-$0.3 $-$ 0.3$^{*}$ & & $\geq$ 10 & Wilhelm et al. (1999)\\
T4    & \tt ANNSR   & $-$0.3 $-$ 0.8 &  G4  & \tt ANNSR   & $-$0.3 $-$ 0.8   & M4   & \tt ANNSR  & $-$0.3 $-$ 0.8 & & $\geq$ 20 & \S 4.3\\
T5    & \tt ANNRR   & $-$0.3 $-$ 1.2 &  G5  & \tt ANNRR   & $-$0.3 $-$ 1.2   & M5   & \tt ANNRR  & $-$0.3 $-$ 1.2 & & $\geq$ 10 & Re Fiorentin et al. (2007)\\
T6    & \tt NGS1    & $-$0.3 $-$ 1.3 &  G6  & \tt NGS1    & $-$0.3 $-$ 1.3   & M6   & \tt NGS1   & $-$0.3 $-$ 1.3 & & $\geq$ 10$^{*}$ & \S 4.4\\
\dots & \dots\dots  & \dots\dots\dots&  G7  & \tt NGS2    &    0.0 $-$ 1.3   & M7   & \tt NGS2   &    0.0 $-$ 1.3 & & $\geq$ 20 & \S 4.4\\
\dots & \dots\dots  & \dots\dots\dots&  G8  & \tt CaI1    &    0.3 $-$ 1.2$^{*}$   & M8   & \tt CaIIK1 & $-$0.3 $-$ 0.8 & & $\geq$ 10 & \S 4.5\\
\dots & \dots\dots  & \dots\dots\dots&\dots &\dots\dots   &\dots\dots\dots   & M9   & \tt CaIIK2 &    0.1 $-$ 0.8 & & $\geq$ 10 & Beers et al. (1999)\\
\dots & \dots\dots  & \dots\dots\dots&\dots &\dots\dots   &\dots\dots\dots   & M10  & \tt CaIIK3 &    0.1 $-$ 0.8 & & $\geq$ 10 & \S 4.6\\
\dots & \dots\dots  & \dots\dots\dots&\dots &\dots\dots   &\dots\dots\dots   & M11  & \tt ACF    &    0.1 $-$ 0.9 & & $\geq$ 15 & Beers et al. (1999)\\
\dots & \dots\dots  & \dots\dots\dots&\dots &\dots\dots   &\dots\dots\dots   & M12  & \tt CaIIT  &    0.1 $-$ 0.7 & & $\geq$ 20 & Cenarro et al. (2001a,b)\\
\dots & \dots\dots  & \dots\dots\dots&  G9  & \tt CaI2    &    0.3 $-$ 1.2$^{*}$   &\dots &\dots\dots  & \dots\dots\dots& & $\geq$ 10 & Morrison et al. (2003)\\
\dots & \dots\dots  & \dots\dots\dots&  G10 & \tt MgH     &    0.3 $-$ 1.2$^{*}$   &\dots &\dots\dots  & \dots\dots\dots& & $\geq$ 10 & Morrison et al. (2003)\\
T7    & \tt HA24    &    0.1 $-$ 0.8$^{*}$ &\dots &\dots\dots   &\dots\dots\dots   &\dots &\dots\dots  & \dots\dots\dots& & $\geq$ 10 & \S 5.1\\
T8    & \tt HD24    &    0.1 $-$ 0.6$^{*}$ &\dots &\dots\dots   &\dots\dots\dots   &\dots &\dots\dots  & \dots\dots\dots& & $\geq$ 10 & \S 5.1\\
T9    & \tt T$_{K}$ & $-$0.3 $-$ 1.3 &\dots &\dots\dots   &\dots\dots\dots   &\dots &\dots\dots  & \dots\dots\dots& &   N/A  & \S 5.1\\
T10   & \tt T$_{G}$ & $-$0.3 $-$ 1.3 &\dots &\dots\dots   &\dots\dots\dots   &\dots &\dots\dots  & \dots\dots\dots& &   N/A  & \S 5.1\\
T11   & \tt T$_{I}$ & $-$0.3 $-$ 1.3 &\dots &\dots\dots   &\dots\dots\dots   &\dots &\dots\dots  & \dots\dots\dots& &   N/A  & Ivezi\'c et al. (2008)
\enddata
\tablecomments{The symbol $*$ indicates that changes have been made in the color or $S/N$ range. The section number
 listed is that from Paper I, and references therein.}\label{tabapptable1}
\end{deluxetable} 

\clearpage
\begin{deluxetable}{lrlccc}
\tabletypesize{\scriptsize}
\tablecolumns{6} \tablewidth{0pc} \tablecaption{Brief Descriptions
of SSPP Flags} \tablehead{\colhead{Position} & \colhead{Flag} & \colhead{Description} & \colhead{Category} & \colhead{Parameter} & \colhead{Status}}
\startdata
First & & & & & \\
 &n & Appears normal  & \dots\dots  & Yes & \dots\dots \\
 &D & Likely white dwarf & Critical & No & \dots\dots \\
 &d & Likely sdO or sdB & Critical & No & \dots\dots \\
 &H & Hot star with $T_{\rm eff} >$ 10,000 K & Critical & No & \dots\dots \\
 &h & Helium line detected, possibly very hot star & Critical & No & \dots\dots \\
 &l & Likely late type solar abundance star & Cautionary & Yes & \dots\dots \\
 &E & Emission lines in spectrum & Critical & No & \dots\dots \\
 &S & Sky spectrum & Critical & No & \dots\dots \\
 &V & No radial velocity information  & Critical& No & \dots\dots \\
 &N & Very noisy spectrum & Cautionary& Yes & \dots\dots \\[+5pt]
\hline\\[-5pt]
Second & & & & \\
 &n & Appears normal  & \dots\dots  & Yes & \dots\dots \\
 &C & The photometric $g-r$ color may be incorrect & Cautionary & Yes & \dots\dots \\
\hline\\[-5pt]
Third & & & & & \\
 &n & Appears normal  & \dots\dots  & Yes & \dots\dots \\
 &B & Unexpected H$\alpha$ strength predicted from H$\delta$ & Cautionary& Yes & \dots\dots \\
 &b & If d or D flag is not raised among stars with B flag & Yes & Add \\
\hline\\[-5pt]
Fourth & & & & & \\
 &n & Appears normal  & \dots\dots  & Yes  & \dots\dots \\
 &G & Strong G-band feature & Cautionary & Yes & \dots\dots \\
 &g & Mild G-band feature & Cautionary & Yes & \dots\dots \\
\hline\\[-5pt]
Fifth & & & & \\ 
 &n & Appears normal  & \dots\dots  & Yes & \dots\dots  \\
 &P & Parameters reported for 5.0 $\le$ $S/N$ $<$ 10.0 &Cautionary & Yes & Drop \\
 &N & No parameters  & Critical  & No & Drop  \\
 &B & Too blue ($(g-r)_{0} < -0.3$) to estimate parameters  & Critical  & No & Add  \\
 &R & Too red ($(g-r)_{0} > 1.3$) to estimate parameters  & Critical  & No & Add  \\
 &X & No parameters estimated  & Critical  & No & Add  \\
 &c & Correlation coefficient $<$ 0.4  & Cautionary & Yes & Add  \\
 &T & Different between adopted $T_{\rm eff}$ and $(g-z)_{0}$-based $T_{\rm eff} > 500$ K  & Cautionary & Yes & Add  \\
 &P & Possible predicted $(g-r)_{0}$ is wrong  & Cautionary & Yes & Add  \\
\hline\\[-5pt]
RV & & & & & \\
 &NORV & No radial velocity information        & \dots \dots & No  & \dots\dots \\
 &ELRV & Radial velocity from ELODIE template  & \dots \dots & Yes  & \dots\dots \\
 &BSRV & Radial velocity from {\tt spectro1d}  & \dots \dots & Yes  & \dots\dots \\
 &RVCAL & Radial velocity calculated from SSPP & \dots \dots & Yes  & \dots\dots \\
\enddata
\tablecomments{No parameters are reported when `Critical' flags are raised.}\label{tabapptable2}
\end{deluxetable}

\clearpage
\begin{deluxetable}{lrccc}
\tablecolumns{5} \tablewidth{0pc}
\tabletypesize{\scriptsize}
\renewcommand{\tabcolsep}{3pt} \tablecaption{Updated List of High-Resolution
Spectra for SDSS and SEGUE Stars}
\tablehead{\colhead{Telescope} & \colhead{Instrument} &
\colhead{Resolving} & \colhead{Wavelength} &
\colhead{Number} \\
\colhead{} & \colhead{} & \colhead{power} & \colhead{coverage (\AA)}
& \colhead{of stars} } \startdata
KECK - I &  HIRES & 45000 &  3800$-$10000  & 11 \\
KECK - II&  ESI   & 6000  &  3800$-$10000  & 51 \\
HET      & HRS    & 15000 &  4400$-$8000  & 110 \\
SUBARU   & HDS    & 45000 &  3200$-$8000  & 151 \\
VLT      & UVES   & 60000 &  3300$-$8000 & 20 \\
\enddata
\label{tabapptable3}
\end{deluxetable}

\clearpage

\clearpage

\clearpage

\clearpage


\end{document}